\documentclass[fleqn,usenatbib]{mnras}

\usepackage{graphicx}
\usepackage{dcolumn}
\usepackage{bm}
\usepackage{graphics, caption}
\usepackage[utf8]{inputenc}
\usepackage[T1]{fontenc}
\usepackage{amsmath,booktabs,braket,cleveref,mathptmx,xcolor,siunitx}

\usepackage{rotating}

\usepackage[version=4]{mhchem} 
\usepackage[normalem]{ulem}
\usepackage{array}

\newcommand{\abinitio}{\emph{ab initio}}
\newcommand{\MARVEL}{{\sc Marvel}}
\newcommand{\Marvel}{{\sc Marvel}}

\newcommand{\Mollist}{{\sc Mollist}}
\newcommand{\MOLLIST}{{\sc Mollist}}

\newcommand{\cm}{cm$^{-1}$}

\def\a0{{$a_{\rm 0}$}}

\newcommand{\alert}[1]{\textcolor{black}{ #1}}

\newcommand{\mc}{\multicolumn}
\newcolumntype{H}{>{\setbox0=\hbox\bgroup}c<{\egroup}@{}}

\newcolumntype{d}{D{.}{.}{-1}}

\usepackage{graphicx} 
\usepackage[T1]{fontenc}
\usepackage{lmodern}
\usepackage[english]{babel}
\usepackage[autostyle]{csquotes}
\usepackage[]{physics}
\usepackage{xcolor}
\usepackage{caption}
\usepackage{siunitx}
\usepackage{booktabs}
\usepackage{multirow}
\usepackage{cleveref}
\usepackage{tabularx}
\usepackage[normalem]{ulem}
\useunder{\uline}{\ul}{}

\newcommand{\X}{X~${}^2\Sigma^+$}
\newcommand{\A}{A~${}^2\Pi$}
\newcommand{\B}{B~${}^2\Sigma^+$}
\newcommand{\D}{D~${}^2\Pi$}
\newcommand{\E}{E~${}^2\Sigma^+$}
\newcommand{\F}{F~${}^2\Delta$}
\newcommand{\Hpi}{H~${}^2\Pi$}
\newcommand{\J}{J~${}^2\Delta$}
\newcommand{\asig}{a~${}^4\Sigma^+$}

\newcommand{\noenergy}{8083}
\newcommand{\notrans}{40,333}
\newcommand{\noelec}{8}
\newcommand{\nospinvibronic}{221}
\newcommand{\novibronic}{80}
\newcommand{\nosources}{22}
\newcommand{\nobands}{9}
\newcommand{\selfc}{self consistent}
\newcommand{\selfcy}{self consistency}
\newcommand{\noSN}{134}

\title[\ce{^12C^14N} \Marvel{}]{Experimental energy levels of \ce{^12C^14N} through  \Marvel{} analysis}

\author[Syme and McKemmish]{
Anna-Maree Syme$^{1}$, Laura K. McKemmish$^{1}$\thanks{E-mail: l.mckemmish@unsw.edu.au}
\\
$^{1}$School of Chemistry, University of New South Wales, 2052, Sydney, Australia
}
\date{Accepted XXX. Received YYY; in original form ZZZ}

\pubyear{2020}

\begin{document}
\label{firstpage}
\pagerange{\pageref{firstpage}--\pageref{lastpage}}
\maketitle

\begin{abstract}
 The cyano radical (CN) is a key molecule across many different factions of astronomy and chemistry. Accurate, empirical rovibronic energy levels with uncertainties are determined for 8 doublet states of CN using the \Marvel{} (Measured Active Rotational-Vibrational Energy Levels) algorithm. \notrans{} transitions were validated from \nosources{} different published sources to generate \noenergy{} spin-rovibronic energy levels. The empirical energy levels obtained from the \Marvel{} analysis are compared to current energy levels from the \Mollist{} line list. The \Mollist{} transition frequencies are updated with \Marvel{} energy level data which brings the frequencies obtained through experimental data up to \alert{77.3\%} from the original 11.3\%, with 92.6\% of the transitions with intensities over 10$^{-23}$ cm/molecule at 1000 K now known from experimental data. At 2000 K, 100.0\% of the partition function is recovered using only \Marvel{} energy levels, while 98.2\% is still recovered at 5000 K.
 
\end{abstract}


\section{Introduction}


The cyano radical (CN) is one of the most important free radicals and is a key molecule in astronomy. CN was one of the first molecules observed in the interstellar medium back in 1940  \citep{40Mc.CN} and was observed extra-galactically in 1988 \citep{88HeMaSc.CN}.  
Relative to molecular hydrogen, CN has an abundance of around $10^{-9}$ \citep{13McWaMa.abun, 03JoBoDi.CN} in molecular clouds, which is comparable to other radicals, exceeded mainly by the OH radical. Since its first observation, CN had been also observed in many other astrophysical environments \citep{13ScKrWe.CN, 16SnCoKo.CN, 17LaBrSt.CN, 18Mc.CN}. The cyano free radical is a significant molecule in cometary science \citep{17ShKaKo.CN} with its presence and origin not yet completely understood \citep{05FrBeCo.CN}. Applications of CN in astronomy have included determining the temperature of the Cosmic Microwave Background  \citep{12Le.CN}, the formation of galaxies \citep{08BeCaVa.CN} and stars \citep{07RiPaRo.CN, 17JuJoHe.CN}, and the abundance of elements \citep{13SmCuSh.CN, 15RiFeLa.CN}. CN is a principal factor in modelling the growth process of Titan's atmosphere \citep{06Wo.CN}. The presence of CN is important for determining the isotopic ratios and abundances for both Carbon and Nitrogen in astrophysical environments \citep{15RiFeLa.CN, 19HaKaKo.CN}. With an ionisation energy of 112,562.7 \cm{}, the CN radical is suggested to have very slow reactions in the cold regions of interstellar media \citep{17GaBoGa.CN}.

Outside of astronomy, the CN radical is important in chemistry, most notably in high energy environments such as plasma \citep{11PeDiXi.CN}, and combustion \citep{19XuDeHe.CN}, but also for its properties of adsorption on boron nitride nanotubes  \citep{13SoMoBa.CN}. The cyano radical is significant in prebiotic chemistry, as CN is a key intermediate in the production of HCN which is considered central to the origin of life  \citep{06Wo.CN, 17FeKuKn.CN}.

Accurately modelling observations of astronomical or other gaseous environments with CN, and thus understanding these environments, requires high accuracy line lists \citep{13SmCuSh.CN, 17ShKaKo.CN, 19HaKaKo.CN} - i.e. details of all the energy levels in CN and the strength of transitions between these levels. For $^{12}$C$^{14}$N, the most accurate available  data is the  MoLLIST line list  \citep{14BrRaWe.CN}, which considers transitions between the three lowest electronic states of CN, the \X{}, \A{} and \B{} states. Full details of how this line list was constructed are deferred until section \ref{subsec:linelist}, but briefly these line list frequencies were computed using the traditional model, i.e. fitting experimental transition frequencies to a model Hamiltonian using PGopher to obtain a set of spectroscopic constants which are then used to predict unobserved line frequencies. The MoLLIST traditional model interpolates very accurately but does not extrapolate well because it is based on perturbation theory  \citep{20Be.CN}.

In our research, we became interested in CN as a potential probe for testing the variation in the proton-to-electron mass ratio  \citep{19SyMoCu.CN} based on the near degeneracy of its vibronic levels in the \A{} and \X{} states. However, testing this prediction accurately requires an \abinitio{} model of the molecule's spectroscopy, i.e. a set of potential energy and coupling curves for which the nuclear motion (vibration-rotation) Schrodinger equation can be solved with slightly different proton masses to determine the sensitivity of different transitions to a variation in the proton-to-electron mass ratio. 
Lacking this model and given the importance of CN astronomically, we decided to embark on the current paper's goal of collating then validating all available experimental data for CN in a \Marvel{} process, enabling the generation of a full set of experimentally-derived energy levels that can be used in the future to create an \abinitio{} ExoMol-style line list for CN. A subset of the data we collect here was used to develop the \Mollist{} line list, but our work here is a much more extensive compilation that also includes high-lying electronic states not considered in the CN MoLLIST data. This \Marvel{} compilation will be particularly important for high-resolution studies which rely on very accurate line positions. 


\begin{figure}
    \centering
    \includegraphics[width = 0.48\textwidth]{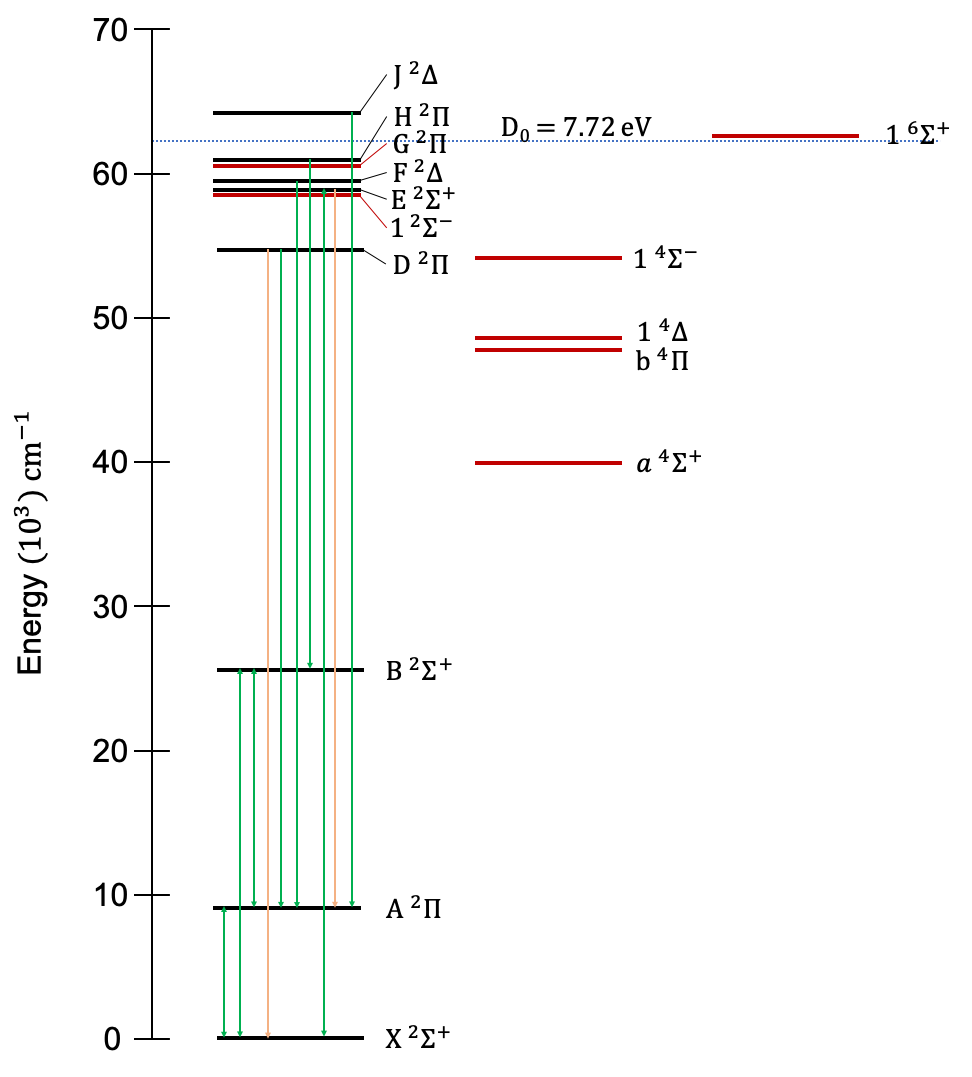}
    \caption{The electronic states and bands in CN. States indicated by the black lines have been experimentally observed. Red lines are electronic states that have been theorised to exist (either by experimental perturbation or theory), but with no available experimental transition data. The vertical green lines indicate electronic bands that have been included in this study, with the orange lines being bands that have been observed, but were not of a suitable nature for this \Marvel{} procedure.  An additional theoretically predicted $1^6\Pi$ state around 96,000 \cm{}  \citep{18YiShSu.CN} was not included in this figure for brevity. The dissociation energy of CN is shown as a blue dashed line  \citep{94PrPaBa.CN}.}
    \label{fig:bands}
\end{figure}

The CN radical is one of the most spectroscopically studied radicals with observations dating back almost 100 years \citep{28Je.CN}. Most studies focus on the a combination of the three lowest electronic states (\X, \A, and \B) that contribute to the visible spectroscopic bands. The full electronic states of CN and observed transition bands can be seen in figure \ref{fig:bands}. Below 30,000 \cm, the electronic structure of CN comprises of a ground state of \X{} followed by the \A{} state around 9145 \cm{}, before a larger gap up to the \B{} state around 25790 \cm. These lower electronic states contribute to the common CN bands; the red band (\A-\X) and the violet band (\B-\A), which have been observed many times \citep{32JeRoMu.CN, 55DoRo.CN, 73Sc.CN,  79ScSu.CN,81GoSa.CN, 89FuAlDa.CN, 92PrBe.CN, 92PrBeFr.CN, 92ReSuMi.CN, 94ItKaKu.CN,   01LiDuLi.CN,  06RaDaWa.CN, 09HaHaSe.CN, 08CiSeKu.CN, 10RaWaBe.CN}. The \B-\A{} band is significantly less studied but is an important band in demonstrating that CN could not be treated as a symmetric molecule with $u$ and $g$ symmetry as originally hypothesised \citep{70Lu.CN, 89FuAlDa.CN}.  The electronic states above 30,000 \cm{} and the associated bands are not as well known. These electronic states are visualised along with the lower electronic states in figure \ref{fig:bands}. Only the doublet states \D, \E, \F, \Hpi, and \J{} have experimentally assigned transitions from older sources \citep{55DoRo.CN, 56Ca.CN, 70Lu.CN} however, none of these transitions are doublet resolved. There are several states of the CN radical that have been computationally \citep{09KuStFi.CN,  11ShLiSu.CN, 18YiShSu.CN} or experimentally \citep{84ItOzNa.CN} predicted that have no observed transitions. These unobserved states include two possible doublet states around 60,000 \cm  \citep{70Lu.CN, 11ShLiSu.CN}, as well as the quartet and sextet states, and are shown in figure \ref{fig:bands}. The \asig{}  state is the most established of these unobserved states due to the perturbations it causes in lower electronic states, such as the \B{} state. 

The electronic structure of CN might appear at first to be straightforward as the low-lying energy levels are well separated; however, there are strong couplings between states that cause large perturbations affecting many bands. These perturbations have been studied extensively in the 1980's  \citep{80KoFiSt.CN, 84ItOzNa.CN, 94ItKaKu.CN, 83OzItSu.CN, 82GoSa.CN, 83OzNaSu.CN, 87ItOzSu.CN} with the Kuchitsu group's many studies on  the high vibrational \B-\X{} bands particularly notable. Near degeneracies of the (\A{} $v$ = 7, \X{} $v$ = 10) and (\B{} $v$ = 0, \A{} $v$ = 10) states cause particularly significant perturbations. Further, the high vibrational levels of the \B{} state are rife with perturbations due to coupling to the high vibrational \A{} states and the 'dark' quartet states \citep{06RaDaWa.CN, 87ItOzSu.CN, 84ItOzNa.CN}. These quartet states, shown in figure \ref{fig:bands} have not been directly observed experimentally, but their properties have been inferred to high precision based on how they perturb observed states \citep{84ItOzNa.CN}.

The astronomical importance of CN and the substantial perturbations in its spectroscopy have lead to significant \abinitio{} investigations into the molecule, with the most accurate calculations being multi-reference configuration interaction (MRCI) calculations with large basis sets, e.g.  \citep{88BaLaTa.CN,89FuAlDa.CN,09KuStFi.CN,11ShLiSu.CN,14BrRaWe.CN,18YiShSu.CN} that cover a range of electronic states. A recent paper  \citep{18YiShSu.CN} provides an excellent summary of \abinitio{} modelling of CN. In the context of this paper, the most important \abinitio{} data to create astronomical line lists are the transition dipole moment and spin-orbit coupling curves; the \Marvel{} experimentally-derived energy levels produced in this paper provide a much better data source for creating spectroscopically-accurate potential energy curves than \abinitio{} data (e.g. see  \citep{Tennyson2016TheMolecules,19McMaHo.CN}). 
New transition dipole moment curves were calculated to enable the production of the  \citep{14BrRaWe.CN} line list. Furthermore, in a still rare recognition of the importance for astronomical applications of \abinitio{} TDM curves over the more easily calculated but far less useful potential energy curves, we would like to highlight and applaud the recent work from  \citep{18YiShSu.CN} which expanded the high-accuracy transition dipole moment (TDM) data for CN from the \X{}, \A{} and \B{} states to five quartet and two doublet states, though unfortunately other doublet states with energies less than the some of the quartet and sextet states were not included presumably due to the congestion of doublet states around $T_e$=60,000 \cm. 

With \abinitio{}, newer calculations usually give the best results, but for experimental spectroscopy, the breadth and coverage of data is most important and the complexity of the data means it has been collated over decades. Bringing all this data together in one place in a usable consistent data format, cross-validating the data and extracting meaningful information from the summation of all experimental data is beneficial and useful. Through a series of papers by many authors, this data compilation, cleaning and validation process has been standardised, routinely using the \Marvel{} software program to convert the collated input assigned transition frequencies with uncertainties into output energy levels with uncertainties, with both input and output files stored online in a single \Marvel{} website which now contains data for 15 molecules. This centralised data repository and consistent format means that future experiments can easily update existing knowledge with their new data and this data can be easily used to create and update line lists used by astronomers.

Within this paper, we bring together all currently available experimental high-resolution spectra of CN to determine accurate empirical energy levels with reliable uncertainties using the Measured Active Rotational-Vibrational Energy Levels (\Marvel) approach  \citep{07FuCsTe.CN}. In section \ref{sct:method}, we review the \Marvel{} procedure and the quantum numbers used as labels. The experimental data was prepared in a standardised format, then validated for self-consistency and processed to produce experimentally-derived energy levels using the \Marvel{} procedure. The transition data is discussed in section \ref{sct:trans} while the resultant energy levels are evaluated in section \ref{sct:energy}. In section \ref{sct:comp}, we compare the \Marvel -derived energy levels with those used in the MoLLIST line list, and improve the current line list by \Marvel ising the MoLLIST states file.








\section{\Marvel{} procedure for CN}\label{sct:method}
The Measured Active Rotational-Vibrational Energy Levels (\Marvel) approach  \citep{07FuCsTe.CN} uses an algorithm to invert experimentally assigned transitions to empirical rovibrationic energy levels. The \Marvel{} procedure uses graph theory, creating Spectroscopic Networks (SN)   \citep{11CsFu.CN} containing all connected energy levels from the experimental transitions.  Uncertainties of assigned transitions are adjusted with a weighted strategy until self-consistent. These uncertainties are then propagated through to the associated energy levels to provide reliable energy level uncertainties. 

The \Marvel{} procedure has been documented many times \citep{07FuCsTe.CN, 12FuCs.CN, 14FuArMe.CN, 13FuSzMa.CN, 13FuSzFa.CN}, and we refer the reader to the original papers for further details. The \Marvel{} procedure is well established (e.g. see  \cite{19FuHoKo.CN, 16FuSzCs.CN, 13FuSzFa.CN, 13FuSzMa.CN, 17McMaSh.CN, 18McBoGo.CN}) and has been used to determine the empirical energy levels of 15 small molecules and their isotopologues. \Marvel{} analysis of small molecules has been used to compute accurate, temperature-dependent ideal-gas thermodynamic data (e.g. \cite{19FuHoKo.CN}), to facilitate the empirical adjustment of potential energy surfaces  (e.g. \cite{18YuSzPy.CN}), and to improve the accuracy of computed line lists  (e.g. \cite{19McMaHo.CN}). The online user interface of \Marvel{} was used for this work and is available at http://kkrk.chem.elte.hu/marvelonline/.

\textbf{Quantum Numbers:} 

\begin{table*}
\centering
\caption{Sample of the \Marvel{} input file, with descriptions of the column headings. The full input file is included in the SI. \label{tab:inp}}
\begin{tabular}{p{2.0cm}p{1.8cm}p{2cm}p{0.8cm}p{0.8cm}p{2cm}p{0.8cm}p{0.8cm}p{2.6cm}}
\toprule
\mc{1}{c}{$\tilde{\nu}$} & \mc{1}{c}{$\Delta\tilde{\nu}$} & \mc{3}{c}{Upper state QNs} & \mc{3}{c}{Lower state QNs} & \mc{1}{c}{ID}\\
\cmidrule(r){3-5}\cmidrule(r){6-8}
& & \mc{1}{c}{State$^\prime$} & \mc{1}{c}{$v^\prime$} & \mc{1}{c}{$J^\prime$} & \mc{1}{c}{State$^{\prime\prime}$} & \mc{1}{c}{$v^{\prime\prime}$} & \mc{1}{c}{$J^{\prime\prime}$}\\ \midrule
9141.6439 & 0.012 & A2Pi\_e1/2 & 0 & 0.5 & X2Sig+\_f & 0 & 0.5 & 10RaWaBe.19241 \\
25838.0341	& 0.02	& B2Sig+\_e	& 0	& 9.5	& X2Sig+\_e	& 0	& 8.5	& 92ReSuMi.8 \\
23748.123 & 0.03 & B2Sig+\_f & 0 & 0.5 & X2Sig+\_f & 1 & 1.5 & 06RaDaWa.5431 \\
2019.211	& 0.0008228	& X2Sig+\_e	& 1	& 5.5 & 	X2Sig+\_e	& 0	& 6.5	& 05HuCaDa.26 \\
7091.7821 & 0.012 & A2Pi\_f1/2 & 0 & 0.5 & X2Sig+\_f & 1 & 1.5 & 10RaWaBe.17995 \\
5075.763 & 0.012 & A2Pi\_f1/2 & 0 & 0.5 & X2Sig+\_f & 2 & 1.5 & 10RaWaBe.17601 \\
52903.51 & 0.06 & E2Sig+\_f & 0 & 0.5 & X2Sig+\_f & 3 & 1.5 & 56Ca.383 \\ \bottomrule
\end{tabular}

\begin{tabular}{ccl}
\\
             Column        &     Notation                  &       \\
\midrule
   1  &    $\tilde{\nu}$              &     Transition frequency (in \cm) \\
   2  &  $\Delta\tilde{\nu}$         &    Estimated uncertainty in transition frequency (in \cm) \\
   3  &   State$^\prime$  &   Electronic state of upper energy level; also includes parity and $\Omega$ for $\Pi$ and $\Delta$ states  \\
    4  &  $v^\prime$  &  Vibrational quantum number  of upper  level \\ 
  5  &   $J^\prime$                &        Total angular momentum of upper  level   \\
   6  &   State$^{\prime\prime}$  &   Electronic state of lower energy level; also includes parity and $\Omega$ for $\Pi$ and $\Delta$ states \\
    7  &  $v^{\prime\prime}$  &  Vibrational quantum number  of lower  level \\
  8  &   $J^{\prime\prime}$                &        Total angular momentum of lower  level   \\
   9  &  ID  &  Unique ID for transition, with reference key for source and counting number \\
\bottomrule
\end{tabular}

\end{table*}

\begin{figure}
    \centering
    \includegraphics[width = 0.4\textwidth]{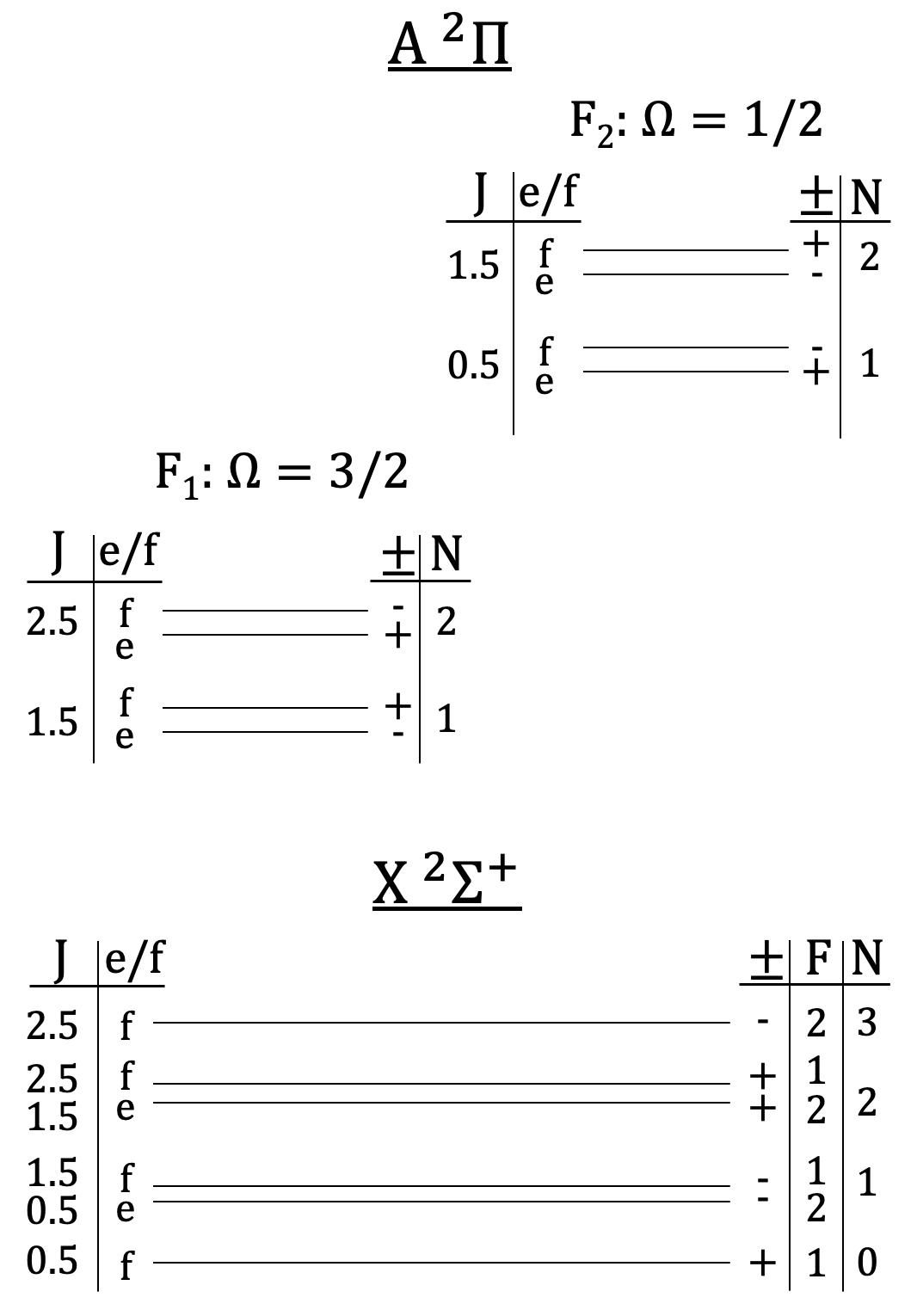}
    \caption{Scheme of quantum number for CN. The quantum numbers ($J$ and $e/f$ ) we have used as labels are given on the left hand side of the lines and the corresponding total parity ($\pm$), $F$ (Spin multiplet component of the state), and $N$ (Angular momentum without the consideration of electron spin) are given on the right hand side. The spaces between the energy levels are representative only and are not to scale. 
    }
    \label{fig:labels}
\end{figure}

The \Marvel{} procedure relies solely on labelling of quantum states. The quantum states are treated as nodes connected by transitions as directed edges in a graph and do not have any physics built in, e.g. model Hamiltonians. 

For our \Marvel{} CN compilation, each quantum state is described uniquely by a set of three quantum numbers: the State, the vibrational quantum number $v$ and the total angular momentum $J$, as shown in the input file extract in  \Cref{tab:inp}. The last two are straightforward, with the only note being that we chose to use Hund's case A labelling $J$ not $N$ (the rotational-only angular momentum) for consistency with most sources and the \Mollist{} line list. The State label, however, is more complicated as it includes not only the electronic state but also the spin splitting through the explicit inclusion of $e/f$ rotational-less parity and $\Omega$ labels when the quantum states were not degenerate. Figure \ref{fig:labels} shows an example of the labelling used for a selection of the energy levels in the \X{} and \A{} states, with both our quantum numbers and other commonly used quantum numbers included for clarity. 

\section{Experimental sources for assigned transitions}\label{sct:trans}
\subsection{Overview}
An extensive literature search was done to find, to our knowledge, an exhaustive list of experimentally assigned rotationally resolved transitions for CN. \nosources{} published sources were identified as containing assigned transitions suitable for the \Marvel{} procedure here. Throughout the literature search, no new experimental assignments were found post 2010 and the higher electronic band have been significantly less studied, with no viable high resolution data since the 1970's. 

The complete transitions input file to \Marvel{} (a sample is shown in table \ref{tab:inp}) consists of \notrans{} assigned transitions, from \nobands{} electronic bands and is included in the Supplementary Information. 
 
To obtain this file, the following procedure was followed. The rotationally-resolved assigned transitions data for each source was converted to \Marvel{} format, with each transition given a starting uncertainty based on discussion in the original paper. This process was not straightforward in many cases, so we give individual notes on each source in \Cref{subsec:individual}. The \Marvel{} procedure was then used to check for self-consistency within the source's assigned transitions, with uncertainties increased and some transitions with very large uncertainties removed until self-consistency was obtained. Then, the \Marvel{} data for each individual source were sequentially put into a single master \Marvel{} file, with uncertainties then adjusted and further transitions removed until self-consistency of the full file was obtained. Note that the transitions removed are retained in the file itself but with a "-" at the beginning of the file to indicate it is not part of the spectroscopic network obtained by the \Marvel{} procedure for the molecule. We refer to the remaining transitions as verified. 

\begin{figure}
    \centering
    \includegraphics[width = 0.48\textwidth]{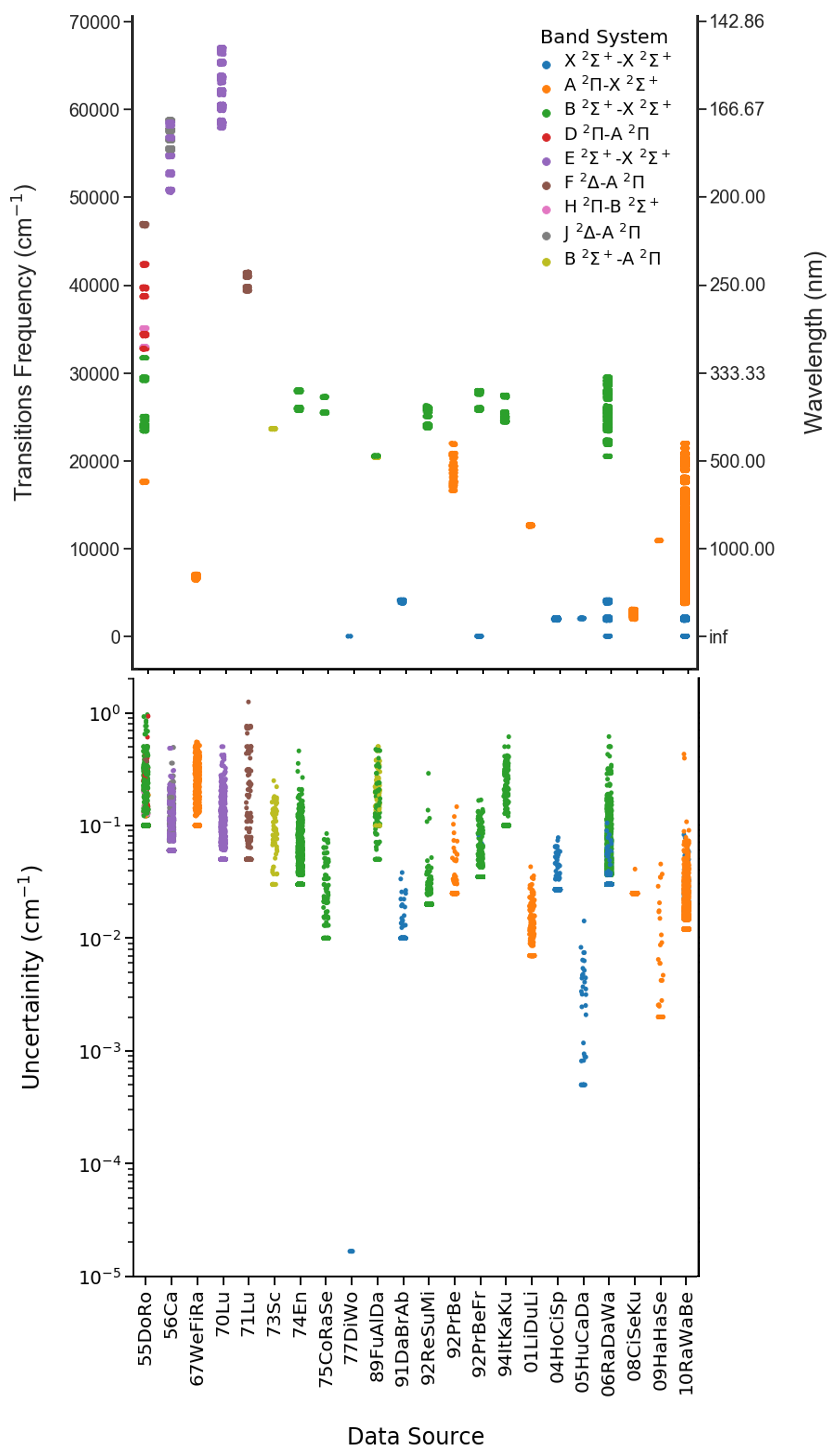}
    \caption{The spread of the frequencies (upper) and uncertainties (lower) of the transitions taken from each data source used in this \Marvel{} procedure for CN. Each electronic band is shown in a unique colour.}
    \label{fig:trans}
\end{figure}

\begin{table*}
    \centering
    \caption{Breakdown of the electronic bands of assigned transitions from 20th century sources used in this study. V is the number of verified transitions and A is the number of available transitions. The mean and maximum uncertainties are given in \cm, these are after \Marvel{} has made the transitions \selfc. }
    \label{tab:trans}
    \resizebox{\textwidth}{!}{%
    \begin{tabular}{p{2.4cm}p{1.8cm}p{2cm}p{2.8cm}p{10cm}p{1.6cm}}
          \toprule
Electronic Band &  V/A &  Mean/Max & Freq Range \cm{} & Vibrational Bands & $J$ Range  \\ \midrule
\mc{6}{l}{\textbf{55DoRo  \citep{55DoRo.CN}}} \\
F $^2\Delta$-A $^2\Pi$  & 58/58 & 0.142/0.5 & 46694.9 - 47062.5 & (0-2)  & 2.5 - 32.5 \\
D $^2\Pi$-A $^2\Pi$  & 335/335 & 0.11/0.931 & 32662.0 - 42482.0 & (0-6), (0-7), (1-4), (2-4), (3-3)  & 1.5 - 27.5 \\
A $^2\Pi$-X $^2\Sigma^+$  & 106/106 & 0.104/0.315 & 17479.4 - 17664.1 & (15-8)  & 0.5 - 21.5 \\
B $^2\Sigma^+$-X $^2\Sigma^+$  & 865/875 & 0.142/0.961 & 23344.2 - 31745.6 & (0-1), (1-2), (14-10), (16-13), (18-17), (18-18), (19-15), (19-18), (2-3), (3-4), (4-5), (5-6)  & 0.5 - 26.5 \\
H $^2\Pi$-B $^2\Sigma^+$  & 140/140 & 0.104/0.249 & 32878.4 - 35165.1 & (0-0), (0-1),  & 0.5 - 17.5 \\
\\ \mc{6}{l}{\textbf{56Ca  \citep{56Ca.CN}}} \\
E $^2\Sigma^+$-X $^2\Sigma^+$  & 529/531 & 0.089/0.484 & 50546.9 - 58595.6 & (0-1), (0-2), (0-3), (0-4), (1-1)  & 0.5 - 32.5 \\
J $^2\Delta$-A $^2\Pi$  & 978/986 & 0.076/0.492 & 55325.7 - 58870.7 & (0-0), (1-0), (2-0), (3-0)  & 1.5 - 30.5 \\
\\ \mc{6}{l}{\textbf{67WeFiRa  \citep{67WeFiRa.CN}}} \\
A $^2\Pi$-X $^2\Sigma^+$  & 377/433 & 0.228/0.55 & 6357.4 - 7107.1 & (0-1), (1-2)  & 3.5 - 55.5 \\
\\ \mc{6}{l}{\textbf{70Lu  \citep{70Lu.CN}}} \\
E $^2\Sigma^+$-X $^2\Sigma^+$  & 1012/1012 & 0.083/0.5 & 57857.7 - 67093.6 & (0-0), (1-0), (1-1), (2-0), (2-1), (3-0), (4-0), (5-0)  & 0.5 - 42.5 \\
\\ \mc{6}{l}{\textbf{71Lu  \citep{71Lu.CN}}} \\
F $^2\Delta$-A $^2\Pi$  & 199/203 & 0.153/1.247 & 39306.2 - 41496.9 & (1-6), (1-7), (2-7), (2-8)  & 0.5 - 16.5 \\
\\ \mc{6}{l}{\textbf{73Sc  \citep{73Sc.CN}}} \\
B $^2\Sigma^+$-A $^2\Pi$  & 70/72 & 0.099/0.249 & 23582.6 - 23708.4 & (7-4)  & 0.5 - 19.5 \\
\\ \mc{6}{l}{\textbf{74En  \citep{74En.CN}}} \\
B $^2\Sigma^+$-X $^2\Sigma^+$  & 1040/1060 & 0.054/0.458 & 25743.4 - 28095.2 & (0-0), (1-0), (1-1), (2-1), (2-2), (3-2), (3-3), (4-3), (4-4), (5-4), (5-5), (6-5), (6-6), (7-6), (7-7)  & 0.5 - 29.5 \\
\\ \mc{6}{l}{\textbf{75CoRaSe  \citep{75CoRaSe.CN}}} \\
B $^2\Sigma^+$-X $^2\Sigma^+$  & 128/128 & 0.025/0.085 & 25399.1 - 27322.6 & (11-10), (11-11)  & 1.5 - 23.5 \\
\\ \mc{6}{l}{\textbf{77DiWo  \citep{77DiWo.CN}}} \\
X $^2\Sigma^+$-X $^2\Sigma^+$  & 4/4 & 0.0/0.0 & 3.8 - 3.8 & (1-1), (2-2)  & 0.5 - 1.5 \\
\\ \mc{6}{l}{\textbf{89FuAlDa  \citep{89FuAlDa.CN}}} \\
B $^2\Sigma^+$-X $^2\Sigma^+$  & 75/77 & 0.148/0.473 & 20461.0 - 20605.0 & (8-11)  & 0.5 - 21.5 \\
B $^2\Sigma^+$-A $^2\Pi$  & 94/115 & 0.198/0.5 & 20352.0 - 20543.5 & (8-7)  & 0.5 - 19.5 \\
\\ \mc{6}{l}{\textbf{91DaBrAb  \citep{91DaBrAb.CN}}} \\
X $^2\Sigma^+$-X $^2\Sigma^+$  & 156/156 & 0.011/0.038 & 3743.5 - 4155.4 & (2-0)  & 2.5 - 57.5 \\
\\ \mc{6}{l}{\textbf{92ReSuMi  \citep{92ReSuMi.CN}}} \\
B $^2\Sigma^+$-X $^2\Sigma^+$  & 850/850 & 0.021/0.29 & 23720.0 - 26299.3 & (0-0), (0-1), (1-1), (1-2), (11-11), (13-13), (2-2), (2-3), (3-3), (3-4), (4-4), (4-5), (6-6), (6-7), (7-7), (7-8), (8-8), (8-9), (9-10), (9-9)  & 0.5 - 63.5 \\
\\ \mc{6}{l}{\textbf{92PrBe  \citep{92PrBe.CN}}} \\
A $^2\Pi$-X $^2\Sigma^+$  & 266/266 & 0.028/0.147 & 16572.9 - 21982.5 & (10-3), (10-4), (11-4), (11-5), (12-4), (12-5), (12-6), (13-5), (13-6), (13-7), (14-5), (14-7), (15-7), (15-8), (16-7), (16-8), (16-9), (19-10), (19-11), (20-10), (20-11), (21-10), (21-11), (8-1), (8-2), (8-3), (9-3)  & 0.5 - 6.5 \\
\\ \mc{6}{l}{\textbf{92PrBeFr  \citep{92PrBeFr.CN}}} \\
B $^2\Sigma^+$-X $^2\Sigma^+$  & 720/720 & 0.041/0.169 & 25745.3 - 28025.3 & (0-0), (1-0), (1-1), (2-1), (2-2), (3-2), (3-3), (4-3), (4-4), (5-4), (5-5), (6-5), (6-6), (7-6), (7-7), (8-7), (9-8)  & 0.5 - 26.5 \\
X $^2\Sigma^+$-X $^2\Sigma^+$  & 54/54 & 0.037/0.083 & 3.5 - 15.1 & (0-0), (1-1), (10-10), (2-2), (3-3), (4-4), (5-5), (6-6), (7-7), (8-8), (9-9)  & 0.5 - 3.5 \\
\\ \mc{6}{l}{\textbf{94ItKaKu  \citep{94ItKaKu.CN}}} \\
B $^2\Sigma^+$-X $^2\Sigma^+$  & 410/419 & 0.139/0.612 & 24383.8 - 27495.8 & (11-11), (14-14), (15-15), (16-14), (18-17), (19-18)  & 0.5 - 41.5 \\
   \bottomrule
   \end{tabular}
   }
\end{table*}

\begin{table*}
    \centering
    \caption{Breakdown of the 21st century sources and electronic bands of assigned transitions used for CN}
    \label{tab:trans2}
    \resizebox{\textwidth}{!}{%
    \begin{tabular}{p{2.4cm}p{1.8cm}p{2cm}p{2.8cm}p{11cm}p{1.6cm}}
          \toprule
Electronic Band &  V/A &  Mean/Max & Freq Range \cm{} & Vibrational Bands &$J$Range  \\ \midrule
\mc{6}{l}{\textbf{01LiDuLi  \citep{01LiDuLi.CN}}} \\
A $^2\Pi$-X $^2\Sigma^+$  & 189/189 & 0.013/0.043 & 12496.8 - 12735.0 & (2-0)  & 0.5 - 22.5 \\
\vspace{21em}\\ \mc{6}{l}{\textbf{04HoCiSp  \citep{04HoCiSp.CN}}} \\
X $^2\Sigma^+$-X $^2\Sigma^+$  & 687/695 & 0.028/0.078 & 1816.7 - 2133.6 & (1-0), (2-1), (3-2), (4-3), (5-4), (6-5), (7-6), (8-7)  & 0.5 - 31.5 \\
\\ \mc{6}{l}{\textbf{05HuCaDa  \citep{05HuCaDa.CN}}} \\
X $^2\Sigma^+$-X $^2\Sigma^+$  & 36/36 & 0.003/0.014 & 1982.1 - 2101.3 & (1-0)  & 4.5 - 17.5 \\
\\ \mc{6}{l}{\textbf{06RaDaWa  \citep{06RaDaWa.CN}}} \\
B $^2\Sigma^+$-X $^2\Sigma^+$  & 5775/5801 & 0.036/0.614 & 20441.5 - 29549.9 & (0-0), (0-1), (1-0), (1-1), (1-2), (10-10), (10-11), (10-12), (10-8), (10-9), (11-10), (11-11), (11-12), (11-13), (11-9), (12-10), (12-11), (12-12), (12-13), (12-14), (13-11), (13-13), (14-14), (15-15), (16-13), (17-14), (17-16), (18-17), (18-18), (19-15), (19-18), (2-1), (2-2), (2-3), (3-2), (3-3), (3-4), (4-3), (4-4), (4-5), (4-6), (5-4), (5-5), (5-6), (5-7), (6-5), (6-6), (6-7), (6-8), (7-10), (7-6), (7-7), (7-8), (7-9), (8-10), (8-7), (8-8), (8-9), (9-10), (9-11), (9-12), (9-7), (9-8), (9-9)  & 0.5 - 63.5 \\
X $^2\Sigma^+$-X $^2\Sigma^+$  & 1788/1788 & 0.031/0.105 & 3.5 - 4155.4 & (0-0), (1-0), (1-1), (10-10), (2-0), (2-1), (2-2), (3-1), (3-2), (3-3), (4-2), (4-3), (4-4), (5-4), (5-5), (6-5), (6-6), (7-6), (7-7), (8-7), (8-8), (9-9)  & 0.5 - 79.5 \\
\\ \mc{6}{l}{\textbf{08CiSeKu  \citep{08CiSeKu.CN}}} \\
A $^2\Pi$-X $^2\Sigma^+$  & 870/870 & 0.025/0.041 & 1905.8 - 3117.2 & (0-3), (1-4), (2-5), (3-6), (4-7), (5-8), (6-9)  & 1.5 - 31.5 \\
\\ \mc{6}{l}{\textbf{09HaHaSe  \citep{09HaHaSe.CN}}} \\
A $^2\Pi$-X $^2\Sigma^+$  & 38/38 & 0.009/0.045 & 10850.0 - 10937.3 & (1-0)  & 0.5 - 7.5 \\
\\ \mc{6}{l}{\textbf{10RaWaBe  \citep{10RaWaBe.CN}}} \\
A $^2\Pi$-X $^2\Sigma^+$  & 20695/20707 & 0.013/0.429 & 3683.4 - 22027.6 & (0-0), (0-1), (0-2), (1-0), (1-1), (1-2), (1-3), (10-5), (10-6), (11-6), (12-7), (13-7), (14-6), (14-7), (15-7), (15-8), (16-7), (16-8), (17-10), (17-8), (18-10), (18-9), (19-10), (19-11), (2-0), (2-1), (2-2), (2-3), (2-4), (20-10), (21-10), (21-11), (22-11), (22-12), (3-0), (3-1), (3-2), (3-3), (3-4), (3-5), (4-0), (4-1), (4-2), (4-4), (4-5), (4-6), (5-1), (5-2), (5-3), (5-5), (5-7), (6-2), (6-3), (6-4), (6-6), (6-7), (6-8), (7-7), (7-8), (8-3), (8-4), (8-9), (9-4)  & 0.5 - 113.5 \\
X $^2\Sigma^+$-X $^2\Sigma^+$  & 1786/1788 & 0.013/0.083 & 3.5 - 4155.4 & (0-0), (1-0), (1-1), (10-10), (2-0), (2-1), (2-2), (3-1), (3-2), (3-3), (4-2), (4-3), (4-4), (5-4), (5-5), (6-5), (6-6), (7-6), (7-7), (8-7), (8-8), (9-9)  & 0.5 - 79.5 \\
   \bottomrule
   \end{tabular}
   }
\end{table*}

The spectroscopic data from the \nosources{} sources and results from the \Marvel{} analysis are summarised in figure \ref{fig:trans} and detailed in tables \ref{tab:trans} and \ref{tab:trans2}. 

Figure \ref{fig:trans} shows the frequency coverage of spectral data below 30,000 \cm{} is quite exhaustive due primarily to 06RaDaWa \citep{06RaDaWa.CN} and 10RaWaBe \citep{10RaWaBe.CN}, but is sparser at higher frequencies. Both 06RaDaWa and 10RaWaBe focus on a comprehensive global fit of a single band (\B-\X{} and \A-\X{} respectively) and equilibrium spectroscopic constants,  performing new analysis of existing but unpublished experimental data. The \Marvel{} approach in this paper provides a different, complementary perspective on this data by using graph theory rather than model Hamiltonian fits. 

The lower part of figure \ref{fig:trans} highlights the spread of uncertainties for each source that are produced through the \Marvel{} procedure. This figure is slightly skewed by the incredibly small uncertainty of the hyperfine transitions from 77DiWo \citep{77DiWo.CN}. Besides the outlier of 77DiWo, most of the data sources from before the 1990s have a much higher starting uncertainty, however, the spread of uncertainties is much the same as we go through more modern assignments. From the transitions between the lower electronic states, we can see in figure \ref{fig:trans} that the \B-\X{} band system generally has a higher uncertainty. This heightened uncertainty is possibly due to its position in the ultraviolet region, or the numerous instances of other states perturbing the \B-\X{} band.

Tables \ref{tab:trans} and \ref{tab:trans2} consider sources from the 20th and 21st century respectively and detail not only the original data (i.e. band, number of assigned transitions, vibrational bands, J-range and frequency range) but also the results of the \Marvel{} procedure (number of validated transitions, mean and maximum uncertainties for the processed data). Vibronic-resolution versions of these tables are provided in the Supplementary Information. Many sources had a small number of unverified transitions due to misassignments and large uncertainties, but none are concerning. Similarly, the maximum uncertainty was always within an order of magnitude of the mean uncertainty, indicating there were no significant problems in the data. 

Some papers were found with CN spectroscopic data that were not suitable for our compilation; these are listed in table \ref{tab:notsource} with justifications for their exclusion. 

\begin{table}
\centering
\caption{Sources considered but not used in the \Marvel{} procedure, with comments for justification.}
\begin{tabular}{p{2.6cm}p{4.8cm}}
\toprule
Reference & Comments \\ \midrule
\cite{28Je.CN} & Very early source with a lot of blended lines, more recent sources cover all the reported bands at higher resolution. \\
\cite{32JeRoMu.CN} & Misassigned bands, unclear which they should be \\
\cite{65IrDa.CN} & Computation of dipole moments and transition probabilities \\
\cite{65PoRi.CN} & Vibrational and rotational constants\\
\cite{68Le.CN} & Only band heads \\
\cite{72Sc.CN} & Franck–Condon factors \\ 
\cite{73PhCh.CN} & Band heads \\
\cite{78CeBaGu.CN} & Analyses rotationally-resolved data, later considered by 10RaWaBe, but doesn't contain the raw experimental assigned transitions in a form suitable for \Marvel{} analysis \\
\cite{78DuErLa.CN} & Lifetimes \\
\cite{79ScSu.CN} & Outdated data, higher resolution of same bands covered by 06RaDaWa \citep{06RaDaWa.CN} \\
\cite{80KoFiSt.CN} & Coupling, no assigned spectra, perturbations \\
\cite{81GoSa.CN} & Significantly different assignments to bands from more recent publications, very small change to main Spectroscopic Network when included. \\
\cite{84ItOzNa.CN} & No assigned spectra \\
 \cite{88ItOzSu.CN} & Coupling, no assigned spectra\\
\cite{01AjKi.CN} & Calculated dipole moments for X and A \\
 \cite{10ShLiZh.CN} & Constants, no assigned spectra \\
\cite{11ShLiSu.CN} & Computation of PEC for 8 low lying states \\
\cite{12CoBe.CN} & Isotopologue data for $^{12}$C$^{15}$N \\
\cite{12RaBe.CN} & Isotopologue data for $^{13}$C$^{14}$N \\
 \cite{14BrRaWe.CN} & Line list creation, calculation of TDM and lifetimes \\
 \cite{14CoBe} & Isotopologue data for $^{13}$C$^{15}$N \\
 \cite{14SnLuRa.CN} & Line list for isotopologues \\
\cite{14WaKiCo.CN} & No rotationally resolved assignments\\
\cite{17FeKuKn.CN} & No assigned spectra published\\
 \cite{17QiZhLi.CN} & Computational \\
\cite{18YiShSu.CN} & Computational \\
\bottomrule
\end{tabular}
\label{tab:notsource}
\end{table}

\subsection{Individual Source Notes}\label{subsec:individual}
Many papers give uncertainties that we adopt unaltered and found to be reasonably consistent with all other CN data. 

Several sources (\cite{55DoRo.CN, 56Ca.CN, 70Lu.CN, 71Lu.CN, 74En.CN, 89FuAlDa.CN, 91DaBrAb.CN, 92PrBeFr.CN, 94ItKaKu.CN, 05HuCaDa.CN}) considering $\Sigma - \Sigma$ transitions did not resolve the doublet splitting at low $J$. We addressed this by including the two unresolved transitions as separate assigned transitions with the same frequency in the \Marvel{} compilation. 

Further comments on individual sources are: 


\textbf{55DoRo:  \citep{55DoRo.CN}} The \D-\X{} band was not included due to a lack of labelling of the spin splitting ($\Omega$) of the \D{} state, as such we were unable to assign reliable quantum numbers. A higher leeway in uncertainties was given for the higher electronic states. Transitions were \alert{excluded from the \Marvel{} procedure} if the \Marvel{} uncertainty grew to be greater than 1 \cm.


\textbf{67WeFiRa:  \citep{67WeFiRa.CN}} Several lines needed to be shifted to become \selfc; 0-1 R(11)(35) was increased 100 \cm, 0-1 Q(11)(24) decreased 10 \cm, 1-2 P(11)(38) increased 10 \cm, and 1-2 P(22)(25) decreased 5 \cm. When adding to the bulk of the transitions several lines were deleted due to increased uncertainty, this was deemed acceptable as there were no new energy levels involved, so newer data took precedent.

\textbf{70Lu:  \citep{70Lu.CN}} Only a few transitions had doublet structure recorded, but with no clear assignment of quantum numbers, these transitions were averaged. 
The transition in the \E-\X{} (3-0) band at N=26 was changed from 809.43 to 689.43 to ensure consistency. 

\textbf{71Lu:  \citep{71Lu.CN}} 
No assigned uncertainty within this paper due to the experimental details being published in a previous work on CN$^+$ \citep{71Lu+.CN}, thus we have taken the relative uncertainty from there. 

\textbf{73Sc:  \citep{73Sc.CN}} One of only 2 sources with rotationally resolved transitions of the \B-\A{} band. Internally \selfc{} but 2 transitions were removed when ensuring \selfcy{} with the rest of the sources, due to a large increase in uncertainty. 

\textbf{74En:  \citep{74En.CN}} For transitions where the same assignment was given with multiple frequencies the ones with higher given intensity was taken. Blended lines with multiple assignments from different vibronic bands were given all assignments. 3 transitions ((0,0) R2(10) was decreased by 1 \cm, (0,0) R1(14) decreased by 1 \cm, and (0,0) P1(16) increased by 0.4 \cm) were shifted in order to become \selfc.


\textbf{77DiWo:  \citep{77DiWo.CN}} Averaged of hyperfine splitting for "method 1" in paper. Uncertainty taken as average difference between the methods. 


\textbf{89FuAlDa:  \citep{89FuAlDa.CN}} 
The uncertainty of the \B-\X{} band was kept at 0.05 \cm{} as the reported relativity uncertainty, however the \B-\A{} band was increased to 0.1 \cm{} to ensure \selfcy. 





\textbf{94ItKaKu:  \citep{94ItKaKu.CN}} For the transitions from the \B-\X{} (11-11) band where the same assignment was given to two frequencies the most intense (reported as `main' in the paper) transition was included and others were not considered.


\textbf{04HoCiSp:  \citep{04HoCiSp.CN}} Reported quantum number given as J, however it is reported as an integer, and since $J$ is half integer for CN, we have taken the reported $J$ as N, and determined $J$ via the reported $F$ quantum number. 

\textbf{05HuCaDa:  \citep{05HuCaDa.CN}}  Reported quantum number given as J, however it is reported as an integer, and since $J$ is half integer for CN, we have taken the reported $J$ as N, and determined $J$ via the reported $F$ quantum number. 

\textbf{06RaDaWa:  \citep{06RaDaWa.CN}} Uncertainty was taken as experimental resolution. 26 transitions were removed for substantially larger uncertainties. 

\textbf{08CiSeKu:  \citep{08CiSeKu.CN}} Uncertainty was taken as experimental resolution.  


\textbf{10RaWaBe:  \citep{10RaWaBe.CN}} Paper uses the $F$ quantum number for \A{} state, with F$_1 \xrightarrow{} \Omega = 1/2$ and F$_2 \xrightarrow{} \Omega = 3/2$. As \A{} is an inverted $\Pi$ state we have converted their use of $F$ quantum number to explicit $\Omega$. 12 transitions were removed for substantially larger uncertainties.

\section{\MARVEL{} empirical energy levels for CN}\label{sct:energy}
\subsection{Spectroscopic Networks}

\begin{figure}
    \centering
      \includegraphics[width=0.4\textwidth]{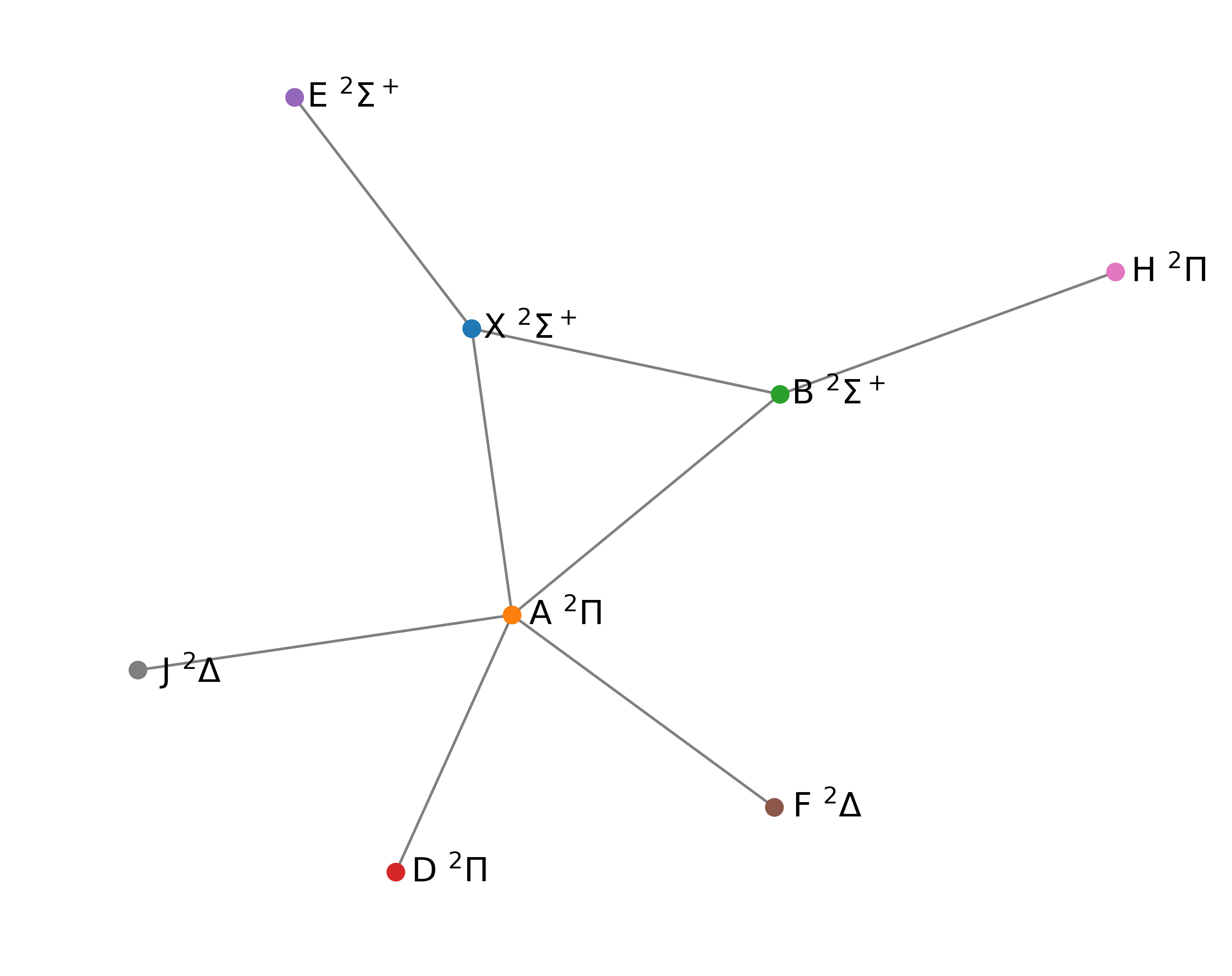}
      \includegraphics[width=0.4\textwidth]{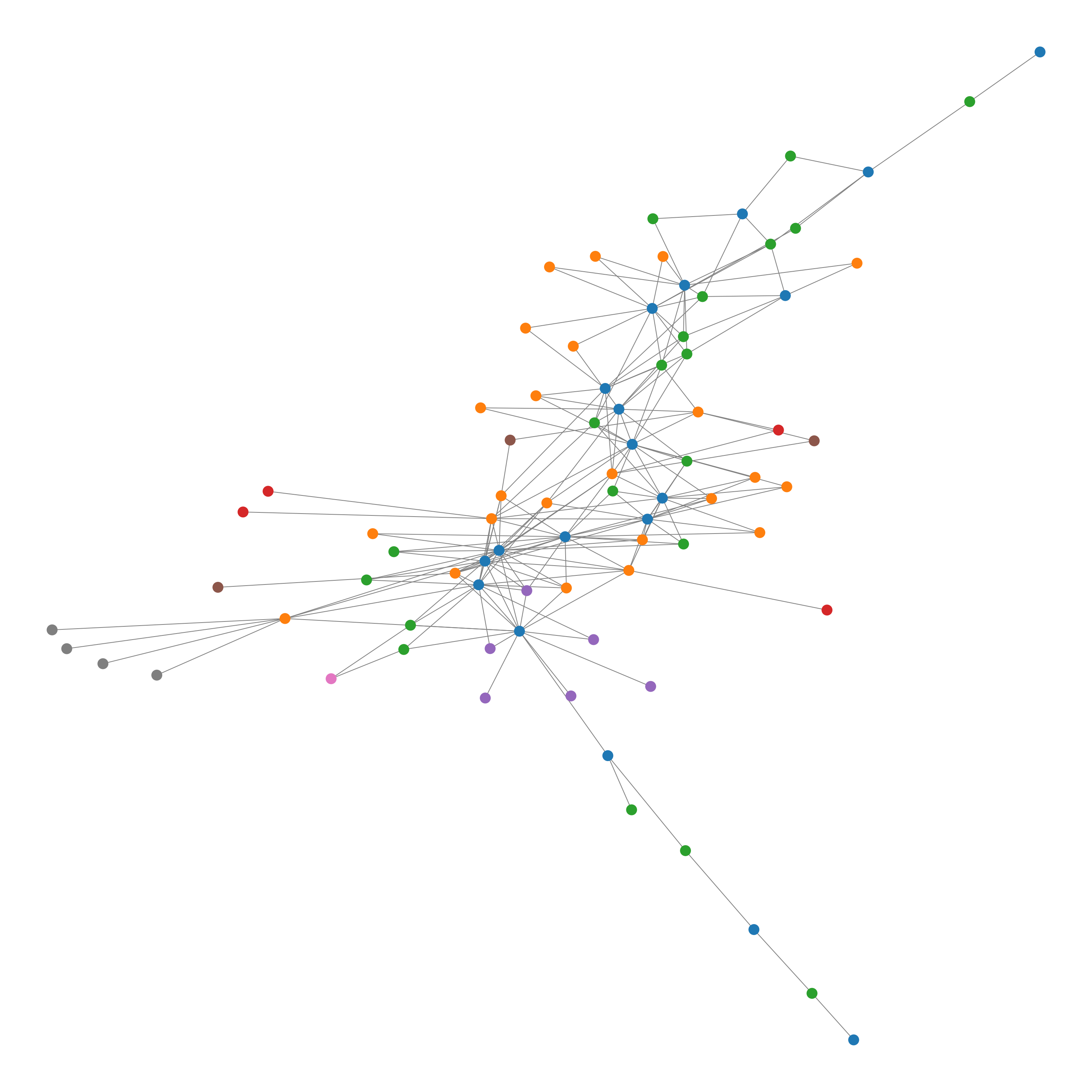}
      \includegraphics[width=0.4\textwidth]{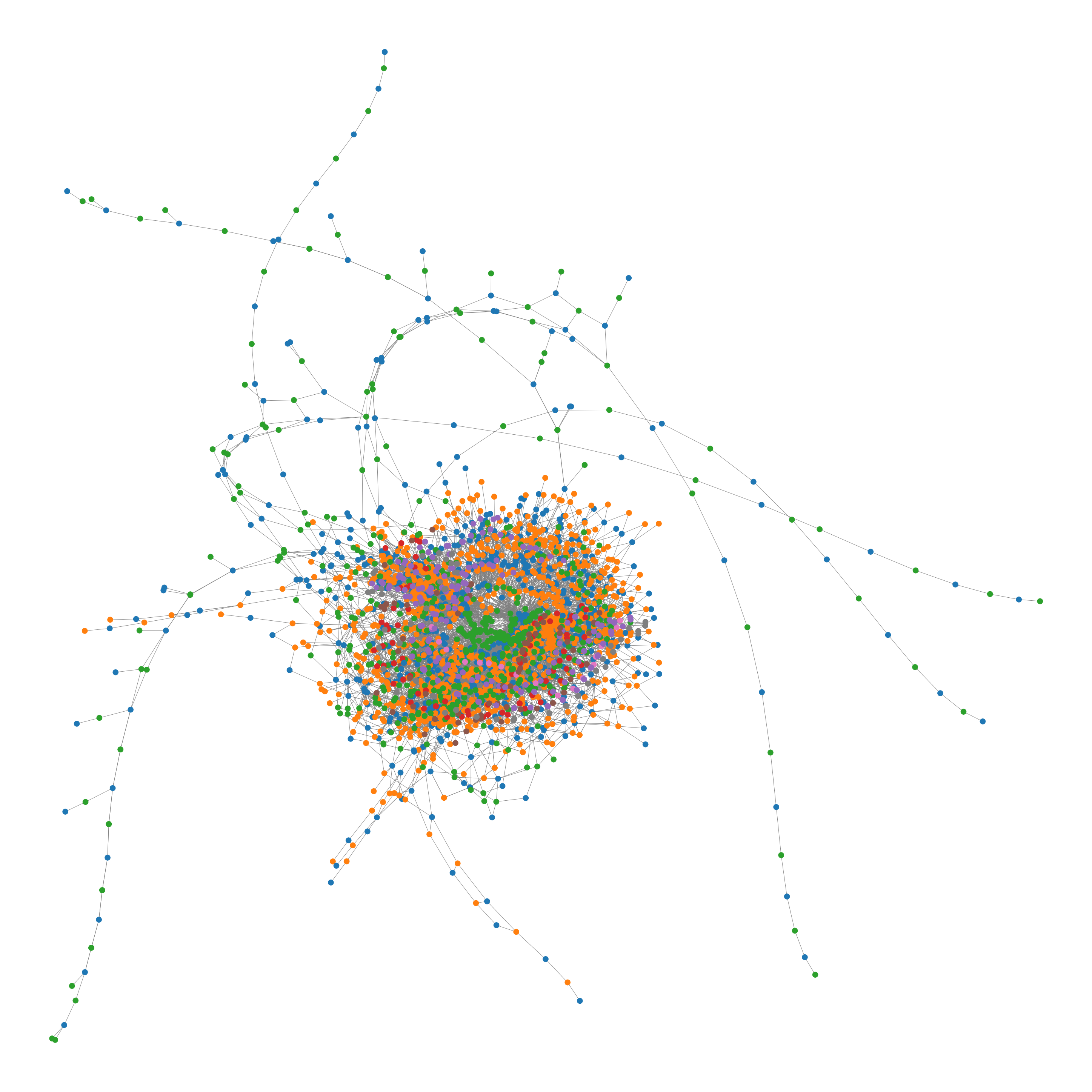}
    \caption{Spectroscopic networks of CN produced using the \Marvel{} input and output data. The energy levels are displayed as nodes and the transitions are the edges that join the nodes. Each electronic state is given a colour which is labelled in subfigure (a) and repeated in subfigures (b) and (c).}
    \label{fig:networks}
\end{figure}

The \Marvel{} process generates Spectroscopic Networks (SN) of interconnected energy levels connected by the input transition data. For the transition data used in this paper \noSN{} SNs were generated with the main SN containing 7779 energy levels spanning \noelec{} electronic states and 74 vibronic states. The other SNs are relatively small and composed of unconnected high rotational transitions. The exception to this is the four SNs that contain between 70 and 80 energy levels from the vibronic states; \X{} ($v$ = 15), \X{} ($v$ = 17), \X{} ($v$ = 18), \B{} ($v$ = 15), \B{} ($v$ = 18), and \B{} ($v$ = 19). \alert{These four mid-sized SNs were joined to the main SN using the `Magic Number' technique. Joining these networks was accomplished by  by adding four artificial transitions to the input file to join the ground state of \X to the J=0.5 and J=1.5 states of the \X v = 15 state. The frequency of these transitions was estimated by a combination of the energies in the main SN and data from the \Mollist{} line list, with an uncertainty of 0.5 \cm. This brings the main SN up to \noenergy{} energy levels from \novibronic{} vibronic states.} 
The rest of the SNs have less than 16 energy levels and no further analysis on these SNs were done. 

The main Spectroscopic Network (SN) is visualised in figure \ref{fig:networks}, by plotting the states as nodes and transitions as edges between them. Figure \ref{fig:networks} is broken up into the electronic bands (subfigure \ref{subfig:el}), vibronic bands (subfigure \ref{subfig:vib}), and all rovibronic transitions in subfigure \ref{subfig:rovib}. Each electronic state is uniquely coloured, as labelled in figure \ref{subfig:el} and this colour scheme flows through the vibronic and rovibronic levels as well. In figure \ref{subfig:el} we can see that the \A{} state is the most connected to the other electronic states due to its $\Pi$ symmetry. The inter-connectivity of the lowest three electronic states gives rise to the clustering that is seen in figures \ref{subfig:vib} and \ref{subfig:rovib}. This inter-connectivity in figures \ref{subfig:vib} and \ref{subfig:rovib} also highlights how an upper state can decay into various different lower states. 
From the vibronic SN (figure, \ref{subfig:vib}), we can see the extension of the higher vibrational \B-\X{} bands from the bulk of the connected states. The rovibronic SN figure (figure \ref{subfig:rovib}) is created from only the main SN of the CN data, with the unconnected isolated rovibronic states excluded for clarity. The high vibrational \B-\X{} bands are much more apparent in figure \ref{subfig:rovib} as the other rovibronic states become more connected and drawn towards the centre collection.

Since there are no observed transitions that involve the quartet states in CN they are not in any of the SNs. If there were observed transitions, they would only join the main SN through spin forbidden transitions, otherwise making a unique SN of their own.  

\begin{table*}
\centering
\caption{Sample of the \Marvel{} output energies file. The full output file is included in the SI. \label{tab:out}}
\begin{tabular}{p{2.0cm}p{1cm}p{1cm}p{2.5cm}p{2cm}p{1cm}}
\toprule
State & $v$ & $J$ & $E$ & $\Delta E$ & \#\\ \midrule
X2Sig+\_e & 3 & 5.5 & 6103.507576 & 0.001771 & 89 \\
A2Pi\_e3/2 & 0 & 4.5 & 9129.091256 & 0.002998 & 27 \\
X2Sig+\_e & 8 & 8.5 & 15726.855299 & 0.002278 & 55 \\
A2Pi\_e3/2 & 5 & 7.5 & 17871.555021 & 0.003399 & 14 \\
X2Sig+\_e & 8 & 10.5 & 15793.319748 & 0.002338 & 53 \\
A2Pi\_e3/2 & 5 & 9.5 & 17928.336351 & 0.003156 & 16 \\ \bottomrule
\end{tabular}

\begin{tabular}{ccl}
\\
             Column        &     Notation                  &       \\
\midrule
   1  &   State  &   Electronic state of the energy level; also includes parity and $\Omega$ for $\Pi$ and $\Delta$ states  \\
    2  &  $v$  &  Vibrational quantum number of the state \\ 
  3  &   $J$                &        Total angular momentum of the state   \\
   4  &   $E$  &   Energy in \cm{} of the state \\
    5  &  $\Delta E$  &  Uncertainty in \cm{} of the state \\
   6  &   \#   &        Number of transitions from the input file that contribute to the energy of the state   \\
\bottomrule
\end{tabular}

\end{table*}

\begin{figure}
\centering
  \includegraphics[width = 0.47\textwidth]{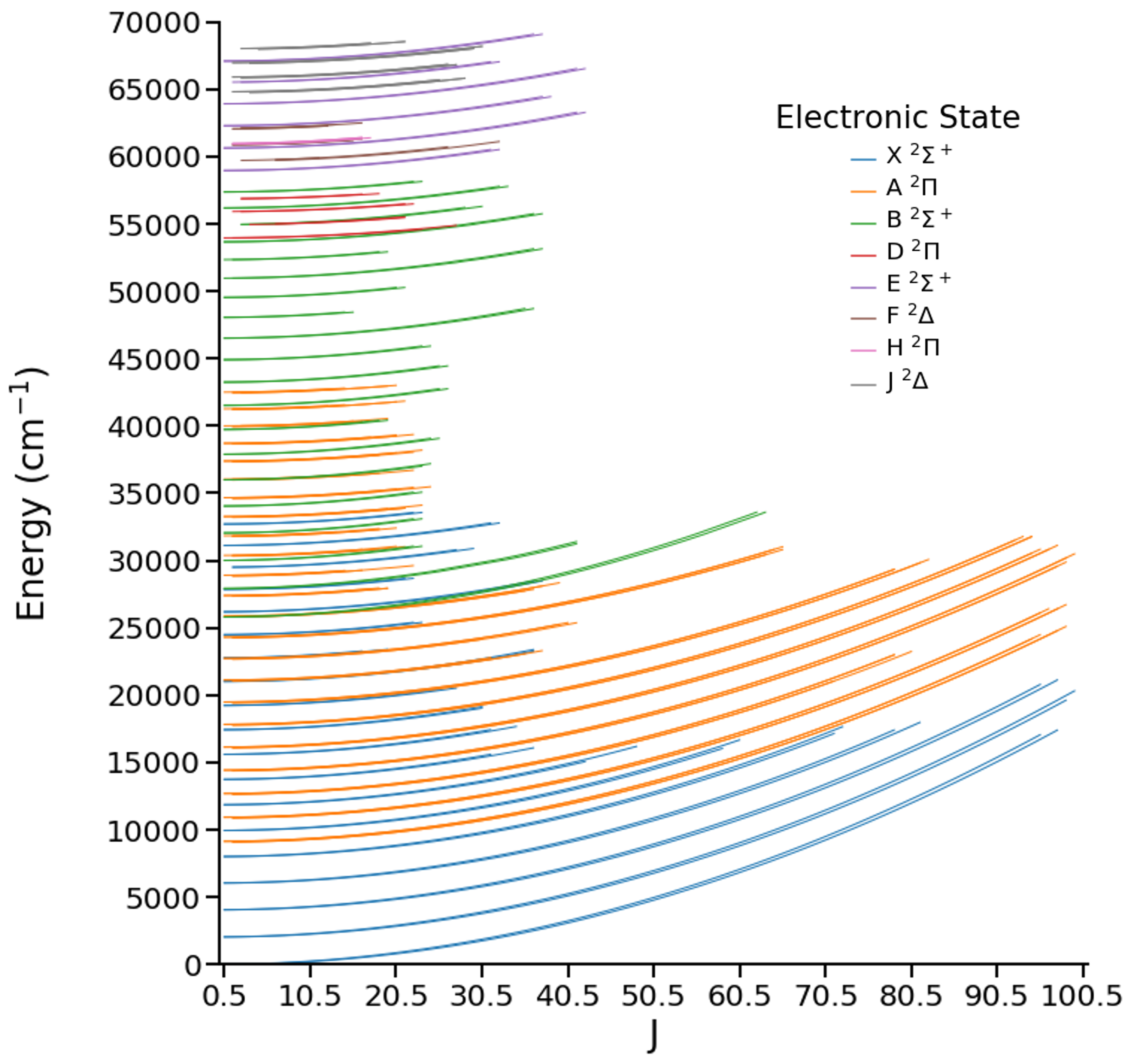}
  \caption{The spin-vibronic energy levels as a function of the total angular momentum, $J$. Each colour is the electronic state that the energy level belongs to.}
  \label{fig:energyJ}
\end{figure}

\begin{figure}
\centering
  \includegraphics[width = 0.47\textwidth]{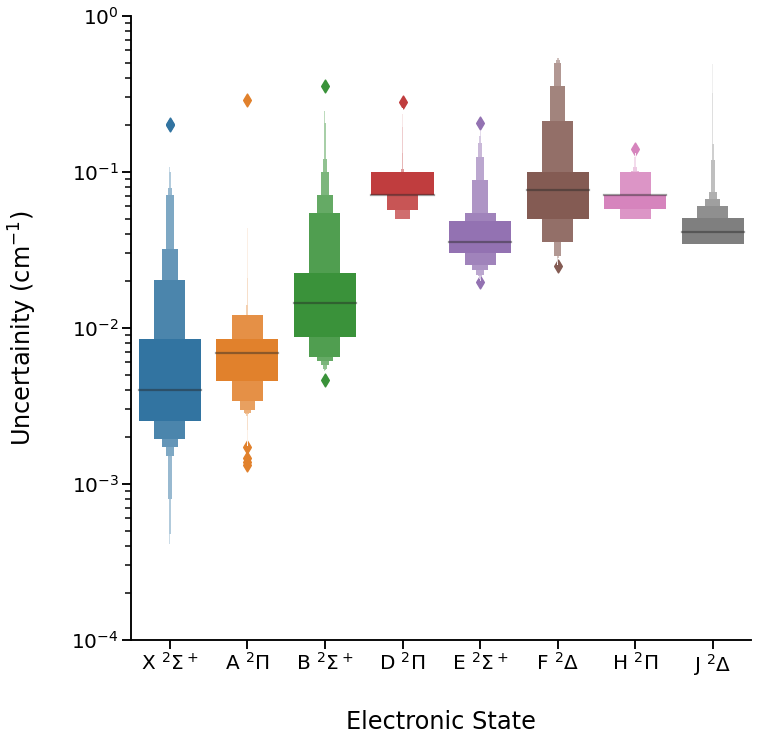}
  \caption{The spread of uncertainties for each electronic state in CN. The solid line across the largest box shows the mean uncertainty, with the diamond points indicating the highest and lowest values for each electronic state. }
  \label{fig:unc-el}
\end{figure}

\begin{figure}
    \centering
    \includegraphics[width = 0.48\textwidth]{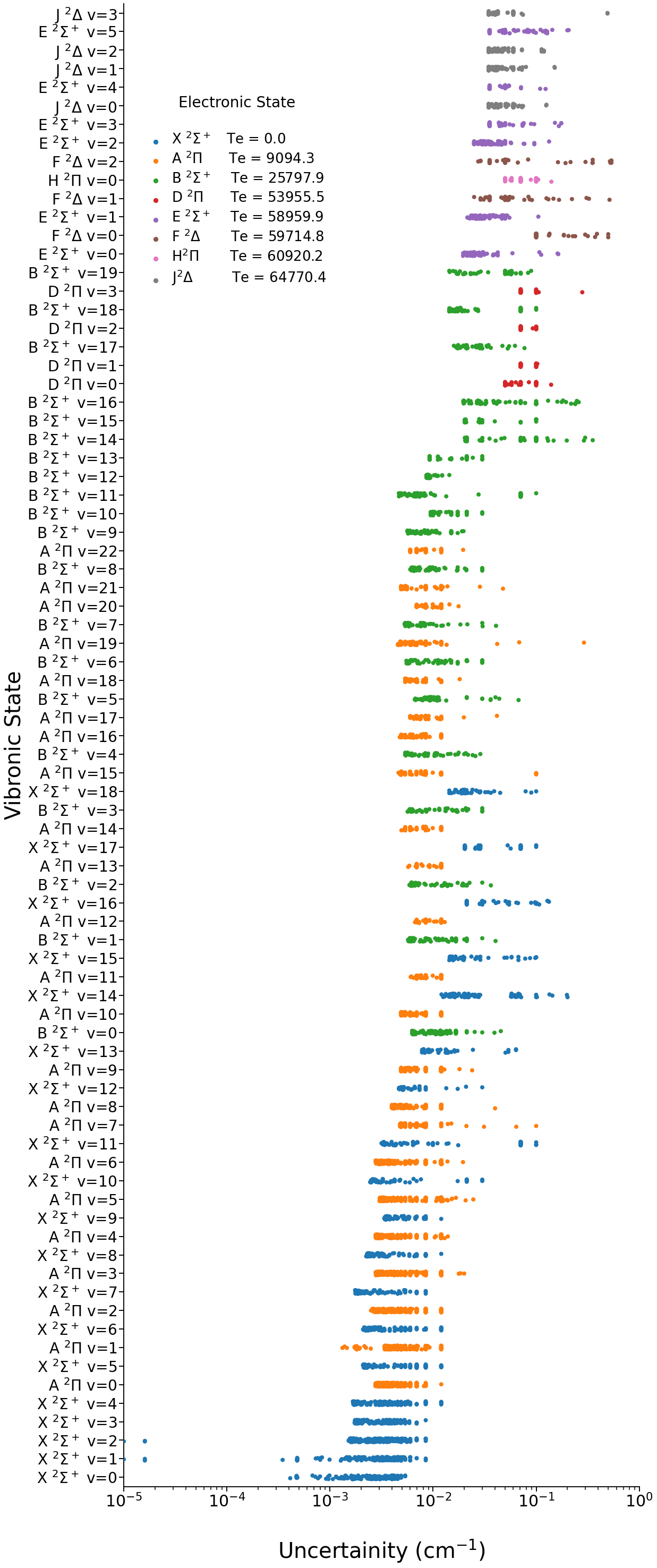}
    \caption{The spread of uncertainties for each vibronic state. The colours represent the electronic states that belong to each vibronic state, and the Te of the electronic states are included in the legend.}
    \label{fig:unc-vib}
\end{figure}

\begin{figure}
\centering
  \includegraphics[width = 0.48\textwidth]{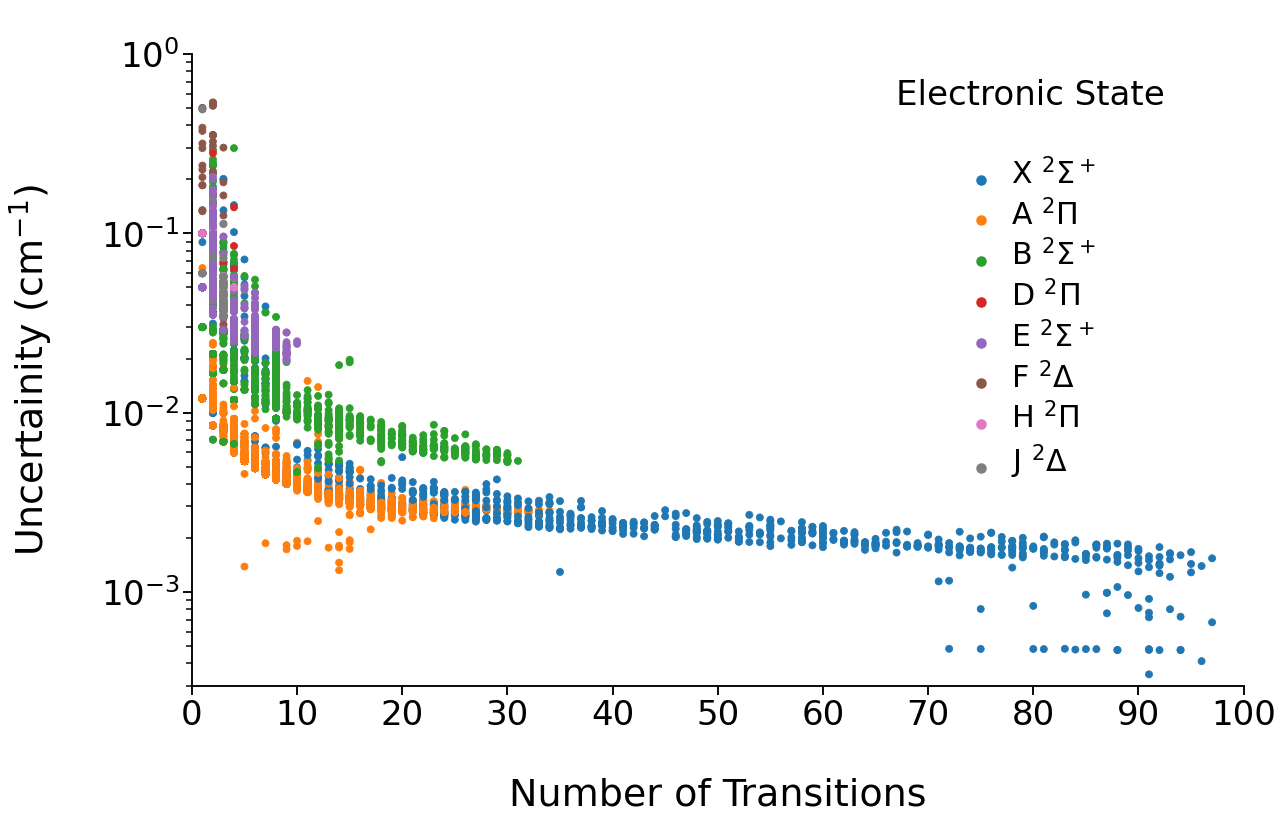}
  \caption{The relationship between the uncertainty of each energy level with the number of experimental transitions that contribute to it.}
  \label{fig:unc-t}
\end{figure}

\subsection{Energy Levels}

All \nospinvibronic{} spin-vibronic states are shown as unique lines in figure \ref{fig:energyJ} as a function of the total angular momentum quantum number, $J$. The visibly smooth quadratic lines show no issues with \noenergy{} empirical energy levels of the main SN.  Figure \ref{fig:energyJ} also shows the range of $J$ for the vibronic states, where low vibrational states in the ground \X{} and \A{} electronic states have significantly higher rotational energy levels.

The distribution of the empirical uncertainties for each electronic state is shown in figure \ref{fig:unc-el}, which allows the median (shown as the flat line inside the largest box) to be visualised alongside increasingly smaller boxes as the distribution thins. Outliers are shown individually. Figure \ref{fig:unc-el} demonstrates the consistently increased uncertainties for higher electronic states, which are about an order of magnitude above the ground electronic state. 

Delving deeper into the vibronic states for CN allows us to examine vibrational levels that have higher uncertainties. Table \ref{tab:el} breaks down the vibronic states for each of the lower electronic states, including: the $J$ range, energy range, number of energy levels, mean and maximum empirical uncertainty, and the sources that contributed to the vibronic state. The \X{} state is very well studied, especially at low vibration, as demonstrated by the large number of sources that contribute to the determination of the energy levels, as well as the small uncertainties. The low vibrational states of the \A{} and \B{} states are fairly well characterised, but we see fewer sources as we increase the vibrational quantum number. The largest mean uncertainty of the lower electronic states is 0.073 \cm{} from \B{} ($v$ = 14). This state is well known to be perturbed by the \A{} ($v$ = 30) vibronic state. Several of the known perturbed vibronic states can be seen to have an increased uncertainty spread in figure \ref{fig:unc-vib}. While there is a general upward trend of uncertainties across vibronic states, the interaction of states can also be seen in the spread of uncertainties.  The well known near degeneracy of the \X($v$ = 11) and \A($v$ = 7) can be seen through this lens, as well as the high vibrational \X and \B states. Many of the high vibrational \B{} states are perturbed by high vibronic \A{} states not seen in this compilation, as well as the `dark' quartet states. One example of this the \B{} ($v$=11) state that is thought to be perturbed by the \asig{} state \citep{75CoRaSe.CN, 83OzItSu.CN}, which has relatively low and clustered uncertainty, with a minimum of 0.005 \cm{} but has several energy levels with increased uncertainties of up to 0.1 \cm.

The higher electronic states have much fewer vibronic states observed. These vibronic states are given the same breakdown as the lower states in table \ref{tab:high-el}. All of these energy levels are only determined from one or two sources. The lack of experimental observation of these bands is highlighted in the \F{} state which has the highest average and mean uncertainties across all vibronic states.

\begin{table*}
\centering
\caption{Summary of  experimentally-derived \Marvel{} energy levels, 
    including uncertainties and data sources, for low-lying electronic states of CN. No is the number of energy levels in that vibronic state. \label{tab:el}}
\resizebox{\textwidth}{!}{%
\begin{tabular}{p{0.6cm}p{1.6cm}p{2.8cm}p{0.6cm}p{2.0cm}p{12cm}}
\toprule
$v$ & $J$ range & E-range (\cm) & No &  Mean/Max & Sources \\ \midrule
\mc{5}{l}{\textbf{X$^2\Sigma^+$}} \\
0 & 0.5 - 97.5 & 0.0 - 17403.4 & 194 & 0.003/0.005 & 06RaDaWa, 10RaWaBe, 92PrBeFr, 01LiDuLi, 04HoCiSp, 05HuCaDa, 09HaHaSe, 70Lu, 74En, 91DaBrAb, 92ReSuMi \\
1 & 0.5 - 99.5 & 2042.4 - 20317.7 & 195 & 0.003/0.008 & 04HoCiSp, 05HuCaDa, 06RaDaWa, 10RaWaBe, 77DiWo, 92PrBeFr, 55DoRo, 56Ca, 67WeFiRa, 70Lu, 74En, 92PrBe, 92ReSuMi \\
2 & 0.5 - 97.5 & 4058.5 - 21127.9 & 194 & 0.003/0.008 & 04HoCiSp, 06RaDaWa, 10RaWaBe, 77DiWo, 91DaBrAb, 92PrBeFr, 55DoRo, 56Ca, 67WeFiRa, 74En, 92PrBe, 92ReSuMi \\
3 & 0.5 - 81.5 & 6048.3 - 17978.5 & 161 & 0.003/0.008 & 04HoCiSp, 06RaDaWa, 10RaWaBe, 92PrBeFr, 08CiSeKu, 55DoRo, 56Ca, 74En, 92PrBe, 92ReSuMi \\
4 & 0.5 - 72.5 & 8011.8 - 17661.4 & 145 & 0.004/0.012 & 04HoCiSp, 06RaDaWa, 10RaWaBe, 92PrBeFr, 08CiSeKu, 55DoRo, 56Ca, 74En, 92PrBe, 92ReSuMi \\
5 & 0.5 - 60.5 & 9948.8 - 16676.8 & 117 & 0.005/0.012 & 04HoCiSp, 06RaDaWa, 10RaWaBe, 92PrBeFr, 08CiSeKu, 55DoRo, 74En, 92PrBe, 92ReSuMi \\
6 & 0.5 - 48.5 & 11859.3 - 16195.3 & 92 & 0.005/0.012 & 04HoCiSp, 06RaDaWa, 10RaWaBe, 92PrBeFr, 08CiSeKu, 55DoRo, 74En, 92PrBe, 92ReSuMi \\
7 & 0.5 - 36.5 & 13743.4 - 16086.8 & 69 & 0.003/0.008 & 04HoCiSp, 06RaDaWa, 10RaWaBe, 92PrBeFr, 08CiSeKu, 74En, 92PrBe, 92ReSuMi \\
8 & 0.5 - 34.5 & 15600.9 - 17674.3 & 67 & 0.004/0.012 & 04HoCiSp, 06RaDaWa, 10RaWaBe, 92PrBeFr, 08CiSeKu, 55DoRo, 92PrBe, 92ReSuMi \\
9 & 0.5 - 30.5 & 17431.8 - 19143.5 & 62 & 0.005/0.012 & 06RaDaWa, 10RaWaBe, 92PrBeFr, 08CiSeKu, 92PrBe, 92ReSuMi \\
10 & 0.5 - 27.5 & 19236.0 - 20528.0 & 54 & 0.007/0.03 & 06RaDaWa, 10RaWaBe, 92PrBeFr, 55DoRo, 75CoRaSe, 92PrBe, 92ReSuMi \\
11 & 0.5 - 36.5 & 21013.3 - 23384.8 & 74 & 0.029/0.1 & 06RaDaWa, 10RaWaBe, 75CoRaSe, 89FuAlDa, 92PrBe, 92ReSuMi, 94ItKaKu \\
12 & 0.5 - 19.5 & 22765.7 - 23402.5 & 37 & 0.007/0.03 & 06RaDaWa, 10RaWaBe \\
13 & 0.5 - 23.5 & 24488.7 - 25403.2 & 47 & 0.016/0.064 & 06RaDaWa, 55DoRo, 92ReSuMi \\
14 & 0.5 - 37.5 & 26185.7 - 28479.7 & 75 & 0.04/0.201 & 06RaDaWa, 94ItKaKu \\
15 & 0.5 - 22.5 & 27856.2 - 28676.2 & 45 & 0.034/0.1 & 06RaDaWa, 55DoRo, 94ItKaKu \\
16 & 1.5 - 29.5 & 29502.0 - 30889.3 & 54 & 0.045/0.133 & 06RaDaWa \\
17 & 0.5 - 32.5 & 31115.1 - 32784.5 & 64 & 0.04/0.1 & 06RaDaWa, 55DoRo, 94ItKaKu \\
18 & 0.5 - 23.5 & 32703.8 - 33566.9 & 47 & 0.027/0.1 & 06RaDaWa, 55DoRo, 94ItKaKu \\
\\\mc{5}{l}{\textbf{A$^2\Pi$}} \\
0 & 0.5 - 98.5 & 9094.3 - 25099.4 & 385 & 0.004/0.012 & 08CiSeKu, 10RaWaBe, 67WeFiRa, 56Ca \\
1 & 0.5 - 98.5 & 10882.0 - 26718.0 & 387 & 0.006/0.012 & 08CiSeKu, 09HaHaSe, 10RaWaBe, 67WeFiRa \\
2 & 0.5 - 80.5 & 12644.2 - 23241.4 & 306 & 0.006/0.012 & 01LiDuLi, 08CiSeKu, 10RaWaBe, 55DoRo \\
3 & 0.5 - 99.5 & 14380.7 - 30495.4 & 377 & 0.006/0.02 & 08CiSeKu, 10RaWaBe, 55DoRo \\
4 & 0.5 - 97.5 & 16091.7 - 31122.4 & 360 & 0.008/0.014 & 08CiSeKu, 10RaWaBe, 55DoRo, 73Sc \\
5 & 0.5 - 94.5 & 17777.1 - 31777.3 & 326 & 0.008/0.025 & 08CiSeKu, 10RaWaBe \\
6 & 0.5 - 82.5 & 19436.8 - 30073.8 & 298 & 0.008/0.02 & 08CiSeKu, 10RaWaBe, 55DoRo, 71Lu \\
7 & 0.5 - 37.5 & 21070.9 - 23301.8 & 136 & 0.008/0.1 & 10RaWaBe, 55DoRo, 71Lu, 89FuAlDa \\
8 & 0.5 - 41.5 & 22679.3 - 25386.9 & 147 & 0.007/0.04 & 10RaWaBe, 92PrBe, 71Lu \\
9 & 0.5 - 65.5 & 24262.0 - 31016.2 & 219 & 0.009/0.024 & 10RaWaBe, 92PrBe \\
10 & 0.5 - 39.5 & 25818.9 - 28341.6 & 146 & 0.008/0.012 & 10RaWaBe, 92PrBe \\
11 & 0.5 - 19.5 & 27350.0 - 27956.4 & 73 & 0.01/0.012 & 10RaWaBe, 92PrBe \\
12 & 0.5 - 22.5 & 28855.1 - 29622.4 & 60 & 0.01/0.013 & 10RaWaBe, 92PrBe \\
13 & 0.5 - 21.5 & 30334.4 - 31047.1 & 76 & 0.011/0.012 & 10RaWaBe, 92PrBe \\
14 & 0.5 - 20.5 & 31787.5 - 32410.5 & 70 & 0.009/0.012 & 10RaWaBe, 92PrBe \\
15 & 0.5 - 23.5 & 33214.4 - 34107.2 & 86 & 0.011/0.1 & 10RaWaBe, 55DoRo, 92PrBe \\
16 & 0.5 - 24.5 & 34615.0 - 35478.6 & 83 & 0.008/0.012 & 10RaWaBe, 92PrBe \\
17 & 0.5 - 22.5 & 35988.9 - 36709.4 & 68 & 0.009/0.042 & 10RaWaBe \\
18 & 0.5 - 23.5 & 37336.3 - 38195.2 & 85 & 0.008/0.018 & 10RaWaBe \\
19 & 0.5 - 22.5 & 38656.7 - 39357.8 & 80 & 0.013/0.291 & 10RaWaBe, 92PrBe \\
20 & 0.5 - 19.5 & 39949.8 - 40545.7 & 66 & 0.011/0.018 & 10RaWaBe, 92PrBe \\
21 & 0.5 - 21.5 & 41215.4 - 41837.7 & 66 & 0.009/0.048 & 10RaWaBe, 92PrBe \\
22 & 0.5 - 20.5 & 42453.0 - 43011.2 & 66 & 0.009/0.02 & 10RaWaBe \\
\\\mc{5}{l}{\textbf{B$^2\Sigma^+$}} \\
0 & 0.5 - 63.5 & 25797.9 - 33588.7 & 114 & 0.012/0.045 & 06RaDaWa, 55DoRo, 74En, 92PrBeFr, 92ReSuMi \\
1 & 0.5 - 41.5 & 27921.5 - 31399.2 & 65 & 0.013/0.04 & 06RaDaWa, 55DoRo, 74En, 92PrBeFr, 92ReSuMi \\
2 & 0.5 - 23.5 & 30004.9 - 31060.8 & 46 & 0.012/0.036 & 06RaDaWa, 55DoRo, 74En, 92PrBeFr, 92ReSuMi \\
3 & 0.5 - 23.5 & 32045.9 - 33095.6 & 47 & 0.012/0.03 & 06RaDaWa, 55DoRo, 74En, 92PrBeFr, 92ReSuMi \\
4 & 0.5 - 23.5 & 34042.0 - 35072.4 & 47 & 0.011/0.029 & 06RaDaWa, 55DoRo, 74En, 92PrBeFr, 92ReSuMi \\
5 & 0.5 - 24.5 & 35990.0 - 37186.5 & 48 & 0.014/0.067 & 06RaDaWa, 55DoRo, 74En, 92PrBeFr \\
6 & 0.5 - 25.5 & 37887.4 - 39066.8 & 51 & 0.013/0.03 & 06RaDaWa, 74En, 92PrBeFr, 92ReSuMi \\
7 & 0.5 - 19.5 & 39730.5 - 40409.0 & 39 & 0.01/0.041 & 06RaDaWa, 73Sc, 74En, 92PrBeFr, 92ReSuMi \\
8 & 0.5 - 26.5 & 41516.6 - 42749.5 & 50 & 0.012/0.03 & 06RaDaWa, 89FuAlDa, 92PrBeFr, 92ReSuMi \\
9 & 0.5 - 26.5 & 43243.0 - 44452.5 & 53 & 0.009/0.02 & 06RaDaWa, 92PrBeFr, 92ReSuMi \\
10 & 0.5 - 24.5 & 44908.8 - 45924.5 & 49 & 0.014/0.03 & 06RaDaWa \\
11 & 0.5 - 36.5 & 46511.4 - 48708.3 & 73 & 0.03/0.1 & 06RaDaWa, 75CoRaSe, 92ReSuMi, 94ItKaKu \\
12 & 0.5 - 15.5 & 48053.7 - 48443.6 & 31 & 0.009/0.014 & 06RaDaWa \\
13 & 0.5 - 21.5 & 49537.3 - 50273.5 & 43 & 0.016/0.03 & 06RaDaWa, 92ReSuMi \\
14 & 0.5 - 37.5 & 50967.7 - 53146.3 & 71 & 0.073/0.354 & 06RaDaWa, 55DoRo, 94ItKaKu \\
15 & 0.5 - 19.5 & 52343.0 - 52921.1 & 38 & 0.038/0.1 & 06RaDaWa, 94ItKaKu \\
16 & 0.5 - 37.5 & 53664.4 - 55753.6 & 75 & 0.069/0.259 & 06RaDaWa, 55DoRo, 94ItKaKu \\
17 & 2.5 - 30.5 & 54955.1 - 56299.8 & 54 & 0.026/0.077 & 06RaDaWa \\
18 & 0.5 - 33.5 & 56178.1 - 57782.9 & 64 & 0.036/0.1 & 06RaDaWa, 55DoRo, 94ItKaKu \\
19 & 0.5 - 23.5 & 57371.3 - 58145.5 & 47 & 0.041/0.089 & 06RaDaWa, 55DoRo, 94ItKaKu \\
\bottomrule
\end{tabular}
}
\end{table*}

\begin{table}
\centering
\caption{Summary of  experimentally-derived \Marvel{} energy levels, 
    including uncertainties and data sources, for higher electronic states of CN. No is the number of energy levels in that vibronic state. \label{tab:high-el}}
\resizebox{\columnwidth}{!}{%
\begin{tabular}{p{0.4cm}p{1.6cm}p{2.8cm}p{0.6cm}p{1.9cm}p{1.8cm}}
\toprule
$v$ & $J$ range & E-range (\cm) & No &  Mean/Max & Sources \\ \midrule
\mc{5}{l}{\textbf{D$^2\Pi$}} \\
0 & 0.5 - 27.5 & 53955.5 - 54888.1 & 51 & 0.07/0.14 & 55DoRo \\
1 & 3.5 - 21.5 & 54954.7 - 55517.2 & 35 & 0.08/0.103 & 55DoRo \\
2 & 1.5 - 22.5 & 55917.5 - 56480.2 & 42 & 0.076/0.1 & 55DoRo \\
3 & 2.5 - 18.5 & 56869.4 - 57243.6 & 32 & 0.085/0.28 & 55DoRo \\
\\\mc{5}{l}{\textbf{E$^2\Sigma^+$}} \\
0 & 0.5 - 32.5 & 58959.9 - 60520.9 & 65 & 0.034/0.164 & 56Ca, 70Lu \\
1 & 0.5 - 42.5 & 60631.3 - 63280.0 & 85 & 0.034/0.105 & 56Ca, 70Lu \\
2 & 0.5 - 38.5 & 62285.0 - 64447.4 & 77 & 0.04/0.133 & 70Lu \\
3 & 0.5 - 42.5 & 63917.3 - 66529.9 & 85 & 0.047/0.175 & 70Lu \\
4 & 1.5 - 32.5 & 65529.3 - 67038.0 & 62 & 0.045/0.123 & 70Lu \\
5 & 0.5 - 37.5 & 67088.5 - 69082.0 & 75 & 0.061/0.207 & 70Lu \\
\\\mc{5}{l}{\textbf{F$^2\Delta$}} \\
0 & 2.5 - 32.5 & 59714.8 - 61112.9 & 58 & 0.142/0.5 & 55DoRo \\
1 & 1.5 - 16.5 & 60870.1 - 61314.8 & 54 & 0.076/0.514 & 71Lu \\
2 & 1.5 - 16.5 & 62058.8 - 62496.4 & 51 & 0.114/0.537 & 71Lu \\
\\\mc{5}{l}{\textbf{H$^2\Pi$}} \\
0 & 1.5 - 17.5 & 60920.2 - 61410.5 & 61 & 0.073/0.14 & 55DoRo \\
\\\mc{5}{l}{\textbf{J$^2\Delta$}} \\
0 & 1.5 - 28.5 & 64770.4 - 65816.3 & 102 & 0.045/0.126 & 56Ca \\
1 & 1.5 - 27.5 & 65856.6 - 66843.1 & 104 & 0.045/0.151 & 56Ca \\
2 & 1.5 - 30.5 & 66932.0 - 68186.6 & 114 & 0.045/0.119 & 56Ca \\
3 & 2.5 - 21.5 & 67984.8 - 68536.0 & 66 & 0.059/0.492 & 56Ca \\
\bottomrule
\end{tabular}
}
\end{table}



The \Marvel{} process determines the uncertainties of each energy level based on the transitions that include that level. Figure \ref{fig:unc-t} demonstrates how the uncertainties of the transitions (shown in figure \ref{fig:trans}) propagate through to the uncertainties of the energy levels (see figures \ref{fig:unc-el} and \ref{fig:unc-vib}). The uncertainty of the energy level is approximately an order of magnitude smaller than the uncertainties of the transitions involving those electronic states. Figure \ref{fig:unc-t} shows the uncertainty of the energy level as a function of the number of transitions that were involved in determining that energy level. The uncertainty of the energy levels decreases with an increase in the number of transitions, as expected. Thus we can see that a larger number of experimental assignments allow for smaller uncertainties in the \Marvel{} procedure and thus a better understanding of the energy levels. The need for high resolution assignments would only further benefit the understanding of energy levels in CN, especially of perturbed or higher electronic states, for which there are fewer data sources available.


\section{Utilising \Marvel{} data to improve the MoLLIST CN Line list} \label{sct:comp}

\label{subsec:linelist}

Currently the most complete line list for CN is from \Mollist{} \citep{14BrRaWe.CN, 20Be.CN} created using the traditional method, i.e. using spectroscopic constants fitted to experimental data. The sources of data used are a subset of those considered in this \Marvel{} compilation, specifically 91DaBrAb, 92PrBeFr, 92PrBe, 04HoCiSp, 05HuCaDa, 06RaDaWa, 10RaWaBe. 
There are two versions of this CN line list currently available: 
\begin{description}
\item[\textbf{Original format, includes observed frequencies:}] Contains only lower energy levels and transition frequencies. 11.3\% of the predicted frequencies are replaced with directly observed transition frequencies. 
\item[\textbf{ExoMol formatted with predicted energies:}] The ExoMol format \citep{20WaTeYu.CN} calculates transition frequencies from energy levels rather than storing frequencies separately, ensuring self-consistency but reducing the quality of the line list frequencies for high-resolution applications. These energy levels are based solely on the predicted frequencies, i.e. the spectroscopic constants. 
\end{description}

Here, we produce a superior CN line list, the \Marvel ised \Mollist{} line list, by using the ExoMol formatted version of the original \Mollist{} line list, but replacing the predicted energy levels with the \Marvel{} experimentally-derived energy levels where available; \alert{6122} of the 7696 energy levels, i.e. \alert{79.5\%}, are replaced. By propagating these improved energies through to the 195,112 transitions, we not only recover the 22,044 observed transitions but accurately predict an additional \alert{128751} transitions entirely from \Marvel{} energies, thus substantially improving the quality of the linelist, especially for high-resolution studies. The updated ExoMol format 12C-14N\_\_Mollist-Marvelised.states file is included in the SI, and is compatible with the 12C-14N\_\_Mollist.trans file on the ExoMol website. 

\begin{figure}
    \centering
    \includegraphics[width = 0.48\textwidth]{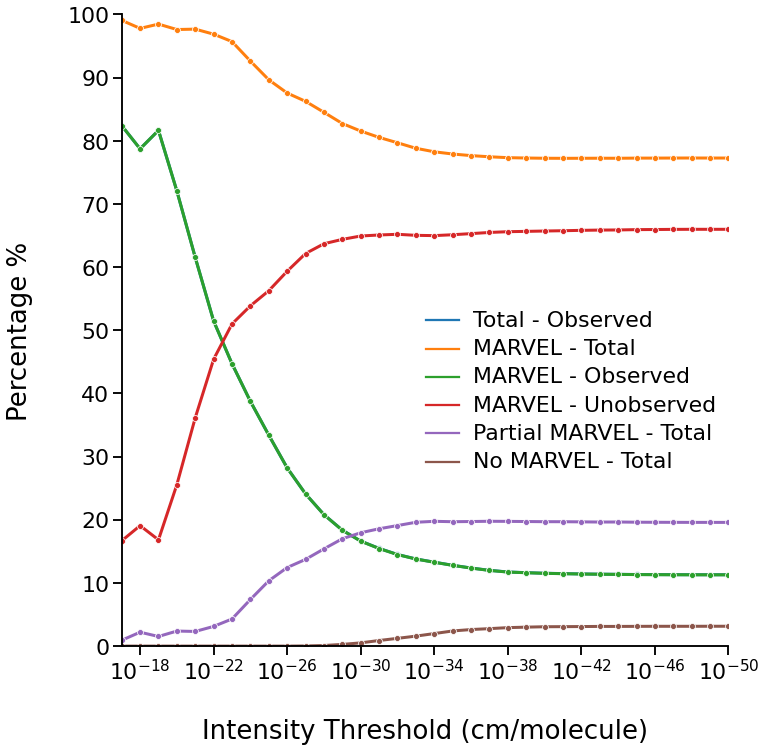}
    \caption{The percentage of transitions with degrees of \Marvel-isation across intensity calculated at 1000 K. The transitions with upper and lower energy levels that have been \Marvel-ised have also been split into observed and unobserved - according to the \Mollist{} SI. The \alert{blue line indicating the} Total - Observed (Observed in the \Mollist{} SI) is indistinguishable from the \Marvel{} - Observed \alert{green} line.}
    \label{fig:MMpercent}
\end{figure}

At 2000 K, \Marvel{} energy levels alone recover all but 0.000026\% of the partition function, while, even at 5000 K and 7000 K, 98.2\% and 95.2\% of the partition function is recovered. 

The spectral coverage of this new \Marvel ised \Mollist{} CN line list is characterised in figure \ref{fig:MMpercent}, which illustrates the source of transition frequency data as a function of minimum transition intensity at 1000 K. 81.6\% of strong transitions (intensities greater than $10^{-18}$ cm/molecule) have been directly observed, while an additional 16.8\% of transitions are unobserved but have frequencies determined completely from \Marvel{} energies. \alert{Of all of the transitions observed in the \MOLLIST{} data set 99.9\% of them are matched with \Marvel{} data for both upper and lower energy levels, this is highlighted in figure \ref{fig:MMpercent} as the green line completely overlays the total observed transitions (blue line) along all intensities.} As we consider weaker transitions, the proportion of directly observed transitions decreases, but the proportion of unobserved but fully \Marvel-ised transitions increases, keeping the number of transitions whose frequency is directly and completely determined by \Marvel{} energies above 77\% even when considering all \Mollist{} transitions. The figure also details two other categories - partial \Marvel-ised transitions for which either the upper or lower state is a \Marvel{} energy but the other is predicted and No \Marvel{} transitions for which both upper and lower state energies are predicted from spectroscopic constants. These two groups contribute very few strong to moderate intensity transitions and are thus errors in the precise frequency of these transitions are unlikely to be important for cross-correlation of high-resolution spectra.

\begin{figure}
    \centering
    \includegraphics[width = 0.47\textwidth]{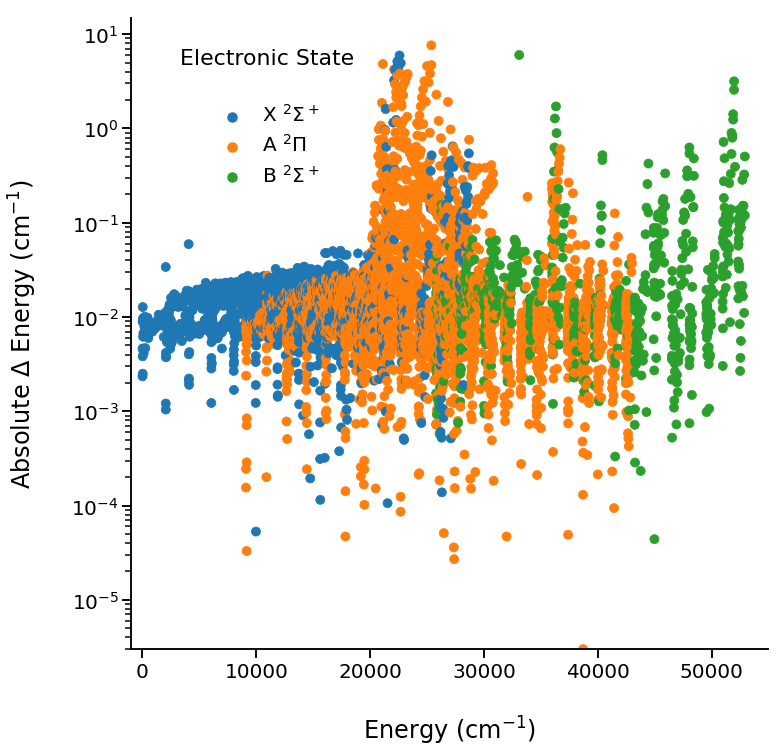}
    \caption{A comparison of the absolute energy difference (\Marvel{} - \Mollist{}) of the 6122 energy levels from the \Mollist{} data and the \Marvel{} procedure. Note the log scale of the y axis.}
    \label{fig:comp-all}
\end{figure}

It is worthwhile to compare the predicted energy levels from the \Mollist{} spectroscopic constants to the \Marvel{} energy levels in order to better understand the challenges of the traditional model approach. Figure \ref{fig:comp-all} shows the energy differences of the 6122 spin-rovibronic energy levels that were matched in both data sets. The log scale is used here to highlight the majority of transitions that have an absolute deviation between the two data sets of less than 0.05 \cm{} (90th percentile). 20.8\% of the \Marvel{} - \Mollist{} energy deviations are within the uncertainty of the \Marvel{} energy levels, with 71.9\% of the other deviations within an order of magnitude of the \Marvel{} uncertainties.

\begin{figure*}
    \centering
    \includegraphics[width = \textwidth]{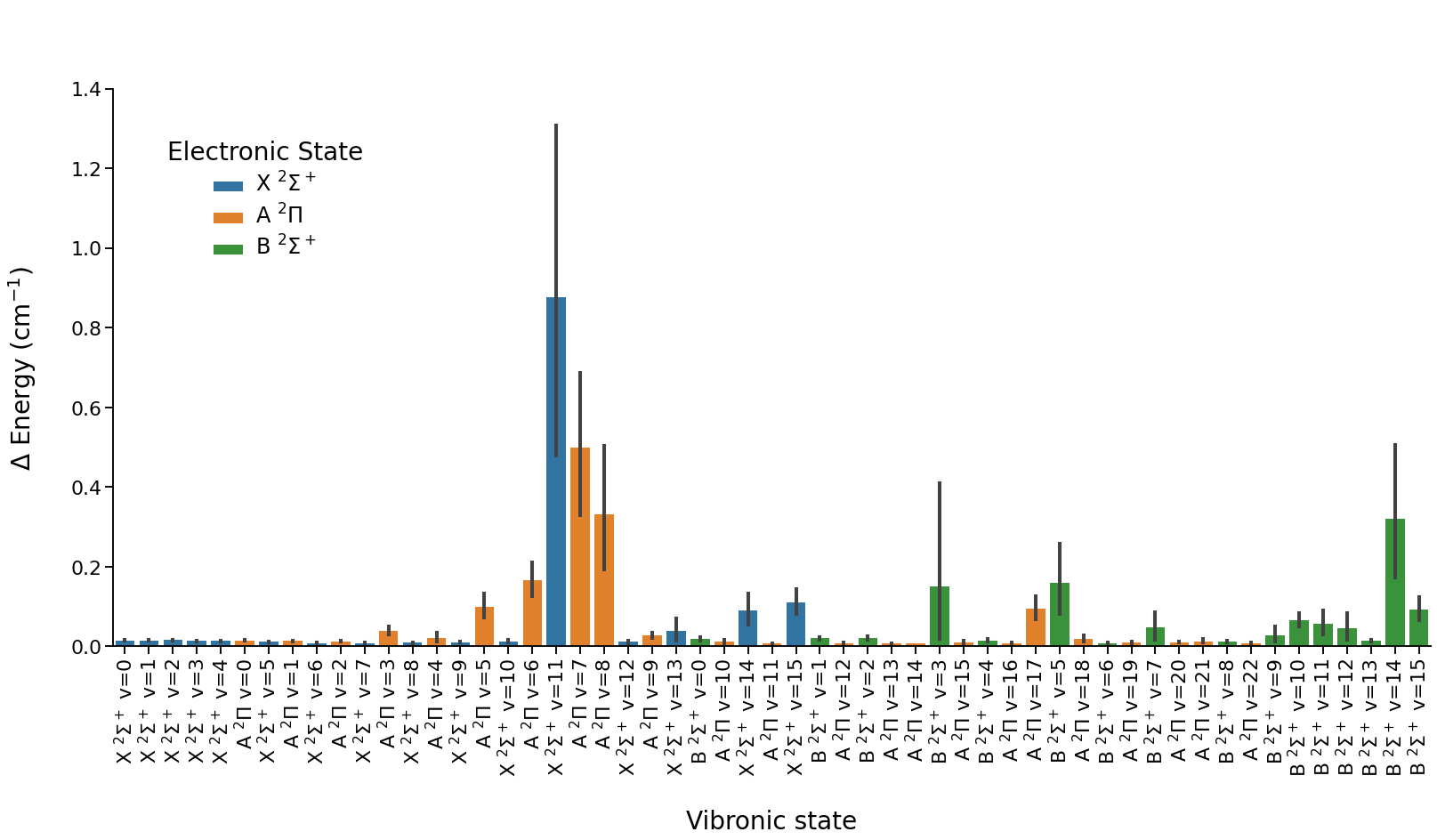}
    \caption{The average deviation between the \Mollist{} and \Marvel{} vibronic energy levels, including standard deviation.}
    \label{fig:compvib}
\end{figure*}

\begin{figure}
    \centering
    \includegraphics[width = 0.48\textwidth]{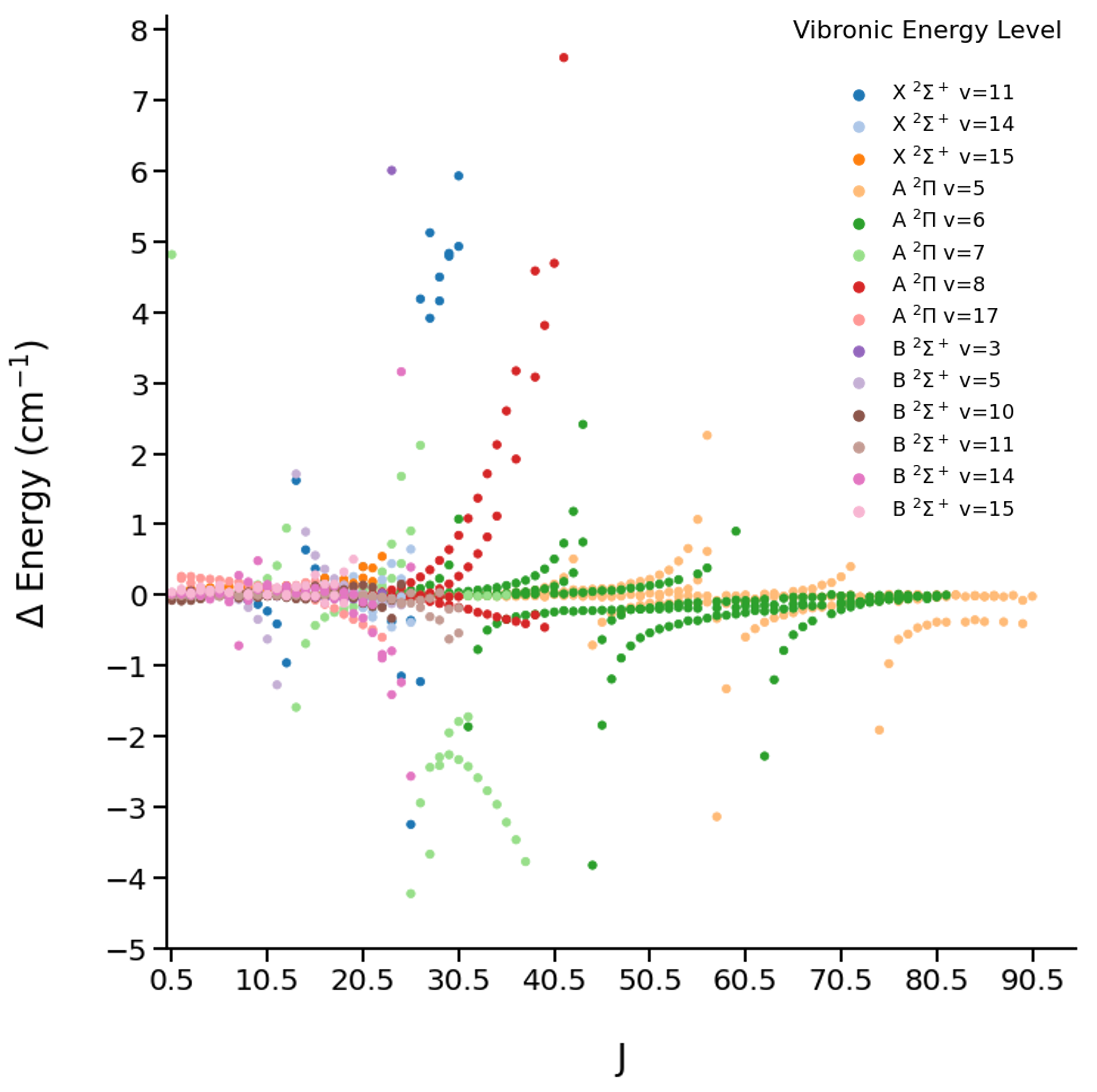}
    \caption{The energy deviations of the vibronic states along $J$, where the mean absolute deviation is greater than 0.05 \cm.}
    \label{fig:comp-MAD05}
\end{figure}

It is well-known that spectroscopic constant fits struggle near perturbations and our results confirm this. Figure \ref{fig:compvib} shows a bar plot of average absolute deviation of the vibronic states, including the standard deviation. While most vibronic states match very closely between \Mollist{} and the empirical energy levels from \Marvel{}, this figure clearly identifies the outliers. These outliers are explored further in figure \ref{fig:comp-MAD05} which plots $J$ vs change in energy for these vibronic states. This figure shows the  systematic deviation between the spectroscopic constant fit and the experimentally-derived energy levels as a function of $J$ that is characteristic of perturbations \citep{83OzNaSu.CN, 94ItKaKu.CN}. Our results show the effects of strong known perturbations due to crossing of states occur between: 
\begin{itemize} 
    \item \A{} ($v$ =  5) and \X ($v$ =  9) at high $J$ \citep{80KoFiSt.CN}
    \item \A{} ($v$ =  6) and \X ($v$ =  10) at high $J$ \citep{80KoFiSt.CN}
    \item \A($v$ =  7) and \X ($v$ =  11) states for $J$ around 12.5 and 27.5 \citep{89FuAlDa.CN, 93DaPaBe.CN, 62FaVaCl.CN, 80KoFiSt.CN, 10RaWaBe.CN}
    \item \A ($v$ =  8) collisional transfer with \X ($v$ =  12) \citep{89FuAlDa.CN}
    \item \B ($v$ =  0) crosses with \X ($v$ =  14) around $J = 29.5$ \citep{80KoFiSt.CN} 
    \item  \B ($v$ =  5) and \A$_{1/2}$ ($v$ =  17)  states for $J$ between  8.5 and 12.5 \citep{86JiAlDa.CN, 10RaWaBe.CN} 
    \item \B ($v$ =  10) perturbed by \A{} ($v$ =  24) \citep{83OzItSu.CN, 83OzNaSu.CN}
    \item \B ($v$ =  11) is perturbed by \A{} ($v$ =  26) \citep{94ItKaKu.CN, 84ItOzNa.CN} and by \asig{} \citep{83OzItSu.CN, 75CoRaSe.CN}
    \item \B ($v$ =  14) is perturbed by \A{} ($v$ =  30) \citep{94ItKaKu.CN, 83OzItSu.CN} and by b $^4\Pi$ \citep{84ItOzNa.CN} or \asig{} \citep{83OzItSu.CN}
    \alert{\item \B ($v$ =  15) is perturbed by \A{} ($v$ =  31) \citep{94ItKaKu.CN}}
\end{itemize}

It is worth noting that these highly perturbed energy levels are very likely to have high sensitivity to variation in the proton-to-electron mass ratio. Low-frequency transitions involving the \A{} ($v$ = 5-8) states and the nearby \X{} ($v$ =  8-11) states are likely to have high sensitivity, though of course the lower state population and thus the transition intensities will be very low. The $\Delta = -4$ transitions, e.g. \X{} ($v$ = 11) to \A{} ($v$ = 7), are likely to be most sensitive, but their intensities are not predicted by \Mollist{}. The $\Delta = -3$ transitions observed experimentally in Earth-based laboratories by 08CiSeKu \citep{08CiSeKu.CN} at 6500 K have frequencies 1905.8- 3117.2 \cm{}. \Mollist{} predicts 
Einstein A coefficients of 6468 \A-\X{} $\Delta = -3$ transitions, but the high initial state energy means that the predicted intensities of these transitions is maximised at $3.0 \cdot 10^{-28}$ cm/molecule at 1000 K (too small for realistic astrophysical study) rising to $5.5 \cdot 10^{-21}$ cm/molecule at 7000 K. The higher temperatures definitely give rise to more realistically observable lines and we know that, due to its high dissociation energy, astrophysical CN can exist at high temperatures and has been found in stars \citep{93BrSm.CN} and our Sun's sunspots \citep{15McKellar.CN}.

\section{Conclusion}

Within this paper, we have collated all available sources of high resolution rotationally resolved experimental transition assignments and determined empirical energy levels with reliable uncertainties for the CN radical through the \Marvel{} procedure. From \nosources{} sources, \notrans{} transitions were collated from 9 different electronic and 204 vibronic bands to generate \noenergy{} energy levels spanning \noelec{} electronic and \novibronic{} vibronic states. The relative lack of observed data of the higher electronic bands is evident. 


The current line list for CN from \Mollist{} has been updated with the empirical energy levels from the \Marvel{} analysis. 78.4\% of the energy levels have been replaced with \Marvel{} energies, recovering all but 0.0026\% of the partition function up to 2000K. The \Mollist{} transitions data has been \Marvel-ised with \alert{77.3\%} of all transitions being fully determined by experimentally-derived \Marvel{} energy levels, compared to 11.3\% of directly observed transitions used in the \Mollist{} line list. Of the strong transitions (intensities greater than $10^{-18}$ cm/molecule) 98.5\% have frequencies determined completely from \Marvel{} energies. The deviations between the \Marvel{} and \Mollist{} energies arise in rovibronic levels with known perturbations, as expected.

The complexity of the CN radical, and the several near degenerate energy levels, shows promise for CN to be a possible molecular probe to test the variation of the proton-to-electron mass ratio. This will require a full spectroscopic model, which will be available in a future publication. 





\section*{Acknowledgements} This research was undertaken with the assistance of resources from the National Computational Infrastructure (NCI Australia), an NCRIS enabled capability supported by the Australian Government.

The authors declare no conflicts of interest. 

\section*{Data Availability Statement}
The data underlying this article are available in the article and in its online supplementary material. These include the following files;
\begin{itemize}
    \item README\_CN.pdf - Explanation of the files within the SI
    \item 12C-14N\_MARVEL.txt - The final \Marvel{} transitions file
    \item 12C-14N\_MARVEL.energies - The final \Marvel{} energies file
    \item 12C-14N\_\_Mollist-Marvelised.states - The updated states file for the \Mollist{} line list, in ExoMol format (to be used with .trans file from the ExoMol website)
    \item SI.pdf - Expanded analysis tables
\end{itemize}


\bibliography{main}
\bibliographystyle{mnras}

\label{lastpage}

\end{document}


\title{Supplementary Information: Experimental energy levels of \ce{^12C^14N} through  \Marvel{} analysis}

\author{Anna-Maree Syme}
\affiliation{School of Chemistry, University of New South Wales, 2052 Sydney}

\author{Laura K. McKemmish}
\email{l.mckemmish@unsw.edu.au}

\affiliation{School of Chemistry, University of New South Wales, 2052 Sydney}

\date{\today}

\maketitle
\onecolumn

\section{MARVEL input transitions} Expansion of table 2 and 3 into vibronic bands 

\begin{center}
\footnotesize
\begin{longtable}{llcccrclllllllllllllllllll}
\multicolumn{8}{c}{{ \tablename\ S.\thetable{}}} \\
\toprule
Band & Vib & J-range & Trans. (V/A) & Freq range (\cm{}) & \mc{4}{c}{\Marvel{} - MoLLIST} \\
\cmidrule(r){6-9}
& & & & & \mc{1}{c}{Min} & \mc{1}{c}{MAD} & \mc{1}{c}{Max} & RMSD\\ 
\midrule
\endfirsthead

\multicolumn{8}{c}%
{{ \tablename\ S.\thetable{} -- continued from previous page}} \\
\toprule
Band & Vib & J-range & Trans. (V/A) & Freq range (\cm{}) & \mc{4}{c}{\Marvel{} - MoLLIST} \\
\cmidrule(r){6-8}
& & & & & \mc{1}{c}{Min} & \mc{1}{c}{MAD} & \mc{1}{c}{Max} & RMSD\\ 
\midrule
\endhead

\bottomrule
\multicolumn{8}{c}{{Continued on next page}} \\
\endfoot

\bottomrule
\endlastfoot
\mc{5}{l}{\textbf{55DoRo }} \\
A $^2\Pi$-X $^2\Sigma^+$ & (15-8) & 0.5 - 21.5 & 106/106 & 17479.4 - 17664.1 & 0.1 & 0.104 & 0.315 & 0.107 \\
B $^2\Sigma^+$-X $^2\Sigma^+$ & (0-1) & 0.5 - 24.5 & 88/88 & 23712.6 - 23896.1 & 0.1 & 0.136 & 0.363 & 0.151 \\
B $^2\Sigma^+$-X $^2\Sigma^+$ & (1-2) & 0.5 - 22.5 & 78/78 & 23820.4 - 23985.7 & 0.1 & 0.132 & 0.318 & 0.146 \\
B $^2\Sigma^+$-X $^2\Sigma^+$ & (14-10) & 0.5 - 9.5 & 35/35 & 31681.0 - 31745.6 & 0.1 & 0.219 & 0.924 & 0.295 \\
B $^2\Sigma^+$-X $^2\Sigma^+$ & (16-13) & 0.5 - 24.5 & 68/68 & 29121.6 - 29191.2 & 0.1 & 0.175 & 0.762 & 0.244 \\
B $^2\Sigma^+$-X $^2\Sigma^+$ & (18-17) & 0.5 - 21.5 & 74/76 & 24923.4 - 25078.3 & 0.1 & 0.112 & 0.432 & 0.127 \\
B $^2\Sigma^+$-X $^2\Sigma^+$ & (18-18) & 0.5 - 21.5 & 84/84 & 23344.2 - 23491.7 & 0.1 & 0.154 & 0.702 & 0.191 \\
B $^2\Sigma^+$-X $^2\Sigma^+$ & (19-15) & 0.5 - 26.5 & 90/90 & 29368.1 - 29525.6 & 0.1 & 0.107 & 0.252 & 0.11 \\
B $^2\Sigma^+$-X $^2\Sigma^+$ & (19-18) & 0.5 - 22.5 & 84/84 & 24514.3 - 24681.3 & 0.1 & 0.128 & 0.323 & 0.139 \\
B $^2\Sigma^+$-X $^2\Sigma^+$ & (2-3) & 0.5 - 19.5 & 76/76 & 23912.6 - 24055.8 & 0.1 & 0.12 & 0.262 & 0.129 \\
B $^2\Sigma^+$-X $^2\Sigma^+$ & (3-4) & 0.5 - 20.5 & 72/75 & 23988.3 - 24130.6 & 0.1 & 0.163 & 0.457 & 0.191 \\
B $^2\Sigma^+$-X $^2\Sigma^+$ & (4-5) & 0.5 - 23.5 & 72/74 & 24043.5 - 24169.1 & 0.1 & 0.125 & 0.38 & 0.14 \\
B $^2\Sigma^+$-X $^2\Sigma^+$ & (5-6) & 0.5 - 12.5 & 44/47 & 24093.0 - 24182.4 & 0.1 & 0.208 & 0.961 & 0.311 \\
D $^2\Pi$-A $^2\Pi$ & (0-6) & 1.5 - 27.5 & 86/86 & 34180.8 - 34523.6 & 0.1 & 0.113 & 0.607 & 0.129 \\
D $^2\Pi$-A $^2\Pi$ & (0-7) & 1.5 - 17.5 & 55/55 & 32662.0 - 32889.5 & 0.1 & 0.128 & 0.931 & 0.176 \\
D $^2\Pi$-A $^2\Pi$ & (1-4) & 2.5 - 21.5 & 60/60 & 38606.4 - 38854.9 & 0.1 & 0.102 & 0.145 & 0.102 \\
D $^2\Pi$-A $^2\Pi$ & (2-4) & 1.5 - 22.5 & 77/77 & 39524.6 - 39825.8 & 0.1 & 0.101 & 0.131 & 0.101 \\
D $^2\Pi$-A $^2\Pi$ & (3-3) & 3.5 - 18.5 & 57/57 & 42223.6 - 42482.0 & 0.1 & 0.112 & 0.396 & 0.125 \\
F $^2\Delta$-A $^2\Pi$ & (0-2) & 2.5 - 32.5 & 58/58 & 46694.9 - 47062.5 & 0.1 & 0.142 & 0.5 & 0.171 \\
H $^2\Pi$-B $^2\Sigma^+$ & (0-0) & 0.5 - 16.5 & 73/73 & 34943.7 - 35165.1 & 0.1 & 0.102 & 0.148 & 0.102 \\
H $^2\Pi$-B $^2\Sigma^+$ & (0-1) & 0.5 - 17.5 & 67/67 & 32878.4 - 33041.7 & 0.1 & 0.106 & 0.249 & 0.108 \\
\\ \mc{5}{l}{\textbf{56Ca }} \\
E $^2\Sigma^+$-X $^2\Sigma^+$ & (0-1) & 2.5 - 26.5 & 86/86 & 56655.3 - 56924.6 & 0.06 & 0.085 & 0.484 & 0.111 \\
E $^2\Sigma^+$-X $^2\Sigma^+$ & (0-2) & 2.5 - 26.5 & 92/92 & 54585.8 - 54908.9 & 0.06 & 0.12 & 0.231 & 0.131 \\
E $^2\Sigma^+$-X $^2\Sigma^+$ & (0-3) & 0.5 - 29.5 & 113/113 & 52517.9 - 52919.2 & 0.06 & 0.072 & 0.162 & 0.075 \\
E $^2\Sigma^+$-X $^2\Sigma^+$ & (0-4) & 1.5 - 32.5 & 119/121 & 50546.9 - 50956.0 & 0.06 & 0.082 & 0.308 & 0.095 \\
E $^2\Sigma^+$-X $^2\Sigma^+$ & (1-1) & 1.5 - 31.5 & 119/119 & 58075.8 - 58595.6 & 0.06 & 0.094 & 0.274 & 0.104 \\
J $^2\Delta$-A $^2\Pi$ & (0-0) & 1.5 - 28.5 & 262/262 & 55325.7 - 55670.3 & 0.06 & 0.076 & 0.247 & 0.084 \\
J $^2\Delta$-A $^2\Pi$ & (1-0) & 1.5 - 27.5 & 280/280 & 56427.6 - 56763.1 & 0.06 & 0.079 & 0.275 & 0.09 \\
J $^2\Delta$-A $^2\Pi$ & (2-0) & 1.5 - 30.5 & 290/290 & 57417.1 - 57829.7 & 0.06 & 0.073 & 0.357 & 0.082 \\
J $^2\Delta$-A $^2\Pi$ & (3-0) & 2.5 - 21.5 & 146/154 & 58642.2 - 58870.7 & 0.06 & 0.074 & 0.492 & 0.092 \\
\\ \mc{5}{l}{\textbf{67WeFiRa }} \\
A $^2\Pi$-X $^2\Sigma^+$ & (0-1) & 3.5 - 55.5 & 207/240 & 6545.8 - 7107.1 & 0.1 & 0.225 & 0.55 & 0.257 \\
A $^2\Pi$-X $^2\Sigma^+$ & (1-2) & 10.5 - 46.5 & 170/193 & 6357.4 - 6876.7 & 0.1 & 0.23 & 0.545 & 0.262 \\
\\ \mc{5}{l}{\textbf{70Lu }} \\
E $^2\Sigma^+$-X $^2\Sigma^+$ & (0-0) & 12.5 - 32.5 & 56/56 & 58547.0 - 58850.9 & 0.05 & 0.093 & 0.244 & 0.109 \\
E $^2\Sigma^+$-X $^2\Sigma^+$ & (1-0) & 0.5 - 35.5 & 132/132 & 60038.6 - 60637.5 & 0.05 & 0.117 & 0.423 & 0.145 \\
E $^2\Sigma^+$-X $^2\Sigma^+$ & (1-1) & 0.5 - 42.5 & 158/158 & 57857.7 - 58595.7 & 0.05 & 0.073 & 0.278 & 0.084 \\
E $^2\Sigma^+$-X $^2\Sigma^+$ & (2-0) & 0.5 - 38.5 & 144/144 & 61650.6 - 62291.8 & 0.05 & 0.085 & 0.5 & 0.11 \\
E $^2\Sigma^+$-X $^2\Sigma^+$ & (2-1) & 2.5 - 31.5 & 104/104 & 59856.6 - 60249.6 & 0.05 & 0.093 & 0.336 & 0.115 \\
E $^2\Sigma^+$-X $^2\Sigma^+$ & (3-0) & 0.5 - 42.5 & 168/168 & 62974.7 - 63923.3 & 0.05 & 0.066 & 0.248 & 0.078 \\
E $^2\Sigma^+$-X $^2\Sigma^+$ & (4-0) & 1.5 - 32.5 & 106/106 & 65157.7 - 65526.7 & 0.05 & 0.059 & 0.174 & 0.064 \\
E $^2\Sigma^+$-X $^2\Sigma^+$ & (5-0) & 0.5 - 37.5 & 144/144 & 66293.3 - 67093.6 & 0.05 & 0.086 & 0.292 & 0.102 \\
\\ \mc{5}{l}{\textbf{71Lu }} \\
F $^2\Delta$-A $^2\Pi$ & (1-6) & 1.5 - 16.5 & 50/51 & 41324.4 - 41496.9 & 0.05 & 0.117 & 0.727 & 0.174 \\
F $^2\Delta$-A $^2\Pi$ & (1-7) & 0.5 - 15.5 & 56/59 & 39660.3 - 39863.9 & 0.05 & 0.112 & 0.727 & 0.173 \\
F $^2\Delta$-A $^2\Pi$ & (2-7) & 0.5 - 12.5 & 49/49 & 40879.5 - 41050.2 & 0.05 & 0.182 & 0.759 & 0.283 \\
F $^2\Delta$-A $^2\Pi$ & (2-8) & 0.5 - 16.5 & 44/44 & 39306.2 - 39443.0 & 0.05 & 0.215 & 1.247 & 0.363 \\
\\ \mc{5}{l}{\textbf{73Sc}} \\
B $^2\Sigma^+$-A $^2\Pi$ & (7-4) & 0.5 - 19.5 & 70/72 & 23582.6 - 23708.4 & 0.03 & 0.099 & 0.249 & 0.111 \\
\\ \mc{5}{l}{\textbf{74En }} \\
B $^2\Sigma^+$-X $^2\Sigma^+$ & (0-0) & 0.5 - 29.5 & 111/115 & 25743.4 - 25973.2 & 0.03 & 0.051 & 0.302 & 0.066 \\
B $^2\Sigma^+$-X $^2\Sigma^+$ & (1-0) & 0.5 - 25.5 & 88/88 & 27860.0 - 28046.1 & 0.03 & 0.054 & 0.177 & 0.062 \\
B $^2\Sigma^+$-X $^2\Sigma^+$ & (1-1) & 0.5 - 25.5 & 96/96 & 25823.9 - 26014.2 & 0.03 & 0.04 & 0.155 & 0.046 \\
B $^2\Sigma^+$-X $^2\Sigma^+$ & (2-1) & 0.5 - 22.5 & 75/76 & 27902.0 - 28066.3 & 0.03 & 0.066 & 0.146 & 0.073 \\
B $^2\Sigma^+$-X $^2\Sigma^+$ & (2-2) & 0.5 - 23.5 & 79/80 & 25893.7 - 26064.6 & 0.03 & 0.045 & 0.134 & 0.05 \\
B $^2\Sigma^+$-X $^2\Sigma^+$ & (3-2) & 0.5 - 19.5 & 68/70 & 27934.5 - 28077.0 & 0.03 & 0.055 & 0.19 & 0.065 \\
B $^2\Sigma^+$-X $^2\Sigma^+$ & (3-3) & 0.5 - 23.5 & 89/89 & 25941.0 - 26118.5 & 0.03 & 0.038 & 0.103 & 0.041 \\
B $^2\Sigma^+$-X $^2\Sigma^+$ & (4-3) & 0.5 - 23.5 & 82/82 & 27929.7 - 28095.2 & 0.03 & 0.068 & 0.183 & 0.076 \\
B $^2\Sigma^+$-X $^2\Sigma^+$ & (4-4) & 0.5 - 22.5 & 76/78 & 25971.3 - 26112.5 & 0.03 & 0.042 & 0.092 & 0.046 \\
B $^2\Sigma^+$-X $^2\Sigma^+$ & (5-4) & 0.5 - 24.5 & 81/85 & 27914.9 - 28084.5 & 0.03 & 0.068 & 0.458 & 0.086 \\
B $^2\Sigma^+$-X $^2\Sigma^+$ & (5-5) & 0.5 - 20.5 & 62/67 & 25983.0 - 26110.3 & 0.03 & 0.056 & 0.355 & 0.074 \\
B $^2\Sigma^+$-X $^2\Sigma^+$ & (6-5) & 0.5 - 18.5 & 61/61 & 27878.4 - 28012.7 & 0.03 & 0.072 & 0.209 & 0.081 \\
B $^2\Sigma^+$-X $^2\Sigma^+$ & (6-6) & 0.5 - 10.5 & 20/20 & 25992.4 - 26063.3 & 0.03 & 0.059 & 0.153 & 0.069 \\
B $^2\Sigma^+$-X $^2\Sigma^+$ & (7-6) & 0.5 - 15.5 & 31/31 & 27828.9 - 27925.6 & 0.03 & 0.064 & 0.209 & 0.073 \\
B $^2\Sigma^+$-X $^2\Sigma^+$ & (7-7) & 2.5 - 13.5 & 21/22 & 25944.6 - 26041.2 & 0.03 & 0.058 & 0.204 & 0.071 \\
\\ \mc{5}{l}{\textbf{75CoRaSe }} \\
B $^2\Sigma^+$-X $^2\Sigma^+$ & (11-10) & 3.5 - 20.5 & 59/59 & 27182.4 - 27322.6 & 0.01 & 0.026 & 0.075 & 0.031 \\
B $^2\Sigma^+$-X $^2\Sigma^+$ & (11-11) & 1.5 - 23.5 & 69/69 & 25399.1 - 25558.5 & 0.01 & 0.024 & 0.085 & 0.03 \\
\\ \mc{5}{l}{\textbf{77DiWo }} \\
X $^2\Sigma^+$-X $^2\Sigma^+$ & (1-1) & 0.5 - 1.5 & 2/2 & 3.8 - 3.8 & 0.0 & 0.0 & 0.0 & 0.0 \\
X $^2\Sigma^+$-X $^2\Sigma^+$ & (2-2) & 0.5 - 1.5 & 2/2 & 3.8 - 3.8 & 0.0 & 0.0 & 0.0 & 0.0 \\
\\ \mc{5}{l}{\textbf{89FuAlDa }} \\
B $^2\Sigma^+$-A $^2\Pi$ & (8-7) & 0.5 - 19.5 & 94/115 & 20352.0 - 20543.5 & 0.1 & 0.198 & 0.5 & 0.226 \\
B $^2\Sigma^+$-X $^2\Sigma^+$ & (8-11) & 0.5 - 21.5 & 75/77 & 20461.0 - 20605.0 & 0.05 & 0.148 & 0.473 & 0.185 \\
\\ \mc{5}{l}{\textbf{91DaBrAb }} \\
X $^2\Sigma^+$-X $^2\Sigma^+$ & (2-0) & 2.5 - 57.5 & 156/156 & 3743.5 - 4155.4 & 0.01 & 0.011 & 0.038 & 0.012 \\
\\ \mc{5}{l}{\textbf{92ReSuMi }} \\
B $^2\Sigma^+$-X $^2\Sigma^+$ & (0-0) & 1.5 - 63.5 & 148/148 & 25745.0 - 26299.3 & 0.02 & 0.02 & 0.029 & 0.02 \\
B $^2\Sigma^+$-X $^2\Sigma^+$ & (0-1) & 0.5 - 13.5 & 31/31 & 23720.0 - 23819.6 & 0.02 & 0.02 & 0.02 & 0.02 \\
B $^2\Sigma^+$-X $^2\Sigma^+$ & (1-1) & 0.5 - 41.5 & 60/60 & 25834.5 - 26149.5 & 0.02 & 0.02 & 0.031 & 0.02 \\
B $^2\Sigma^+$-X $^2\Sigma^+$ & (1-2) & 0.5 - 11.5 & 40/40 & 23829.2 - 23914.6 & 0.02 & 0.022 & 0.037 & 0.022 \\
B $^2\Sigma^+$-X $^2\Sigma^+$ & (11-11) & 0.5 - 8.5 & 26/26 & 25469.3 - 25523.1 & 0.02 & 0.024 & 0.044 & 0.025 \\
B $^2\Sigma^+$-X $^2\Sigma^+$ & (13-13) & 0.5 - 11.5 & 37/37 & 25000.6 - 25075.0 & 0.02 & 0.02 & 0.02 & 0.02 \\
B $^2\Sigma^+$-X $^2\Sigma^+$ & (2-2) & 0.5 - 14.5 & 46/46 & 25901.4 - 26000.3 & 0.02 & 0.021 & 0.049 & 0.021 \\
B $^2\Sigma^+$-X $^2\Sigma^+$ & (2-3) & 0.5 - 12.5 & 44/44 & 23920.9 - 24012.9 & 0.02 & 0.021 & 0.034 & 0.021 \\
B $^2\Sigma^+$-X $^2\Sigma^+$ & (3-3) & 0.5 - 15.5 & 46/46 & 25958.4 - 26071.2 & 0.02 & 0.022 & 0.108 & 0.026 \\
B $^2\Sigma^+$-X $^2\Sigma^+$ & (3-4) & 0.5 - 11.5 & 39/39 & 24002.1 - 24084.0 & 0.02 & 0.021 & 0.031 & 0.021 \\
B $^2\Sigma^+$-X $^2\Sigma^+$ & (4-4) & 0.5 - 14.5 & 51/51 & 25988.2 - 26091.5 & 0.02 & 0.02 & 0.033 & 0.021 \\
B $^2\Sigma^+$-X $^2\Sigma^+$ & (4-5) & 0.5 - 11.5 & 40/40 & 24060.9 - 24141.8 & 0.02 & 0.023 & 0.137 & 0.03 \\
B $^2\Sigma^+$-X $^2\Sigma^+$ & (6-6) & 0.5 - 10.5 & 32/32 & 25998.3 - 26067.6 & 0.02 & 0.02 & 0.02 & 0.02 \\
B $^2\Sigma^+$-X $^2\Sigma^+$ & (6-7) & 0.5 - 12.5 & 40/40 & 24106.0 - 24185.1 & 0.02 & 0.02 & 0.02 & 0.02 \\
B $^2\Sigma^+$-X $^2\Sigma^+$ & (7-7) & 0.5 - 12.5 & 24/24 & 25944.7 - 26025.0 & 0.02 & 0.022 & 0.039 & 0.022 \\
B $^2\Sigma^+$-X $^2\Sigma^+$ & (7-8) & 0.5 - 13.5 & 28/28 & 24088.0 - 24182.4 & 0.02 & 0.022 & 0.039 & 0.023 \\
B $^2\Sigma^+$-X $^2\Sigma^+$ & (8-8) & 0.5 - 10.5 & 32/32 & 25878.5 - 25951.9 & 0.02 & 0.023 & 0.116 & 0.029 \\
B $^2\Sigma^+$-X $^2\Sigma^+$ & (8-9) & 0.5 - 9.5 & 31/31 & 24052.8 - 24118.6 & 0.02 & 0.03 & 0.29 & 0.056 \\
B $^2\Sigma^+$-X $^2\Sigma^+$ & (9-10) & 1.5 - 9.5 & 29/29 & 23974.2 - 24039.3 & 0.02 & 0.02 & 0.02 & 0.02 \\
B $^2\Sigma^+$-X $^2\Sigma^+$ & (9-9) & 0.5 - 11.5 & 26/26 & 25769.4 - 25842.2 & 0.02 & 0.02 & 0.02 & 0.02 \\
\\ \mc{5}{l}{\textbf{92PrBe}} \\
A $^2\Pi$-X $^2\Sigma^+$ & (10-3) & 0.5 - 5.5 & 9/9 & 19751.6 - 19774.3 & 0.025 & 0.025 & 0.025 & 0.025 \\
A $^2\Pi$-X $^2\Sigma^+$ & (10-4) & 0.5 - 4.5 & 10/10 & 17792.7 - 17816.5 & 0.025 & 0.025 & 0.025 & 0.025 \\
A $^2\Pi$-X $^2\Sigma^+$ & (11-4) & 0.5 - 4.5 & 11/11 & 19314.5 - 19344.9 & 0.025 & 0.025 & 0.025 & 0.025 \\
A $^2\Pi$-X $^2\Sigma^+$ & (11-5) & 0.5 - 5.5 & 12/12 & 17382.8 - 17410.5 & 0.025 & 0.026 & 0.033 & 0.026 \\
A $^2\Pi$-X $^2\Sigma^+$ & (12-4) & 0.5 - 4.5 & 10/10 & 20824.4 - 20849.9 & 0.025 & 0.025 & 0.025 & 0.025 \\
A $^2\Pi$-X $^2\Sigma^+$ & (12-5) & 0.5 - 5.5 & 11/11 & 18887.8 - 18913.0 & 0.025 & 0.025 & 0.025 & 0.025 \\
A $^2\Pi$-X $^2\Sigma^+$ & (12-6) & 0.5 - 4.5 & 11/11 & 16972.8 - 17002.6 & 0.025 & 0.025 & 0.025 & 0.025 \\
A $^2\Pi$-X $^2\Sigma^+$ & (13-5) & 0.5 - 5.5 & 12/12 & 20361.7 - 20392.0 & 0.025 & 0.028 & 0.049 & 0.029 \\
A $^2\Pi$-X $^2\Sigma^+$ & (13-6) & 0.5 - 3.5 & 7/7 & 18456.6 - 18475.0 & 0.025 & 0.026 & 0.031 & 0.026 \\
A $^2\Pi$-X $^2\Sigma^+$ & (13-7) & 0.5 - 4.5 & 10/10 & 16572.9 - 16597.5 & 0.025 & 0.035 & 0.12 & 0.045 \\
A $^2\Pi$-X $^2\Sigma^+$ & (14-5) & 0.5 - 2.5 & 6/6 & 21824.2 - 21842.2 & 0.025 & 0.025 & 0.025 & 0.025 \\
A $^2\Pi$-X $^2\Sigma^+$ & (14-7) & 0.5 - 5.5 & 11/11 & 18025.8 - 18050.5 & 0.025 & 0.025 & 0.025 & 0.025 \\
A $^2\Pi$-X $^2\Sigma^+$ & (15-7) & 0.5 - 5.5 & 11/11 & 19452.5 - 19477.2 & 0.025 & 0.026 & 0.035 & 0.026 \\
A $^2\Pi$-X $^2\Sigma^+$ & (15-8) & 0.5 - 6.5 & 13/13 & 17595.4 - 17622.0 & 0.025 & 0.025 & 0.031 & 0.025 \\
A $^2\Pi$-X $^2\Sigma^+$ & (16-7) & 0.5 - 5.5 & 11/11 & 20857.3 - 20879.5 & 0.025 & 0.025 & 0.025 & 0.025 \\
A $^2\Pi$-X $^2\Sigma^+$ & (16-8) & 0.5 - 4.5 & 10/10 & 18990.7 - 19017.5 & 0.025 & 0.028 & 0.059 & 0.03 \\
A $^2\Pi$-X $^2\Sigma^+$ & (16-9) & 0.5 - 3.5 & 6/6 & 17169.4 - 17186.7 & 0.025 & 0.025 & 0.025 & 0.025 \\
A $^2\Pi$-X $^2\Sigma^+$ & (19-10) & 0.5 - 5.5 & 12/12 & 19402.4 - 19428.1 & 0.025 & 0.026 & 0.032 & 0.026 \\
A $^2\Pi$-X $^2\Sigma^+$ & (19-11) & 0.5 - 3.5 & 9/9 & 17625.4 - 17649.2 & 0.025 & 0.025 & 0.025 & 0.025 \\
A $^2\Pi$-X $^2\Sigma^+$ & (20-10) & 1.5 - 3.5 & 7/7 & 20699.8 - 20719.3 & 0.025 & 0.032 & 0.073 & 0.036 \\
A $^2\Pi$-X $^2\Sigma^+$ & (20-11) & 0.5 - 3.5 & 6/6 & 18922.7 - 18939.7 & 0.025 & 0.025 & 0.025 & 0.025 \\
A $^2\Pi$-X $^2\Sigma^+$ & (21-10) & 0.5 - 4.5 & 9/9 & 21955.3 - 21982.5 & 0.025 & 0.057 & 0.147 & 0.071 \\
A $^2\Pi$-X $^2\Sigma^+$ & (21-11) & 0.5 - 3.5 & 8/8 & 20183.7 - 20207.5 & 0.025 & 0.032 & 0.063 & 0.034 \\
A $^2\Pi$-X $^2\Sigma^+$ & (8-1) & 0.5 - 3.5 & 7/7 & 20625.6 - 20643.9 & 0.025 & 0.034 & 0.055 & 0.036 \\
A $^2\Pi$-X $^2\Sigma^+$ & (8-2) & 0.5 - 6.5 & 12/12 & 18601.9 - 18627.9 & 0.025 & 0.03 & 0.074 & 0.033 \\
A $^2\Pi$-X $^2\Sigma^+$ & (8-3) & 0.5 - 4.5 & 11/11 & 16607.8 - 16638.2 & 0.025 & 0.028 & 0.056 & 0.029 \\
A $^2\Pi$-X $^2\Sigma^+$ & (9-3) & 0.5 - 5.5 & 14/14 & 18184.6 - 18223.2 & 0.025 & 0.028 & 0.051 & 0.029 \\
\\ \mc{5}{l}{\textbf{92PrBeFr }} \\
B $^2\Sigma^+$-X $^2\Sigma^+$ & (0-0) & 0.5 - 26.5 & 92/92 & 25745.3 - 25943.3 & 0.035 & 0.035 & 0.046 & 0.035 \\
B $^2\Sigma^+$-X $^2\Sigma^+$ & (1-0) & 0.5 - 21.5 & 78/78 & 27860.0 - 28005.4 & 0.035 & 0.047 & 0.167 & 0.053 \\
B $^2\Sigma^+$-X $^2\Sigma^+$ & (1-1) & 0.5 - 11.5 & 42/42 & 25844.9 - 25928.8 & 0.035 & 0.035 & 0.048 & 0.035 \\
B $^2\Sigma^+$-X $^2\Sigma^+$ & (2-1) & 0.5 - 9.5 & 30/30 & 27938.0 - 28000.1 & 0.035 & 0.043 & 0.108 & 0.046 \\
B $^2\Sigma^+$-X $^2\Sigma^+$ & (2-2) & 0.5 - 10.5 & 36/36 & 25912.1 - 25980.4 & 0.035 & 0.036 & 0.054 & 0.036 \\
B $^2\Sigma^+$-X $^2\Sigma^+$ & (3-2) & 0.5 - 9.5 & 36/36 & 27953.6 - 28024.3 & 0.035 & 0.04 & 0.099 & 0.041 \\
B $^2\Sigma^+$-X $^2\Sigma^+$ & (3-3) & 0.5 - 8.5 & 30/30 & 25971.2 - 26031.1 & 0.035 & 0.035 & 0.044 & 0.035 \\
B $^2\Sigma^+$-X $^2\Sigma^+$ & (4-3) & 0.5 - 8.5 & 32/32 & 27962.8 - 28025.3 & 0.035 & 0.057 & 0.139 & 0.067 \\
B $^2\Sigma^+$-X $^2\Sigma^+$ & (4-4) & 0.5 - 7.5 & 26/26 & 26003.8 - 26054.2 & 0.035 & 0.035 & 0.043 & 0.035 \\
B $^2\Sigma^+$-X $^2\Sigma^+$ & (5-4) & 0.5 - 6.5 & 18/18 & 27960.6 - 28001.3 & 0.035 & 0.06 & 0.135 & 0.071 \\
B $^2\Sigma^+$-X $^2\Sigma^+$ & (5-5) & 0.5 - 5.5 & 17/17 & 26024.2 - 26060.7 & 0.035 & 0.035 & 0.035 & 0.035 \\
B $^2\Sigma^+$-X $^2\Sigma^+$ & (6-5) & 0.5 - 11.5 & 38/38 & 27903.9 - 27980.5 & 0.035 & 0.045 & 0.169 & 0.052 \\
B $^2\Sigma^+$-X $^2\Sigma^+$ & (6-6) & 0.5 - 6.5 & 21/21 & 26004.5 - 26047.0 & 0.035 & 0.036 & 0.058 & 0.036 \\
B $^2\Sigma^+$-X $^2\Sigma^+$ & (7-6) & 0.5 - 18.5 & 60/60 & 27804.6 - 27914.7 & 0.035 & 0.039 & 0.091 & 0.04 \\
B $^2\Sigma^+$-X $^2\Sigma^+$ & (7-7) & 0.5 - 7.5 & 10/10 & 25960.1 - 26005.5 & 0.035 & 0.035 & 0.035 & 0.035 \\
B $^2\Sigma^+$-X $^2\Sigma^+$ & (8-7) & 0.5 - 17.5 & 68/68 & 27707.5 - 27831.4 & 0.035 & 0.037 & 0.085 & 0.038 \\
B $^2\Sigma^+$-X $^2\Sigma^+$ & (9-8) & 0.5 - 22.5 & 86/86 & 27551.0 - 27705.5 & 0.035 & 0.044 & 0.132 & 0.047 \\
X $^2\Sigma^+$-X $^2\Sigma^+$ & (0-0) & 0.5 - 3.5 & 9/9 & 3.8 - 15.1 & 0.035 & 0.035 & 0.035 & 0.035 \\
X $^2\Sigma^+$-X $^2\Sigma^+$ & (1-1) & 0.5 - 3.5 & 7/7 & 3.7 - 11.2 & 0.035 & 0.035 & 0.035 & 0.035 \\
X $^2\Sigma^+$-X $^2\Sigma^+$ & (10-10) & 1.5 - 3.5 & 2/2 & 10.3 - 10.3 & 0.035 & 0.035 & 0.035 & 0.035 \\
X $^2\Sigma^+$-X $^2\Sigma^+$ & (2-2) & 0.5 - 3.5 & 6/6 & 3.7 - 11.1 & 0.035 & 0.049 & 0.083 & 0.052 \\
X $^2\Sigma^+$-X $^2\Sigma^+$ & (3-3) & 0.5 - 2.5 & 5/5 & 3.7 - 11.0 & 0.035 & 0.035 & 0.035 & 0.035 \\
X $^2\Sigma^+$-X $^2\Sigma^+$ & (4-4) & 0.5 - 2.5 & 4/4 & 3.6 - 7.3 & 0.035 & 0.035 & 0.035 & 0.035 \\
X $^2\Sigma^+$-X $^2\Sigma^+$ & (5-5) & 0.5 - 2.5 & 4/4 & 3.6 - 7.2 & 0.035 & 0.035 & 0.035 & 0.035 \\
X $^2\Sigma^+$-X $^2\Sigma^+$ & (6-6) & 0.5 - 2.5 & 4/4 & 3.6 - 7.1 & 0.035 & 0.035 & 0.035 & 0.035 \\
X $^2\Sigma^+$-X $^2\Sigma^+$ & (7-7) & 0.5 - 2.5 & 4/4 & 3.5 - 7.1 & 0.035 & 0.035 & 0.035 & 0.035 \\
X $^2\Sigma^+$-X $^2\Sigma^+$ & (8-8) & 0.5 - 3.5 & 6/6 & 3.5 - 10.5 & 0.035 & 0.035 & 0.035 & 0.035 \\
X $^2\Sigma^+$-X $^2\Sigma^+$ & (9-9) & 0.5 - 3.5 & 3/3 & 3.5 - 10.4 & 0.035 & 0.035 & 0.035 & 0.035 \\
\\ \mc{5}{l}{\textbf{94ItKaKu }} \\
B $^2\Sigma^+$-X $^2\Sigma^+$ & (11-11) & 22.5 - 41.5 & 60/60 & 25296.4 - 25570.5 & 0.1 & 0.1 & 0.1 & 0.1 \\
B $^2\Sigma^+$-X $^2\Sigma^+$ & (14-14) & 17.5 - 37.5 & 50/56 & 24551.7 - 24814.8 & 0.1 & 0.148 & 0.5 & 0.18 \\
B $^2\Sigma^+$-X $^2\Sigma^+$ & (15-15) & 0.5 - 19.5 & 66/66 & 24383.8 - 24510.8 & 0.1 & 0.12 & 0.376 & 0.132 \\
B $^2\Sigma^+$-X $^2\Sigma^+$ & (16-14) & 0.5 - 37.5 & 133/136 & 27206.9 - 27495.8 & 0.1 & 0.19 & 0.612 & 0.218 \\
B $^2\Sigma^+$-X $^2\Sigma^+$ & (18-17) & 16.5 - 39.5 & 61/61 & 24813.5 - 25068.9 & 0.1 & 0.106 & 0.355 & 0.112 \\
B $^2\Sigma^+$-X $^2\Sigma^+$ & (19-18) & 19.5 - 38.5 & 40/40 & 24400.8 - 24658.9 & 0.1 & 0.1 & 0.1 & 0.1 \\
\\ \mc{5}{l}{\textbf{01LiDuLi}} \\
A $^2\Pi$-X $^2\Sigma^+$ & (2-0) & 0.5 - 22.5 & 189/189 & 12496.8 - 12735.0 & 0.007 & 0.013 & 0.043 & 0.014 \\
\\ \mc{5}{l}{\textbf{04HoCiSp }} \\
X $^2\Sigma^+$-X $^2\Sigma^+$ & (1-0) & 0.5 - 28.5 & 104/108 & 1928.6 - 2133.6 & 0.027 & 0.028 & 0.057 & 0.029 \\
X $^2\Sigma^+$-X $^2\Sigma^+$ & (2-1) & 0.5 - 31.5 & 106/110 & 1912.4 - 2114.2 & 0.027 & 0.029 & 0.078 & 0.03 \\
X $^2\Sigma^+$-X $^2\Sigma^+$ & (3-2) & 0.5 - 30.5 & 94/94 & 1913.4 - 2084.1 & 0.027 & 0.029 & 0.065 & 0.03 \\
X $^2\Sigma^+$-X $^2\Sigma^+$ & (4-3) & 0.5 - 27.5 & 92/92 & 1879.1 - 2048.9 & 0.027 & 0.028 & 0.065 & 0.029 \\
X $^2\Sigma^+$-X $^2\Sigma^+$ & (5-4) & 0.5 - 26.5 & 92/92 & 1853.4 - 2018.9 & 0.027 & 0.028 & 0.065 & 0.029 \\
X $^2\Sigma^+$-X $^2\Sigma^+$ & (6-5) & 0.5 - 25.5 & 76/76 & 1844.5 - 1988.8 & 0.027 & 0.027 & 0.027 & 0.027 \\
X $^2\Sigma^+$-X $^2\Sigma^+$ & (7-6) & 0.5 - 21.5 & 67/67 & 1830.9 - 1950.6 & 0.027 & 0.027 & 0.033 & 0.027 \\
X $^2\Sigma^+$-X $^2\Sigma^+$ & (8-7) & 0.5 - 18.5 & 56/56 & 1816.7 - 1914.9 & 0.027 & 0.029 & 0.074 & 0.029 \\
\\ \mc{5}{l}{\textbf{05HuCaDa }} \\
X $^2\Sigma^+$-X $^2\Sigma^+$ & (1-0) & 4.5 - 17.5 & 36/36 & 1982.1 - 2101.3 & 0.0 & 0.003 & 0.014 & 0.004 \\
\\ \mc{5}{l}{\textbf{06RaDaWa }} \\
B $^2\Sigma^+$-X $^2\Sigma^+$ & (0-0) & 0.5 - 63.5 & 326/326 & 25745.0 - 26299.3 & 0.03 & 0.031 & 0.097 & 0.031 \\
B $^2\Sigma^+$-X $^2\Sigma^+$ & (0-1) & 0.5 - 15.5 & 89/89 & 23717.1 - 23831.9 & 0.03 & 0.031 & 0.051 & 0.031 \\
B $^2\Sigma^+$-X $^2\Sigma^+$ & (1-0) & 0.5 - 21.5 & 78/78 & 27860.0 - 28005.4 & 0.03 & 0.047 & 0.128 & 0.053 \\
B $^2\Sigma^+$-X $^2\Sigma^+$ & (1-1) & 0.5 - 41.5 & 154/156 & 25829.9 - 26149.5 & 0.03 & 0.035 & 0.137 & 0.039 \\
B $^2\Sigma^+$-X $^2\Sigma^+$ & (1-2) & 0.5 - 13.5 & 86/86 & 23829.2 - 23926.1 & 0.03 & 0.037 & 0.139 & 0.042 \\
B $^2\Sigma^+$-X $^2\Sigma^+$ & (10-10) & 0.5 - 24.5 & 94/94 & 25576.5 - 25741.9 & 0.03 & 0.03 & 0.045 & 0.03 \\
B $^2\Sigma^+$-X $^2\Sigma^+$ & (10-11) & 0.5 - 11.5 & 41/41 & 23858.0 - 23933.4 & 0.03 & 0.03 & 0.03 & 0.03 \\
B $^2\Sigma^+$-X $^2\Sigma^+$ & (10-12) & 1.5 - 15.5 & 45/45 & 22094.1 - 22182.5 & 0.03 & 0.032 & 0.095 & 0.033 \\
B $^2\Sigma^+$-X $^2\Sigma^+$ & (10-8) & 0.5 - 14.5 & 54/54 & 29243.9 - 29346.1 & 0.03 & 0.03 & 0.03 & 0.03 \\
B $^2\Sigma^+$-X $^2\Sigma^+$ & (10-9) & 0.5 - 21.5 & 75/75 & 27385.0 - 27525.3 & 0.03 & 0.03 & 0.03 & 0.03 \\
B $^2\Sigma^+$-X $^2\Sigma^+$ & (11-10) & 0.5 - 20.5 & 80/80 & 27182.3 - 27322.6 & 0.03 & 0.04 & 0.185 & 0.05 \\
B $^2\Sigma^+$-X $^2\Sigma^+$ & (11-11) & 0.5 - 23.5 & 114/114 & 25403.9 - 25559.4 & 0.03 & 0.048 & 0.5 & 0.086 \\
B $^2\Sigma^+$-X $^2\Sigma^+$ & (11-12) & 0.5 - 17.5 & 63/63 & 23688.6 - 23801.9 & 0.03 & 0.037 & 0.154 & 0.046 \\
B $^2\Sigma^+$-X $^2\Sigma^+$ & (11-13) & 0.5 - 20.5 & 80/80 & 21954.9 - 22090.5 & 0.03 & 0.035 & 0.14 & 0.039 \\
B $^2\Sigma^+$-X $^2\Sigma^+$ & (11-9) & 0.5 - 19.5 & 76/76 & 28984.4 - 29119.3 & 0.03 & 0.038 & 0.175 & 0.048 \\
B $^2\Sigma^+$-X $^2\Sigma^+$ & (12-10) & 0.5 - 15.5 & 56/56 & 28741.5 - 28846.7 & 0.03 & 0.031 & 0.062 & 0.031 \\
B $^2\Sigma^+$-X $^2\Sigma^+$ & (12-11) & 0.5 - 15.5 & 54/54 & 26969.3 - 27073.2 & 0.03 & 0.031 & 0.072 & 0.032 \\
B $^2\Sigma^+$-X $^2\Sigma^+$ & (12-12) & 0.5 - 15.5 & 115/115 & 25221.4 - 25326.6 & 0.03 & 0.03 & 0.03 & 0.03 \\
B $^2\Sigma^+$-X $^2\Sigma^+$ & (12-13) & 0.5 - 15.5 & 56/56 & 23512.5 - 23606.7 & 0.03 & 0.031 & 0.072 & 0.032 \\
B $^2\Sigma^+$-X $^2\Sigma^+$ & (12-14) & 0.5 - 15.5 & 60/60 & 21811.6 - 21913.6 & 0.03 & 0.042 & 0.166 & 0.05 \\
B $^2\Sigma^+$-X $^2\Sigma^+$ & (13-11) & 0.5 - 13.5 & 49/49 & 28464.9 - 28550.4 & 0.03 & 0.031 & 0.058 & 0.032 \\
B $^2\Sigma^+$-X $^2\Sigma^+$ & (13-13) & 0.5 - 21.5 & 117/117 & 24958.4 - 25088.8 & 0.03 & 0.03 & 0.059 & 0.03 \\
B $^2\Sigma^+$-X $^2\Sigma^+$ & (14-14) & 0.5 - 18.5 & 70/70 & 24689.5 - 24810.6 & 0.03 & 0.054 & 0.614 & 0.1 \\
B $^2\Sigma^+$-X $^2\Sigma^+$ & (15-15) & 1.5 - 15.5 & 52/52 & 24409.8 - 24509.6 & 0.03 & 0.031 & 0.068 & 0.032 \\
B $^2\Sigma^+$-X $^2\Sigma^+$ & (16-13) & 0.5 - 24.5 & 68/68 & 29121.6 - 29191.2 & 0.03 & 0.046 & 0.231 & 0.058 \\
B $^2\Sigma^+$-X $^2\Sigma^+$ & (17-14) & 2.5 - 30.5 & 78/85 & 28559.0 - 28772.5 & 0.03 & 0.079 & 0.428 & 0.113 \\
B $^2\Sigma^+$-X $^2\Sigma^+$ & (17-16) & 1.5 - 34.5 & 98/100 & 25327.6 - 25462.7 & 0.03 & 0.061 & 0.188 & 0.077 \\
B $^2\Sigma^+$-X $^2\Sigma^+$ & (18-17) & 0.5 - 21.5 & 74/74 & 24923.4 - 25078.3 & 0.03 & 0.035 & 0.143 & 0.038 \\
B $^2\Sigma^+$-X $^2\Sigma^+$ & (18-18) & 0.5 - 21.5 & 74/84 & 23344.8 - 23491.7 & 0.03 & 0.048 & 0.343 & 0.071 \\
B $^2\Sigma^+$-X $^2\Sigma^+$ & (19-15) & 0.5 - 25.5 & 42/44 & 29379.7 - 29525.6 & 0.03 & 0.053 & 0.168 & 0.064 \\
B $^2\Sigma^+$-X $^2\Sigma^+$ & (19-18) & 0.5 - 21.5 & 38/41 & 24514.3 - 24681.3 & 0.03 & 0.06 & 0.19 & 0.074 \\
B $^2\Sigma^+$-X $^2\Sigma^+$ & (2-1) & 0.5 - 9.5 & 29/29 & 27938.0 - 28000.1 & 0.03 & 0.044 & 0.119 & 0.049 \\
B $^2\Sigma^+$-X $^2\Sigma^+$ & (2-2) & 0.5 - 23.5 & 158/158 & 25892.6 - 26064.6 & 0.03 & 0.032 & 0.066 & 0.033 \\
B $^2\Sigma^+$-X $^2\Sigma^+$ & (2-3) & 0.5 - 18.5 & 106/106 & 23913.3 - 24030.4 & 0.03 & 0.031 & 0.087 & 0.032 \\
B $^2\Sigma^+$-X $^2\Sigma^+$ & (3-2) & 0.5 - 9.5 & 57/57 & 27953.6 - 28024.3 & 0.03 & 0.034 & 0.1 & 0.036 \\
B $^2\Sigma^+$-X $^2\Sigma^+$ & (3-3) & 0.5 - 18.5 & 114/114 & 25954.3 - 26082.7 & 0.03 & 0.031 & 0.065 & 0.031 \\
B $^2\Sigma^+$-X $^2\Sigma^+$ & (3-4) & 0.5 - 17.5 & 82/82 & 23991.0 - 24084.0 & 0.03 & 0.031 & 0.066 & 0.031 \\
B $^2\Sigma^+$-X $^2\Sigma^+$ & (4-3) & 0.5 - 21.5 & 98/98 & 27945.1 - 28085.2 & 0.03 & 0.04 & 0.141 & 0.048 \\
B $^2\Sigma^+$-X $^2\Sigma^+$ & (4-4) & 0.5 - 23.5 & 149/149 & 25976.1 - 26140.8 & 0.03 & 0.032 & 0.258 & 0.038 \\
B $^2\Sigma^+$-X $^2\Sigma^+$ & (4-5) & 0.5 - 18.5 & 104/104 & 24047.5 - 24180.9 & 0.03 & 0.03 & 0.03 & 0.03 \\
B $^2\Sigma^+$-X $^2\Sigma^+$ & (4-6) & 0.5 - 17.5 & 61/61 & 22144.3 - 22256.5 & 0.03 & 0.03 & 0.03 & 0.03 \\
B $^2\Sigma^+$-X $^2\Sigma^+$ & (5-4) & 0.5 - 21.5 & 100/100 & 27910.2 - 28066.3 & 0.03 & 0.038 & 0.14 & 0.045 \\
B $^2\Sigma^+$-X $^2\Sigma^+$ & (5-5) & 0.5 - 23.5 & 133/133 & 25983.4 - 26147.5 & 0.03 & 0.033 & 0.127 & 0.035 \\
B $^2\Sigma^+$-X $^2\Sigma^+$ & (5-6) & 0.5 - 20.5 & 74/74 & 24082.2 - 24227.6 & 0.03 & 0.031 & 0.09 & 0.032 \\
B $^2\Sigma^+$-X $^2\Sigma^+$ & (5-7) & 0.5 - 19.5 & 71/71 & 22206.3 - 22343.8 & 0.03 & 0.03 & 0.041 & 0.03 \\
B $^2\Sigma^+$-X $^2\Sigma^+$ & (6-5) & 0.5 - 24.5 & 113/113 & 27857.7 - 28000.6 & 0.03 & 0.034 & 0.147 & 0.037 \\
B $^2\Sigma^+$-X $^2\Sigma^+$ & (6-6) & 0.5 - 25.5 & 127/127 & 25978.9 - 26138.3 & 0.03 & 0.03 & 0.058 & 0.03 \\
B $^2\Sigma^+$-X $^2\Sigma^+$ & (6-7) & 0.5 - 21.5 & 108/108 & 24097.8 - 24241.6 & 0.03 & 0.03 & 0.038 & 0.03 \\
B $^2\Sigma^+$-X $^2\Sigma^+$ & (6-8) & 0.5 - 18.5 & 46/46 & 22257.7 - 22373.1 & 0.03 & 0.03 & 0.043 & 0.03 \\
B $^2\Sigma^+$-X $^2\Sigma^+$ & (7-10) & 0.5 - 16.5 & 60/60 & 20459.1 - 20569.8 & 0.03 & 0.034 & 0.099 & 0.036 \\
B $^2\Sigma^+$-X $^2\Sigma^+$ & (7-6) & 0.5 - 18.5 & 128/128 & 27804.6 - 27929.2 & 0.03 & 0.035 & 0.105 & 0.037 \\
B $^2\Sigma^+$-X $^2\Sigma^+$ & (7-7) & 0.5 - 17.5 & 108/108 & 25933.0 - 26053.7 & 0.03 & 0.031 & 0.078 & 0.032 \\
B $^2\Sigma^+$-X $^2\Sigma^+$ & (7-8) & 0.5 - 19.5 & 100/100 & 24078.7 - 24210.4 & 0.03 & 0.034 & 0.173 & 0.04 \\
B $^2\Sigma^+$-X $^2\Sigma^+$ & (7-9) & 0.5 - 17.5 & 64/64 & 22257.3 - 22375.0 & 0.03 & 0.033 & 0.101 & 0.035 \\
B $^2\Sigma^+$-X $^2\Sigma^+$ & (8-10) & 0.5 - 16.5 & 56/56 & 22235.2 - 22333.7 & 0.03 & 0.031 & 0.055 & 0.031 \\
B $^2\Sigma^+$-X $^2\Sigma^+$ & (8-7) & 0.5 - 26.5 & 159/159 & 27672.8 - 27847.4 & 0.03 & 0.032 & 0.087 & 0.033 \\
B $^2\Sigma^+$-X $^2\Sigma^+$ & (8-8) & 0.5 - 22.5 & 97/97 & 25840.8 - 25982.7 & 0.03 & 0.03 & 0.048 & 0.03 \\
B $^2\Sigma^+$-X $^2\Sigma^+$ & (8-9) & 0.5 - 18.5 & 82/82 & 24029.0 - 24148.0 & 0.03 & 0.03 & 0.042 & 0.03 \\
B $^2\Sigma^+$-X $^2\Sigma^+$ & (9-10) & 0.5 - 26.5 & 120/120 & 23924.5 - 24081.7 & 0.03 & 0.033 & 0.109 & 0.035 \\
B $^2\Sigma^+$-X $^2\Sigma^+$ & (9-11) & 0.5 - 21.5 & 82/82 & 22172.6 - 22320.5 & 0.03 & 0.054 & 0.5 & 0.101 \\
B $^2\Sigma^+$-X $^2\Sigma^+$ & (9-12) & 0.5 - 15.5 & 51/51 & 20441.5 - 20540.7 & 0.03 & 0.033 & 0.122 & 0.036 \\
B $^2\Sigma^+$-X $^2\Sigma^+$ & (9-7) & 0.5 - 18.5 & 70/70 & 29424.3 - 29549.9 & 0.03 & 0.035 & 0.134 & 0.04 \\
B $^2\Sigma^+$-X $^2\Sigma^+$ & (9-8) & 0.5 - 26.5 & 190/190 & 27532.1 - 27716.8 & 0.03 & 0.041 & 0.169 & 0.049 \\
B $^2\Sigma^+$-X $^2\Sigma^+$ & (9-9) & 0.5 - 26.5 & 122/122 & 25714.8 - 25885.4 & 0.03 & 0.032 & 0.111 & 0.035 \\
X $^2\Sigma^+$-X $^2\Sigma^+$ & (0-0) & 0.5 - 9.5 & 19/19 & 3.8 - 34.0 & 0.03 & 0.03 & 0.03 & 0.03 \\
X $^2\Sigma^+$-X $^2\Sigma^+$ & (1-0) & 0.5 - 66.5 & 344/344 & 1797.7 - 2207.5 & 0.03 & 0.031 & 0.062 & 0.031 \\
X $^2\Sigma^+$-X $^2\Sigma^+$ & (1-1) & 0.5 - 8.5 & 13/13 & 3.7 - 30.0 & 0.03 & 0.032 & 0.042 & 0.032 \\
X $^2\Sigma^+$-X $^2\Sigma^+$ & (10-10) & 1.5 - 3.5 & 2/2 & 10.3 - 10.3 & 0.03 & 0.03 & 0.03 & 0.03 \\
X $^2\Sigma^+$-X $^2\Sigma^+$ & (2-0) & 2.5 - 57.5 & 156/156 & 3743.5 - 4155.4 & 0.03 & 0.03 & 0.045 & 0.03 \\
X $^2\Sigma^+$-X $^2\Sigma^+$ & (2-1) & 0.5 - 53.5 & 278/278 & 1815.2 - 2160.8 & 0.03 & 0.031 & 0.078 & 0.031 \\
X $^2\Sigma^+$-X $^2\Sigma^+$ & (2-2) & 0.5 - 7.5 & 11/11 & 3.7 - 29.7 & 0.03 & 0.038 & 0.083 & 0.042 \\
X $^2\Sigma^+$-X $^2\Sigma^+$ & (3-1) & 0.5 - 59.5 & 144/144 & 3796.6 - 4100.6 & 0.03 & 0.031 & 0.105 & 0.032 \\
X $^2\Sigma^+$-X $^2\Sigma^+$ & (3-2) & 0.5 - 79.5 & 228/228 & 1830.4 - 2159.2 & 0.03 & 0.03 & 0.059 & 0.03 \\
X $^2\Sigma^+$-X $^2\Sigma^+$ & (3-3) & 0.5 - 6.5 & 7/7 & 3.7 - 22.1 & 0.03 & 0.03 & 0.03 & 0.03 \\
X $^2\Sigma^+$-X $^2\Sigma^+$ & (4-2) & 0.5 - 39.5 & 82/82 & 3751.5 - 4026.4 & 0.03 & 0.03 & 0.03 & 0.03 \\
X $^2\Sigma^+$-X $^2\Sigma^+$ & (4-3) & 0.5 - 46.5 & 182/182 & 1852.3 - 2092.0 & 0.03 & 0.03 & 0.036 & 0.03 \\
X $^2\Sigma^+$-X $^2\Sigma^+$ & (4-4) & 0.5 - 6.5 & 6/6 & 3.6 - 21.9 & 0.03 & 0.03 & 0.03 & 0.03 \\
X $^2\Sigma^+$-X $^2\Sigma^+$ & (5-4) & 0.5 - 26.5 & 92/92 & 1853.4 - 2018.9 & 0.03 & 0.031 & 0.065 & 0.031 \\
X $^2\Sigma^+$-X $^2\Sigma^+$ & (5-5) & 0.5 - 6.5 & 6/6 & 3.6 - 21.6 & 0.03 & 0.03 & 0.03 & 0.03 \\
X $^2\Sigma^+$-X $^2\Sigma^+$ & (6-5) & 0.5 - 25.5 & 76/76 & 1844.5 - 1988.8 & 0.03 & 0.03 & 0.03 & 0.03 \\
X $^2\Sigma^+$-X $^2\Sigma^+$ & (6-6) & 0.5 - 6.5 & 6/6 & 3.6 - 21.4 & 0.03 & 0.03 & 0.03 & 0.03 \\
X $^2\Sigma^+$-X $^2\Sigma^+$ & (7-6) & 0.5 - 21.5 & 67/67 & 1830.9 - 1950.6 & 0.03 & 0.03 & 0.03 & 0.03 \\
X $^2\Sigma^+$-X $^2\Sigma^+$ & (7-7) & 0.5 - 6.5 & 6/6 & 3.5 - 21.2 & 0.03 & 0.03 & 0.03 & 0.03 \\
X $^2\Sigma^+$-X $^2\Sigma^+$ & (8-7) & 0.5 - 18.5 & 56/56 & 1816.7 - 1914.9 & 0.03 & 0.031 & 0.074 & 0.032 \\
X $^2\Sigma^+$-X $^2\Sigma^+$ & (8-8) & 0.5 - 3.5 & 4/4 & 3.5 - 10.5 & 0.03 & 0.03 & 0.03 & 0.03 \\
X $^2\Sigma^+$-X $^2\Sigma^+$ & (9-9) & 0.5 - 3.5 & 3/3 & 3.5 - 10.4 & 0.03 & 0.03 & 0.03 & 0.03 \\
\\ \mc{5}{l}{\textbf{08CiSeKu }} \\
A $^2\Pi$-X $^2\Sigma^+$ & (0-3) & 1.5 - 25.5 & 113/113 & 2950.1 - 3117.2 & 0.025 & 0.025 & 0.025 & 0.025 \\
A $^2\Pi$-X $^2\Sigma^+$ & (1-4) & 1.5 - 31.5 & 141/141 & 2699.2 - 2929.4 & 0.025 & 0.025 & 0.041 & 0.025 \\
A $^2\Pi$-X $^2\Sigma^+$ & (2-5) & 1.5 - 27.5 & 145/145 & 2549.5 - 2779.0 & 0.025 & 0.025 & 0.025 & 0.025 \\
A $^2\Pi$-X $^2\Sigma^+$ & (3-6) & 1.5 - 25.5 & 137/137 & 2411.7 - 2580.0 & 0.025 & 0.025 & 0.025 & 0.025 \\
A $^2\Pi$-X $^2\Sigma^+$ & (4-7) & 1.5 - 28.5 & 118/118 & 2204.4 - 2406.7 & 0.025 & 0.025 & 0.025 & 0.025 \\
A $^2\Pi$-X $^2\Sigma^+$ & (5-8) & 1.5 - 25.5 & 123/123 & 2068.2 - 2234.3 & 0.025 & 0.025 & 0.025 & 0.025 \\
A $^2\Pi$-X $^2\Sigma^+$ & (6-9) & 1.5 - 22.5 & 93/93 & 1905.8 - 2062.9 & 0.025 & 0.025 & 0.025 & 0.025 \\
\\ \mc{5}{l}{\textbf{09HaHaSe }} \\
A $^2\Pi$-X $^2\Sigma^+$ & (1-0) & 0.5 - 7.5 & 38/38 & 10850.0 - 10937.3 & 0.002 & 0.009 & 0.045 & 0.015 \\
\\ \mc{5}{l}{\textbf{10RaWaBe}} \\
A $^2\Pi$-X $^2\Sigma^+$ & (0-0) & 0.5 - 104.5 & 2092/2092 & 7043.6 - 9195.8 & 0.012 & 0.012 & 0.072 & 0.013 \\
A $^2\Pi$-X $^2\Sigma^+$ & (0-1) & 0.5 - 95.5 & 898/898 & 5334.5 - 7159.2 & 0.012 & 0.012 & 0.047 & 0.013 \\
A $^2\Pi$-X $^2\Sigma^+$ & (0-2) & 0.5 - 85.5 & 887/888 & 4119.2 - 5150.4 & 0.012 & 0.012 & 0.051 & 0.013 \\
A $^2\Pi$-X $^2\Sigma^+$ & (1-0) & 0.5 - 106.5 & 636/636 & 8611.7 - 10977.7 & 0.012 & 0.013 & 0.051 & 0.013 \\
A $^2\Pi$-X $^2\Sigma^+$ & (1-1) & 0.5 - 28.5 & 227/227 & 8623.2 - 8937.7 & 0.012 & 0.015 & 0.057 & 0.017 \\
A $^2\Pi$-X $^2\Sigma^+$ & (1-2) & 0.5 - 95.5 & 783/783 & 5117.4 - 6929.2 & 0.012 & 0.013 & 0.078 & 0.014 \\
A $^2\Pi$-X $^2\Sigma^+$ & (1-3) & 0.5 - 75.5 & 644/644 & 3724.8 - 4946.9 & 0.012 & 0.013 & 0.046 & 0.014 \\
A $^2\Pi$-X $^2\Sigma^+$ & (10-5) & 0.5 - 39.5 & 281/281 & 15413.9 - 15946.3 & 0.012 & 0.012 & 0.025 & 0.012 \\
A $^2\Pi$-X $^2\Sigma^+$ & (10-6) & 0.5 - 33.5 & 205/205 & 13663.8 - 14037.5 & 0.012 & 0.012 & 0.043 & 0.013 \\
A $^2\Pi$-X $^2\Sigma^+$ & (11-6) & 0.5 - 19.5 & 134/134 & 15362.7 - 15566.1 & 0.012 & 0.012 & 0.021 & 0.012 \\
A $^2\Pi$-X $^2\Sigma^+$ & (12-7) & 0.5 - 22.5 & 93/93 & 14986.1 - 15186.3 & 0.012 & 0.012 & 0.026 & 0.013 \\
A $^2\Pi$-X $^2\Sigma^+$ & (13-7) & 0.5 - 21.5 & 116/116 & 16468.5 - 16641.7 & 0.012 & 0.012 & 0.037 & 0.013 \\
A $^2\Pi$-X $^2\Sigma^+$ & (14-6) & 0.5 - 16.5 & 57/57 & 19848.8 - 19978.9 & 0.012 & 0.012 & 0.019 & 0.012 \\
A $^2\Pi$-X $^2\Sigma^+$ & (14-7) & 0.5 - 20.5 & 110/110 & 17881.7 - 18095.0 & 0.012 & 0.012 & 0.015 & 0.012 \\
A $^2\Pi$-X $^2\Sigma^+$ & (15-7) & 0.5 - 20.5 & 105/105 & 19344.9 - 19521.6 & 0.012 & 0.012 & 0.016 & 0.012 \\
A $^2\Pi$-X $^2\Sigma^+$ & (15-8) & 0.5 - 23.5 & 142/142 & 17458.5 - 17664.2 & 0.012 & 0.012 & 0.015 & 0.012 \\
A $^2\Pi$-X $^2\Sigma^+$ & (16-7) & 0.5 - 19.5 & 99/99 & 20751.0 - 20921.4 & 0.012 & 0.012 & 0.023 & 0.012 \\
A $^2\Pi$-X $^2\Sigma^+$ & (16-8) & 0.5 - 24.5 & 134/134 & 18830.0 - 19064.4 & 0.012 & 0.013 & 0.066 & 0.014 \\
A $^2\Pi$-X $^2\Sigma^+$ & (17-10) & 0.5 - 20.5 & 79/80 & 16632.3 - 16803.2 & 0.012 & 0.015 & 0.059 & 0.017 \\
A $^2\Pi$-X $^2\Sigma^+$ & (17-8) & 0.5 - 22.5 & 90/91 & 20224.6 - 20437.6 & 0.012 & 0.014 & 0.059 & 0.016 \\
A $^2\Pi$-X $^2\Sigma^+$ & (18-10) & 0.5 - 22.5 & 150/150 & 17945.5 - 18166.5 & 0.012 & 0.014 & 0.051 & 0.015 \\
A $^2\Pi$-X $^2\Sigma^+$ & (18-9) & 0.5 - 23.5 & 115/115 & 19717.9 - 19954.2 & 0.012 & 0.013 & 0.028 & 0.013 \\
A $^2\Pi$-X $^2\Sigma^+$ & (19-10) & 0.5 - 22.5 & 127/128 & 19256.1 - 19469.9 & 0.012 & 0.018 & 0.429 & 0.042 \\
A $^2\Pi$-X $^2\Sigma^+$ & (19-11) & 0.5 - 19.5 & 137/141 & 17491.2 - 17708.6 & 0.012 & 0.018 & 0.395 & 0.039 \\
A $^2\Pi$-X $^2\Sigma^+$ & (2-0) & 0.5 - 36.5 & 338/339 & 12249.0 - 12735.0 & 0.012 & 0.013 & 0.06 & 0.014 \\
A $^2\Pi$-X $^2\Sigma^+$ & (2-1) & 0.5 - 32.5 & 283/283 & 10358.9 - 10696.4 & 0.012 & 0.013 & 0.045 & 0.013 \\
A $^2\Pi$-X $^2\Sigma^+$ & (2-2) & 0.5 - 24.5 & 140/140 & 8442.3 - 8679.9 & 0.012 & 0.012 & 0.025 & 0.012 \\
A $^2\Pi$-X $^2\Sigma^+$ & (2-3) & 0.5 - 82.5 & 624/624 & 5292.1 - 6700.6 & 0.012 & 0.012 & 0.023 & 0.012 \\
A $^2\Pi$-X $^2\Sigma^+$ & (2-4) & 0.5 - 72.5 & 588/588 & 3806.3 - 4744.3 & 0.012 & 0.012 & 0.026 & 0.012 \\
A $^2\Pi$-X $^2\Sigma^+$ & (20-10) & 0.5 - 19.5 & 105/105 & 20565.1 - 20762.6 & 0.012 & 0.014 & 0.052 & 0.015 \\
A $^2\Pi$-X $^2\Sigma^+$ & (21-10) & 0.5 - 16.5 & 87/87 & 21883.1 - 22027.6 & 0.012 & 0.014 & 0.067 & 0.016 \\
A $^2\Pi$-X $^2\Sigma^+$ & (21-11) & 0.5 - 21.5 & 99/100 & 20058.2 - 20250.4 & 0.012 & 0.015 & 0.067 & 0.017 \\
A $^2\Pi$-X $^2\Sigma^+$ & (22-11) & 0.5 - 19.5 & 37/37 & 21302.5 - 21483.5 & 0.012 & 0.013 & 0.059 & 0.015 \\
A $^2\Pi$-X $^2\Sigma^+$ & (22-12) & 0.5 - 20.5 & 122/122 & 19556.5 - 19735.1 & 0.012 & 0.013 & 0.03 & 0.013 \\
A $^2\Pi$-X $^2\Sigma^+$ & (3-0) & 0.5 - 36.5 & 270/270 & 14076.6 - 14467.6 & 0.012 & 0.013 & 0.047 & 0.014 \\
A $^2\Pi$-X $^2\Sigma^+$ & (3-1) & 0.5 - 113.5 & 656/656 & 9483.7 - 12428.2 & 0.012 & 0.012 & 0.043 & 0.013 \\
A $^2\Pi$-X $^2\Sigma^+$ & (3-2) & 0.5 - 85.5 & 472/472 & 8995.6 - 10415.7 & 0.012 & 0.013 & 0.091 & 0.014 \\
A $^2\Pi$-X $^2\Sigma^+$ & (3-3) & 1.5 - 32.5 & 288/288 & 8062.4 - 8430.4 & 0.012 & 0.012 & 0.033 & 0.012 \\
A $^2\Pi$-X $^2\Sigma^+$ & (3-4) & 0.5 - 62.5 & 436/436 & 5597.1 - 6472.5 & 0.012 & 0.012 & 0.027 & 0.012 \\
A $^2\Pi$-X $^2\Sigma^+$ & (3-5) & 0.5 - 61.5 & 421/421 & 3854.4 - 4542.4 & 0.012 & 0.012 & 0.015 & 0.012 \\
A $^2\Pi$-X $^2\Sigma^+$ & (4-0) & 0.5 - 30.5 & 216/216 & 15859.3 - 16174.1 & 0.012 & 0.013 & 0.049 & 0.014 \\
A $^2\Pi$-X $^2\Sigma^+$ & (4-1) & 0.5 - 39.5 & 314/314 & 13691.9 - 14135.3 & 0.012 & 0.013 & 0.042 & 0.013 \\
A $^2\Pi$-X $^2\Sigma^+$ & (4-2) & 0.5 - 109.5 & 642/642 & 9467.0 - 12122.1 & 0.012 & 0.012 & 0.073 & 0.013 \\
A $^2\Pi$-X $^2\Sigma^+$ & (4-4) & 0.5 - 38.5 & 314/314 & 7780.0 - 8176.2 & 0.012 & 0.012 & 0.036 & 0.012 \\
A $^2\Pi$-X $^2\Sigma^+$ & (4-5) & 0.5 - 27.5 & 238/238 & 5943.0 - 6242.9 & 0.012 & 0.012 & 0.023 & 0.012 \\
A $^2\Pi$-X $^2\Sigma^+$ & (4-6) & 0.5 - 36.5 & 303/303 & 3975.5 - 4341.3 & 0.012 & 0.012 & 0.015 & 0.012 \\
A $^2\Pi$-X $^2\Sigma^+$ & (5-1) & 0.5 - 94.5 & 525/525 & 13365.7 - 15817.3 & 0.012 & 0.013 & 0.045 & 0.013 \\
A $^2\Pi$-X $^2\Sigma^+$ & (5-2) & 0.5 - 41.5 & 318/318 & 13253.5 - 13803.6 & 0.012 & 0.013 & 0.072 & 0.015 \\
A $^2\Pi$-X $^2\Sigma^+$ & (5-3) & 0.5 - 35.5 & 298/298 & 11409.6 - 11816.8 & 0.012 & 0.012 & 0.036 & 0.012 \\
A $^2\Pi$-X $^2\Sigma^+$ & (5-5) & 0.5 - 30.5 & 287/287 & 7575.4 - 7924.1 & 0.012 & 0.012 & 0.012 & 0.012 \\
A $^2\Pi$-X $^2\Sigma^+$ & (5-7) & 0.5 - 32.5 & 263/263 & 3825.6 - 4140.9 & 0.012 & 0.012 & 0.015 & 0.012 \\
A $^2\Pi$-X $^2\Sigma^+$ & (6-2) & 0.5 - 40.5 & 292/292 & 15023.2 - 15460.1 & 0.012 & 0.013 & 0.071 & 0.013 \\
A $^2\Pi$-X $^2\Sigma^+$ & (6-3) & 0.5 - 88.5 & 533/533 & 11524.3 - 13472.7 & 0.012 & 0.012 & 0.04 & 0.012 \\
A $^2\Pi$-X $^2\Sigma^+$ & (6-4) & 0.5 - 28.5 & 221/221 & 11248.4 - 11512.2 & 0.012 & 0.012 & 0.021 & 0.012 \\
A $^2\Pi$-X $^2\Sigma^+$ & (6-6) & 0.5 - 48.5 & 338/338 & 7157.6 - 7672.2 & 0.012 & 0.012 & 0.035 & 0.012 \\
A $^2\Pi$-X $^2\Sigma^+$ & (6-7) & 0.5 - 31.5 & 245/245 & 5483.6 - 5793.0 & 0.012 & 0.012 & 0.012 & 0.012 \\
A $^2\Pi$-X $^2\Sigma^+$ & (6-8) & 0.5 - 28.5 & 206/206 & 3683.4 - 3941.3 & 0.012 & 0.012 & 0.027 & 0.012 \\
A $^2\Pi$-X $^2\Sigma^+$ & (7-7) & 0.5 - 37.5 & 282/282 & 6985.8 - 7421.3 & 0.012 & 0.012 & 0.032 & 0.013 \\
A $^2\Pi$-X $^2\Sigma^+$ & (7-8) & 0.5 - 34.5 & 284/284 & 5181.2 - 5569.0 & 0.012 & 0.013 & 0.064 & 0.014 \\
A $^2\Pi$-X $^2\Sigma^+$ & (8-3) & 0.5 - 22.5 & 198/200 & 16451.2 - 16707.3 & 0.012 & 0.013 & 0.041 & 0.014 \\
A $^2\Pi$-X $^2\Sigma^+$ & (8-4) & 0.5 - 41.5 & 250/250 & 14246.2 - 14745.9 & 0.012 & 0.013 & 0.05 & 0.013 \\
A $^2\Pi$-X $^2\Sigma^+$ & (8-9) & 0.5 - 36.5 & 282/282 & 5016.8 - 5343.6 & 0.012 & 0.012 & 0.018 & 0.012 \\
A $^2\Pi$-X $^2\Sigma^+$ & (9-4) & 0.5 - 65.5 & 539/539 & 15022.5 - 16327.1 & 0.012 & 0.012 & 0.034 & 0.013 \\
X $^2\Sigma^+$-X $^2\Sigma^+$ & (0-0) & 0.5 - 9.5 & 19/19 & 3.8 - 34.0 & 0.012 & 0.012 & 0.015 & 0.012 \\
X $^2\Sigma^+$-X $^2\Sigma^+$ & (1-0) & 0.5 - 66.5 & 344/344 & 1797.7 - 2207.5 & 0.012 & 0.013 & 0.054 & 0.014 \\
X $^2\Sigma^+$-X $^2\Sigma^+$ & (1-1) & 0.5 - 8.5 & 13/13 & 3.7 - 30.0 & 0.012 & 0.017 & 0.042 & 0.02 \\
X $^2\Sigma^+$-X $^2\Sigma^+$ & (10-10) & 1.5 - 3.5 & 2/2 & 10.3 - 10.3 & 0.012 & 0.012 & 0.012 & 0.012 \\
X $^2\Sigma^+$-X $^2\Sigma^+$ & (2-0) & 2.5 - 57.5 & 156/156 & 3743.5 - 4155.4 & 0.012 & 0.013 & 0.043 & 0.014 \\
X $^2\Sigma^+$-X $^2\Sigma^+$ & (2-1) & 0.5 - 53.5 & 278/278 & 1815.2 - 2160.8 & 0.012 & 0.014 & 0.078 & 0.015 \\
X $^2\Sigma^+$-X $^2\Sigma^+$ & (2-2) & 0.5 - 7.5 & 11/11 & 3.7 - 29.7 & 0.012 & 0.025 & 0.083 & 0.035 \\
X $^2\Sigma^+$-X $^2\Sigma^+$ & (3-1) & 0.5 - 59.5 & 142/144 & 3796.6 - 4100.6 & 0.012 & 0.013 & 0.035 & 0.013 \\
X $^2\Sigma^+$-X $^2\Sigma^+$ & (3-2) & 0.5 - 79.5 & 228/228 & 1830.4 - 2159.2 & 0.012 & 0.013 & 0.059 & 0.014 \\
X $^2\Sigma^+$-X $^2\Sigma^+$ & (3-3) & 0.5 - 6.5 & 7/7 & 3.7 - 22.1 & 0.012 & 0.012 & 0.012 & 0.012 \\
X $^2\Sigma^+$-X $^2\Sigma^+$ & (4-2) & 0.5 - 39.5 & 82/82 & 3751.5 - 4026.4 & 0.012 & 0.013 & 0.027 & 0.013 \\
X $^2\Sigma^+$-X $^2\Sigma^+$ & (4-3) & 0.5 - 46.5 & 182/182 & 1852.3 - 2092.0 & 0.012 & 0.013 & 0.033 & 0.013 \\
X $^2\Sigma^+$-X $^2\Sigma^+$ & (4-4) & 0.5 - 6.5 & 6/6 & 3.6 - 21.9 & 0.012 & 0.012 & 0.012 & 0.012 \\
X $^2\Sigma^+$-X $^2\Sigma^+$ & (5-4) & 0.5 - 26.5 & 92/92 & 1853.4 - 2018.9 & 0.012 & 0.014 & 0.061 & 0.016 \\
X $^2\Sigma^+$-X $^2\Sigma^+$ & (5-5) & 0.5 - 6.5 & 6/6 & 3.6 - 21.6 & 0.012 & 0.012 & 0.012 & 0.012 \\
X $^2\Sigma^+$-X $^2\Sigma^+$ & (6-5) & 0.5 - 25.5 & 76/76 & 1844.5 - 1988.8 & 0.012 & 0.012 & 0.023 & 0.012 \\
X $^2\Sigma^+$-X $^2\Sigma^+$ & (6-6) & 0.5 - 6.5 & 6/6 & 3.6 - 21.4 & 0.012 & 0.012 & 0.012 & 0.012 \\
X $^2\Sigma^+$-X $^2\Sigma^+$ & (7-6) & 0.5 - 21.5 & 67/67 & 1830.9 - 1950.6 & 0.012 & 0.013 & 0.03 & 0.013 \\
X $^2\Sigma^+$-X $^2\Sigma^+$ & (7-7) & 0.5 - 6.5 & 6/6 & 3.5 - 21.2 & 0.012 & 0.012 & 0.012 & 0.012 \\
X $^2\Sigma^+$-X $^2\Sigma^+$ & (8-7) & 0.5 - 18.5 & 56/56 & 1816.7 - 1914.9 & 0.012 & 0.016 & 0.069 & 0.019 \\
X $^2\Sigma^+$-X $^2\Sigma^+$ & (8-8) & 0.5 - 3.5 & 4/4 & 3.5 - 10.5 & 0.012 & 0.012 & 0.012 & 0.012 \\
X $^2\Sigma^+$-X $^2\Sigma^+$ & (9-9) & 0.5 - 3.5 & 3/3 & 3.5 - 10.4 & 0.012 & 0.012 & 0.012 & 0.012 \\
\end{longtable}
\end{center}

\begin{table*}[]
\centering
\caption{Vibronic comparison of the difference between the energy levels in both MoLLIST and \Marvel{}}
\begin{tabular}{p{1.2cm}p{.8cm}p{1.8cm}p{2.8cm}p{1.cm}p{1.cm}p{1.cm}p{1.cm}}
\toprule
State & Vib & J range & E range & \mc{4}{c}{\Marvel{} - MoLLIST} \\
\cmidrule(r){5-8}
& & & & \mc{1}{c}{Min} & \mc{1}{c}{MAD} & \mc{1}{c}{Max} & RMSD\\ \midrule
\X  & 0 & 0.5-97.5 & 0.0-17403.4 & 0.0 & 0.015 & 0.05 & 0.017 \\
\X  & 1 & 0.5-99.5 & 2042.4-20317.7 & 0.0 & 0.015 & 0.047 & 0.017 \\
\X  & 4 & 0.5-72.5 & 8011.8-17661.4 & 0.002 & 0.013 & 0.046 & 0.015 \\
\X  & 5 & 0.5-60.5 & 9948.8-16676.8 & 0.0 & 0.011 & 0.028 & 0.013 \\
\X  & 6 & 0.5-48.5 & 11859.3-16195.3 & 0.001 & 0.008 & 0.048 & 0.009 \\
\X  & 7 & 0.5-36.5 & 13743.4-16086.8 & 0.001 & 0.008 & 0.031 & 0.009 \\
\X  & 8 & 0.5-34.5 & 15600.9-17674.3 & 0.0 & 0.009 & 0.035 & 0.01 \\
\X  & 9 & 0.5-30.5 & 17431.8-19143.5 & 0.001 & 0.01 & 0.02 & 0.011 \\
\X  & 10 & 0.5-27.5 & 19236.0-20528.0 & 0.001 & 0.013 & 0.069 & 0.017 \\
\X  & 11 & 0.5-30.5 & 21013.3-22689.9 & 0.0 & 0.876 & 5.934 & 1.888 \\
\X  & 12 & 0.5-19.5 & 22765.7-23402.5 & 0.0 & 0.011 & 0.051 & 0.016 \\
\X  & 13 & 0.5-23.5 & 24488.7-25403.2 & 0.001 & 0.039 & 0.518 & 0.1 \\
\X  & 14 & 0.5-25.5 & 26185.7-27333.8 & 0.0 & 0.091 & 0.646 & 0.173 \\
\X  & 15 & 0.5-22.5 & 27856.2-28676.2 & 0.002 & 0.11 & 0.543 & 0.155 \\
\X  & 2 & 0.5-97.5 & 4058.5-21127.9 & 0.0 & 0.015 & 0.068 & 0.018 \\
\X  & 3 & 0.5-81.5 & 6048.3-17978.5 & 0.001 & 0.014 & 0.035 & 0.015 \\
\\
\A  & 0 & 0.5-98.5 & 9094.3-25099.4 & 0.0 & 0.015 & 0.055 & 0.017 \\
\A  & 1 & 0.5-71.5 & 10882.0-19619.9 & 0.0 & 0.014 & 0.028 & 0.015 \\
\A  & 2 & 0.5-71.5 & 12644.2-21290.6 & 0.001 & 0.013 & 0.041 & 0.015 \\
\A  & 3 & 0.5-99.5 & 14380.7-30495.4 & 0.0 & 0.039 & 1.137 & 0.107 \\
\A  & 4 & 0.5-56.5 & 16091.7-21464.5 & 0.0 & 0.022 & 0.98 & 0.084 \\
\A  & 5 & 0.5-90.5 & 17777.1-30656.7 & 0.0 & 0.1 & 3.142 & 0.302 \\
\A  & 6 & 0.5-81.5 & 19436.8-29827.2 & 0.0 & 0.166 & 3.83 & 0.409 \\
\A  & 7 & 0.5-37.5 & 21070.9-23301.8 & 0.001 & 0.5 & 4.819 & 1.152 \\
\A  & 8 & 0.5-41.5 & 22679.3-25386.9 & 0.0 & 0.332 & 7.608 & 1.047 \\
\A  & 9 & 0.5-65.5 & 24262.0-31016.2 & 0.0 & 0.027 & 0.334 & 0.06 \\
\A  & 10 & 0.5-39.5 & 25818.9-28341.6 & 0.0 & 0.011 & 0.279 & 0.028 \\
\A  & 11 & 0.5-19.5 & 27350.0-27956.4 & 0.0 & 0.007 & 0.018 & 0.008 \\
\A  & 12 & 0.5-22.5 & 28855.1-29622.4 & 0.0 & 0.008 & 0.025 & 0.009 \\
\A  & 13 & 0.5-21.5 & 30334.4-31047.1 & 0.0 & 0.007 & 0.027 & 0.009 \\
\A  & 14 & 0.5-20.5 & 31787.5-32410.5 & 0.0 & 0.007 & 0.022 & 0.009 \\
\A  & 15 & 0.5-23.5 & 33214.4-34107.2 & 0.0 & 0.009 & 0.188 & 0.022 \\
\A  & 16 & 0.5-24.5 & 34615.0-35478.6 & 0.0 & 0.008 & 0.042 & 0.011 \\
\A  & 17 & 0.5-22.5 & 35988.9-36709.4 & 0.0 & 0.094 & 0.602 & 0.157 \\
\A  & 18 & 0.5-23.5 & 37336.3-38195.2 & 0.0 & 0.019 & 0.267 & 0.042 \\
\A  & 19 & 0.5-22.5 & 38656.7-39357.8 & 0.0 & 0.01 & 0.058 & 0.013 \\
\A  & 20 & 0.5-19.5 & 39949.8-40545.7 & 0.0 & 0.01 & 0.036 & 0.012 \\
\A  & 21 & 0.5-21.5 & 41215.4-41837.7 & 0.0 & 0.013 & 0.125 & 0.022 \\
\A  & 22 & 0.5-20.5 & 42453.0-43011.2 & 0.0 & 0.008 & 0.043 & 0.011 \\
\\
\B  & 0 & 0.5-63.5 & 25797.9-33588.7 & 0.001 & 0.019 & 0.155 & 0.029 \\
\B  & 1 & 0.5-41.5 & 27921.5-31399.2 & 0.001 & 0.02 & 0.066 & 0.025 \\
\B  & 2 & 0.5-23.5 & 30004.9-31060.8 & 0.001 & 0.021 & 0.065 & 0.027 \\
\B  & 3 & 0.5-23.5 & 32045.9-33095.6 & 0.002 & 0.151 & 6.011 & 0.877 \\
\B  & 4 & 0.5-23.5 & 34042.0-35072.4 & 0.002 & 0.015 & 0.046 & 0.018 \\
\B  & 5 & 0.5-24.5 & 35990.0-37186.5 & 0.001 & 0.159 & 1.711 & 0.368 \\
\B  & 6 & 0.5-25.5 & 37887.4-39066.8 & 0.002 & 0.008 & 0.023 & 0.009 \\
\B  & 7 & 0.5-19.5 & 39730.5-40409.0 & 0.001 & 0.047 & 0.52 & 0.119 \\
\B  & 8 & 0.5-26.5 & 41516.6-42749.5 & 0.0 & 0.012 & 0.036 & 0.013 \\
\B  & 9 & 0.5-26.5 & 43243.0-44452.5 & 0.0 & 0.028 & 0.424 & 0.075 \\
\B  & 10 & 0.5-24.5 & 44908.8-45924.5 & 0.0 & 0.065 & 0.333 & 0.088 \\
\B  & 11 & 0.5-30.5 & 46511.4-48152.7 & 0.001 & 0.057 & 0.628 & 0.133 \\
\B  & 12 & 0.5-15.5 & 48053.7-48443.6 & 0.001 & 0.045 & 0.48 & 0.111 \\
\B  & 13 & 0.5-21.5 & 49537.3-50273.5 & 0.001 & 0.014 & 0.047 & 0.018 \\
\B  & 14 & 0.5-25.5 & 50967.7-52057.0 & 0.003 & 0.321 & 3.16 & 0.685 \\
\B  & 15 & 0.5-19.5 & 52343.0-52921.1 & 0.003 & 0.092 & 0.504 & 0.134 \\
\bottomrule
\end{tabular}
\label{tab:comp}
\end{table*}